%% file: paper20230509.tex
\documentclass[fleqn,usenatbib]{mnras}

\usepackage{newtxtext,newtxmath}
\usepackage[T1]{fontenc}

\DeclareRobustCommand{\VAN}[3]{#2}
\let\VANthebibliography\thebibliography
\def\thebibliography{\DeclareRobustCommand{\VAN}[3]{##3}\VANthebibliography}

\usepackage{acronym}
\usepackage{graphicx}	% Including figure files
\usepackage{xcolor}
\usepackage{ulem} % sout % REMOVE FOR FINAL VERSION!

\newcommand{\cdummy}{\cdot}
\newcommand{\mathd}{\mathrm{d}}
\newcommand{\tmabbr}[1]{#1}

\newcommand{\tmem}[1]{{\textit{#1}}}
\newcommand{\tmmathbf}[1]{\ensuremath{\boldsymbol{#1}}}

\newcommand{\tmop}[1]{\ensuremath{\operatorname{#1}}}
\newcommand{\tmtextit}[1]{\text{{\itshape{#1}}}}
\newcommand{\tmtexttt}[1]{\text{{\ttfamily{#1}}}}

\title[\texttt{THC\_M1}: Methods and First NS Merger Simulations]{%
A New Moment-Based General-Relativistic Neutrino-Radiation Transport
Code: Methods and First Applications to Neutron Star Mergers}

\author[D.~Radice et al.]{
David Radice$^{1,2,3}$\thanks{E-mail: david.radice@psu.edu},
Sebastiano Bernuzzi$^{4}$,
Albino Perego$^{5,6}$,
and Roland Haas$^{7}$
\\
$^{1}$ Institute for Gravitation \& the Cosmos, The Pennsylvania State University, University Park PA 16802, USA\\
$^{2}$ Department of Physics, The Pennsylvania State University, University Park PA 16802, USA\\
$^{3}$ Department of Astronomy \& Astrophysics,  The Pennsylvania State University, University Park PA 16802, USA\\
$^{4}$ Theoretisch-Physikalisches Institut, Friedrich-SchillerUniversit{\"a}t Jena, 07743, Jena, Germany\\
$^{5}$ Dipartimento di Fisica, Universit\`{a} di Trento, Via Sommarive 14,
  38123 Trento, Italy \\
$^{6}$ INFN-TIFPA, Trento Institute for Fundamental Physics and Applications, via Sommarive 14, I-38123 Trento, Italy\\
$^{7}$ National Center for Supercomputing Applications \& Department of Physics, University of Illinois, Urbana, IL 61801, USA\\
}

\date{Accepted XXX. Received YYY; in original form ZZZ}

\pubyear{2021}

\begin{document}
\label{firstpage}
\pagerange{\pageref{firstpage}--\pageref{lastpage}}
\maketitle

% Abstract of the paper - 250 words maximum
\begin{abstract}
We present a new moment-based energy-integrated neutrino transport
code for neutron star merger simulations in general relativity. In the
merger context, ours is the first code to include Doppler effects at all
orders in $\upsilon/c$, retaining all nonlinear neutrino-matter coupling
terms.  The code is validated with a stringent series of tests.  We show
that the inclusion of full neutrino-matter coupling terms is necessary
to correctly capture the trapping of neutrinos in relativistically
moving media, such as in differentially rotating merger remnants.  We
perform preliminary simulations proving the robustness of the scheme in
simulating ab-initio mergers to black hole collapse and long-term
neutron star remnants up to ${\sim}70\,$ms.  The latter is the longest
dynamical spacetime, 3D, general relativistic simulations with full
neutrino transport to date.  We compare results obtained at different
resolutions and using two different closures for the moment scheme.  We
do not find evidences of significant out-of-thermodynamic equilibrium
effects, such as bulk viscosity, on the postmerger dynamics or
gravitational wave emission. Neutrino luminosities and average energies
are in good agreement with theory expectations and previous simulations
by other groups using similar schemes.  We compare dynamical and early
wind ejecta properties obtained with M1 and with our older neutrino
treatment.  We find that the M1 results have systematically larger
proton fractions.  However, the differences in the nucleosynthesis
yields are modest.  This work sets the basis for future detailed studies
spanning a wider set of neutrino reactions, binaries and equations of
state.
\end{abstract}

\begin{keywords}
neutrinos -- methods: numerical -- stars: neutron
\end{keywords}

%%%%%%%%%%%%%%%%%%%%%%%%%%%%%%%%%%%%%%%%%%%%%%%%%%

%%%%%%%%%%%%%%%
% ------------------------------------------------
\section{Introduction}
% ------------------------------------------------
Neutrinos mediate the transport of energy and lepton number in dense and
hot environments. As such, neutrinos play a crucial role in powering the
explosion of massive stars as core-collapse supernovae
\citep{Melson:2015spa, Lentz:2015nxa, OConnor:2018tuw,Burrows:2019zce,
Burrows:2020qrp, Bollig:2020phc, Mezzacappa:2020oyq}, in the cooling of
the protoneutron star \citep{Roberts:2016rsf} and in the synthesis of
heavy elements neutrino-driven winds \citep{Arcones:2012wj}.  Neutrinos
also determine the composition and the final r-process nucleosynthesis
yields of the dynamical ejecta from \ac{NS} mergers
\citep{Sekiguchi:2015dma, Foucart:2015gaa, Radice:2016dwd,
Sekiguchi:2016bjd,Perego:2017wtu}. Neutrinos directly drive winds from
\ac{NS} merger remnants \citep{Dessart:2008zd, Perego:2014fma,
Fujibayashi:2017xsz} and impact the composition of outflows driven by
hydrodynamic or magnetic torques and nuclear processes
\citep{Metzger:2014ila, Fujibayashi:2017puw, Fernandez:2018kax,
Miller:2019dpt, Nedora:2019jhl, Fujibayashi:2020qda,
Fujibayashi:2020dvr, Just:2021cls, Li:2021vqj}.  Finally, neutrinos
might participate in the launching of gamma-ray burst jets from these
systems \citep{Eichler:1989ve, Rosswog:2002rt, Zalamea:2010ax,
Just:2015dba, Perego:2017fho}.

In the context of \ac{NS} mergers, the most popular approach to include
neutrinos in simulations is the so-called neutrino leakage scheme. This
method was first proposed by \citet{vanRiper:1981mko} in the context of
core-collapse supernovae, and then used to perform Newtonian simulations
of \ac{NS} mergers by \citet{Ruffert:1995fs} and \citet{Rosswog:2002rt}.
A \ac{GR} extension of the leakage scheme was first proposed in
\citet{Sekiguchi:2010ep} and was subsequently applied to \ac{NS} mergers
in \citet{Sekiguchi:2011zd}. Publicly available implementations of the
relativistic leakage scheme are available in \texttt{GR1D} and
\texttt{Zelmani} \citep{OConnor:2009iuz} and in the \texttt{THC} code
\citep{Radice:2016dwd}. The latter uses a methodology first proposed by
\citet{Neilsen:2014hha} to compute the optical depth, which is able to
capture the complex geometries of neutron star merger remnants
\citep{Endrizzi:2019trv}. This approach has also been used by
\citet{Siegel:2017nub} and \citet{Murguia-Berthier:2021tnt}. More
sophisticated implementations include the Advanced Spectral Leakage
scheme of \citet{Perego:2015agy, Gizzi:2021ssk} and the Improved
Leakage-Equilibration-Absorption scheme of
\citet{Ardevol-Pulpillo:2018btx, Kullmann:2021gvo}.  Leakage schemes do
not attempt to simulate the transport of neutrinos, but instead
parametrize the rate of cooling of the remnant using of phenomenological
formulas based on the optical depth. Specifically, they replace the
emission rate of neutrinos with a scaling factor $O(e^{-\tau})$, where
$\tau$ is the optical depth.  As such, leakage schemes avoid stiff
source terms in the hydrodynamics equations and are computationally
inexpensive. Standard leakage schemes ignore the reabsorption of
neutrinos, so they cannot model the deposition of heat and lepton number
in the ejecta by neutrinos.  Moreover, leakage schemes are not accurate
over timescales comparable with the cooling timescale of optically thick
source, that is several hundreds of milliseconds for \ac{NS} merger
remnants \citep{Sekiguchi:2011zd}.

To include the effect of neutrino reabsorption, several groups have
coupled leakage scheme, used to treat the optically thick regions, with
schemes designed to treat the free streaming neutrinos
\citep{Perego:2014fma,Sekiguchi:2015dma, Radice:2016dwd,
Fujibayashi:2017xsz, Radice:2018pdn, Ardevol-Pulpillo:2018btx}. This
approach is likely inspired by the isotropic diffusion source
approximation developed in the context of core-collapse supernovae
\citep{Liebendoerfer:2007dz}. The combination of leakage and transport
schemes addresses some of the limitations of the formers, namely
the inability to model reabsorption, while preserving the overall
computational efficiency of the method, since no stiff source terms are
present.  However, the use of these methods is questionable when
modeling optically thick sources on timescales comparable to their
cooling timescale. This is an important limitation, since it is now well
established that secular ejecta, launched on timescales of several
seconds, likely dominate the kilonova signal and the nucleosynthesis
yield from mergers \citep{Siegel:2019mlp, Shibata:2019wef,
Radice:2020ddv, Nedora:2020hxc, Shibata:2021xmo}.  Moreover, most of
these methods cannot model out-of-weak-equilibrium effects, which might
impact the postmerger evolution and the \ac{GW} signal of binary \ac{NS}
systems \citep{Alford:2017rxf, Alford:2020lla, Most:2021zvc,
Hammond:2021vtv}.

On the opposite end of the spectrum, the most sophisticated \ac{GR}
radiation-(magneto)hydrodynamics simulations of \ac{NS} mergers and
their postmerger evolution use Monte Carlo schemes
\citep{Miller:2019gig, Miller:2019dpt, Foucart:2020qjb,
Foucart:2021mcb}. These schemes directly attempt to solve the
seven-dimensional Boltzmann equation by sampling the distribution
function of neutrinos at random points in phase space. While these
methods can be very accurate, they become prohibitively expensive when
optically thick media are present.  This is because, in order to
correctly capture the thermodynamic equilibrium of matter and radiation,
Monte Carlo schemes need to resolve the mean free path of the neutrinos.
To avoid this issue, the method of \citet{Foucart:2021mcb} artificially
alters emission, absorption, and scattering rates at high optical depth
in a way that does not impact the energy distribution of neutrinos close
to the neutrino sphere.  This approach can accurately predict the
neutrino distribution outside of the remnant, but it is only valid for
short times compared to the diffusion timescale. Moreover, this method
does not correctly capture out-of-thermodynamic equilibrium effects for
matter and neutrinos.

Other methods solving the full-Boltzmann equation of radiation transport
equations in seven dimensions include the short characteristic method
\citep{Davis:2012yv}, the $S_N$ schemes of \citet{Nagakura:2014nta} and
\citet{Chan:2020quo}, the $F\!P_N$ approach \citep{2010JCoPh.229.5597M,
Radice:2012tg}, the lattice Boltzmann method \citep{Weih:2020qyh}, and
the recently proposed method of characteristics moment closure (MOCMC)
method \citep{Ryan:2019hor}. All of these approaches can, in principle,
model the full range of conditions and effects encountered in \ac{NS}
mergers. In practice, these methods are extremely computationally
intensive, because high angular resolutions is required to obtain
solutions that are competitive with those of moment based schemes
\citep{Richers:2017awc}. So while the continued development of such
methods is important and full-Boltzmann simulations are necessary to
validate \ac{NS} merger models, simplified neutrino transport methods
are necessary to perform systematic surveys of the binary and \ac{EOS}
parameter space.

The moment formalism casts the Boltzmann equation for classical neutrino
transport in a form resembling the hydrodynamics equations
\citep{1981MNRAS.194..439T, Shibata:2011kx}. The main advantage of
moment-based approaches is that they reduce the seven-dimensional
Boltzmann equation to a system of 3+1 equations. Unlike the
hydrodynamics equations, however, the moment equations for radiative
transfer cannot be closed with an \ac{EOS}, because, in general, there
is no frame in which radiation can be assumed to be isotropic.
Consequently, although moment-based approaches can model all effects
arising from the interaction between matter and radiation, their
accuracy is limited by the accuracy of the adopted closures
\citep{Richers:2020ntq}. Moment-based approaches are currently becoming
very popular in the context of core-collapse supernovae
\citep{Obergaulinger:2014nna, OConnor:2014sgn, Kuroda:2015bta,
OConnor:2015rwy, Roberts:2016lzn, Skinner:2018iti, Glas:2018oyz,
Rahman:2019yxy, Laiu:2021pha}. Moment-based methods have been first
introduced by \citet{Foucart:2015vpa}, \citet{Foucart:2015gaa}, and
\citet{Foucart:2016rxm} in the context of \ac{NS} mergers.

Here, we introduce \texttt{THC\_M1}: a new moment-based radiation
transport code designed to perform long-term merger and postmerger
simulations of binary \ac{NS}. We adopt a formalism similar to that of
\citet{Foucart:2016rxm}, but with two important differences. First, we
introduce a new numerical scheme able to capture the diffusion limit of
radiative transfer without resorting to the use of the relativistic
heat-transfer equation, which is known to be ill posed
\citep{Hiscock:1985zz, Andersson:2011fd}.  Second, we retain all terms
appearing in the coupling of matter and radiation. To the best of our
knowledge, the only other codes to include these terms are that of
\citet{2020ApJ...900...71A} and \citet{Kuroda:2015bta}, which have not
been applied to \ac{NS} mergers. We demonstrate that these terms are
necessary to correctly capture the trapping of neutrinos in
relativistically moving media.  After having validated our code with a
series of tests, we use it to perform inspiral, merger, and postmerger
simulations of two binary \ac{NS} systems, and we study the impact of
neutrinos on their dynamics, \ac{GW} signal, and nucleosynthesis yields.

The rest of this paper is organized as follows. We introduce the
mathematical formalism for the moment-based treatment of radiation in
Sec.~\ref{sec:math}. We give the details of our numerical implementation
in Sec.~\ref{sec:methods}. We validate our approach with a series of
tests in Sec.~\ref{sec:tests}. We present a first application to the
study of the merger and postmerger evolution of binary \ac{NS} systems
in Secs.~\ref{sec:nsmerger} and \ref{sec:postmerger}. Finally,
Sec.~\ref{sec:conclusions} is dedicated to discussion and conclusions.
Unless otherwise specified, we use a system of units in which $G = c =
1$.

% ------------------------------------------------
\section{Mathematical Formalism}
\label{sec:math}
% ------------------------------------------------
The M1 scheme describes the neutrino fields in term of their associated
(energy integrated) stress energy tensors $T^{\alpha \beta}_{(\nu)
}$, where $\nu  \in \{ \nu_e, \bar{\nu}_e , \nu_x \}$ and $\alpha, \beta
\in \{ 0, 1, 2, 3 \}$. Since the formalism we are going to discuss
applies in the same way to all neutrino species, we will omit the
$\cdummy_{(\nu) }$ subscript in the following discussion.

We decompose the (neutrino) radiation stress energy tensor along and
orthogonally to $n^{\alpha}$, the future-oriented unit normal to the $t
=$const hypersurfaces, as
\begin{equation}
  \label{tmunu.lab} T^{\alpha \beta} = En^{\alpha} n^{\beta} + F^{\alpha}
  n^{\beta} + n^{\alpha} F^{\beta} + P^{\alpha \beta},
\end{equation}
with $F^{\alpha} n_{\alpha} = 0$ and $P^{\alpha \beta} n_{\alpha} = 0.$ The
quantities $E, F^{\alpha},$and $P^{\alpha \beta} $appearing in this
decomposition are the radiation energy density, the radiation
flux, and the radiation pressure tensor in the Eulerian frame, respectively.

In an analogous way, we can decompose the radiation stress energy tensor
using the fluid four-velocity $u^{\alpha}$:
\begin{equation}
  \label{tmunu.fluid} T^{\alpha \beta} = Ju^{\alpha} u^{\beta} + H^{\alpha}
  u^{\beta} + u^{\alpha} H^{\beta} + K^{\alpha \beta},
\end{equation}
with $H^{\alpha} u_{\alpha} = 0$ and $K^{\alpha \beta} u_{\alpha} = 0.$ The
new quantities $J, H^{\alpha},$and $K^{\alpha \beta}$ are, respectively, the
radiation energy density, the radiation flux, and the radiation pressure
tensor in the fluid rest frame.

Conservation of energy and angular momentum reads
\begin{equation}
  \label{divT} \nabla_{\beta} T^{\alpha \beta} = - \nabla_\beta
  T^{\alpha\beta}_{{\rm HD}},
\end{equation}
where $\nabla$ is the covariant derivative operator compatible with the
spacetime metric and $T^{\alpha\beta}_{{\rm HD}}$ is the matter
stress-energy tensor. In 3+1 form Eq.~(\ref{divT}) reads \citep{Shibata:2011kx}
\begin{equation}
  \begin{split}
    \label{3p1.system}
    &\partial_t  \left( \sqrt{\gamma} E \right) + \partial_i
    \left[ \sqrt{\gamma}  (\alpha F^i - \beta^i E) \right] =\\
    &\qquad \alpha \sqrt{\gamma}  [P^{i k} K_{i k} - F^i \partial_i \log \alpha
    -\mathcal{S}^{\mu} n_{\mu}],\\
    &\partial_t  \left( \sqrt{\gamma} F_i \right) + \partial_k  \left[
    \sqrt{\gamma}  \left( \alpha P^k_{\phantom{k} i} - \beta^k F_i \right)
    \right] =\\
    &\qquad \sqrt{\gamma}  \left[ - E \partial_i \alpha + F_k \partial_i
    \beta^k + \frac{\alpha}{2} P^{j k} \partial_i \gamma_{j k} + \alpha
    \mathcal{S}^{\mu} \gamma_{i \mu} \right] ,
  \end{split}
\end{equation}
where $\gamma_{i k}$ is the three metric and $\gamma$ is its determinant,
$\alpha$ is the lapse function, $\beta^i$ is the shift vector, and $K_{i k} $
is the extrinsic curvature, not to be confused with the fluid frame radiation
pressure tensor. $\mathcal{S}^{\mu}$ is the term representing the interaction
between the neutrino radiation and the fluid. It can be written as
\begin{equation}
  \label{matter.sources} \texttt{} \mathcal{S}^{\mu} = (\eta - \kappa_a J)
  u^{\mu} - (\kappa_a + \kappa_s) H^{\mu},
\end{equation}
where $\eta, \kappa_a,$ and $\kappa_s$ are the neutrino emissivity, and
absorption and scattering coefficients. Scattering is assumed to be isotropic
and elastic. Inelastic scattering effects could in principle be treated within
this formalism as absorption events immediately followed by emission.

It is important to remark that Eqs.~\eqref{3p1.system} are exact, but
they are not closed, since $P^{i k}$ cannot be expressed in terms of $E$ and
$F^i$. The key idea of the M1 scheme is to introduce an (approximate) analytic
closure for these equations, that is a relation $P^{i k} = f (E, F^i) .$
Clearly, if $P^{i k}$ were known, then the M1 scheme would provide an exact
solution of the transport equation. However, because $P^{i k}$ depends on the
global geometry of the radiation field, no closure in the form $P^{i k} = f
(E, F^i)$ can be exact in general.

\tmtexttt{THC\_M1} adopts the so-called \tmtextit{Minerbo closure},
which is exact in two limits: 1) the optically thick limit in which
matter and radiation and in thermodynamic equilibrium and 2) the
propagation of radiation from a point source (at large distances) in a
transparent medium. We consider these two cases separately below.

\subsection{Optically thick limit}

In the optically thick limit, in which matter and radiation are in
equilibrium, the radiation pressure tensor is isotropic in the fluid frame
\begin{equation}
  K_{\alpha \beta} = \frac{1}{3} J (g_{\alpha \beta} + u_{\alpha} u_{\beta}),
\end{equation}
where $g_{\alpha \beta}$ is the spacetime metric. The stress energy tensor
reads
\begin{equation}
  \label{tmunu.thick} T_{\phantom{a} \beta}^{\alpha} = \frac{4}{3} Ju^{\alpha}
  u_{\beta} + H^{\alpha} u_{\beta} + H_{\beta} u^{\alpha} + \frac{1}{3}
  J \delta_{\phantom{a} \beta}^{\alpha},
\end{equation}
where $\delta_{\phantom{\alpha} \beta}^{\alpha}$ is the Kronecker delta. The
radiation pressure tensor in the laboratory frame is written as
\begin{equation}
\begin{split}
  \label{pab.thick}
    P_{\alpha \beta} = \gamma_{\alpha \gamma}& \gamma_{\beta \delta} T^{\gamma
    \delta} = \frac{4}{3} JW^2 \upsilon_{\alpha} \upsilon_{\beta} + \\ &\gamma_{\alpha \gamma}
    H^{\gamma} \upsilon_{\beta} W + \gamma_{\gamma \beta} H^{\gamma} \upsilon_{\alpha} W +
    \frac{1}{3} J \gamma_{\alpha \beta},
\end{split}
\end{equation}
where $W = - u^{\alpha} n_{\alpha}$ is the fluid Lorentz factor and
$\upsilon^\alpha = \frac{1}{W} \gamma^\alpha_{\phantom{\alpha}\beta}
u^\beta$ is the fluid three velocity.
Since M1 evolves $(E, F^i)$, it is necessary to reformulate Eq.
\eqref{pab.thick} in terms of these variables. To this aim, we exploit the
decomposition of Eq. \eqref{tmunu.lab} to write
\begin{align}
  & E = T_{\alpha \beta} n^{\alpha} n^{\beta} = \frac{4}{3} JW^2 - 2
  H_{\alpha} n^{\alpha} W - \frac{1}{3} J, &  \label{energy.thick}\\
  & F_{\alpha} = - \gamma_{\alpha \beta} n_{\mu} T^{\beta \mu} = \frac{4}{3}
  JW^2 \upsilon_{\alpha} + WH_{\alpha} + WH^{\beta} n_{\beta} (n_{\alpha} -
  \upsilon_{\alpha}). &  \label{flux.thick}
\end{align}
Since $H^{\alpha}$ is orthogonal to $u_{\alpha}$, it is possible to
project Eq. \eqref{flux.thick} to find
\begin{equation}
\begin{split}
  F_{\alpha} u^{\alpha} = & \frac{4}{3} JW (W^2 - 1) - WH^{\beta} n_{\beta}
  (W + W^{- 1}  (W^2 - 1)) &  \nonumber\\
  = & \frac{4}{3} JW (W^2 - 1) - H^{\beta} n_{\beta}  (2 W^2 - 1) & \\
  = & \left[ \frac{4}{3} JW^2 - 2 H^{\beta} n_{\beta} W - \frac{1}{3} J
  \right] W - JW + H^{\alpha} n_{\alpha} . &  \nonumber
\end{split}
\end{equation}
The term in parenthesis in the last expression is the RHS of Eq.
\eqref{energy.thick}, so we conclude that
\begin{equation}
  H^{\alpha} n_{\alpha} = F_{\alpha} u^{\alpha} - EW + JW.
\end{equation}
Substituting this into Eq. \eqref{energy.thick} we find
\begin{equation}
  \label{J.thick} \left( \frac{2}{3} W^2 + \frac{1}{3} \right) J = E (2 W^2 -
  1) - 2 W^2 F_{\alpha} \upsilon^{\alpha} .
\end{equation}
This equation can be used to evaluate $J$ given the evolved fluid and
radiation quantities. Determining $H^{\alpha}$ is more complex, but
fortunately only its projection on the $t = \tmop{const}$ hypersurface is
required. To find it, we use Eq. \eqref{flux.thick} to write
\begin{equation}
  WH^{\alpha} = F^{\alpha} - \frac{4}{3} JW^2 \upsilon^{\alpha} - WH^{\beta}
  n_{\beta}  (n^{\alpha} - \upsilon^{\alpha})
\end{equation}
and
\begin{equation}
\begin{split}
  \label{h.thick}
    \gamma_{\phantom{a} \beta}^{\alpha} H^{\beta} = & \frac{F^{\alpha}}{W} -
    \frac{4}{3} JW\upsilon^{\alpha} + \upsilon^{\alpha} H^{\beta} n_{\beta}\\
    = & \frac{F^{\alpha}}{W} - \frac{4}{3} JW\upsilon^{\alpha} + \upsilon^{\alpha} [W
    F^{\beta} \upsilon_{\beta} - EW + JW] .
\end{split}
\end{equation}
We can thus evaluate the radiation pressure tensor by combining Eqs.
\eqref{pab.thick}, \eqref{J.thick} and \eqref{h.thick}.

\subsection{Optically thin limit}

In the optically thin limit we assume that radiation is streaming at the speed
of light in the direction of the radiation flux. This ansatz is well verified
for radiation propagating at large distances from a central source. In this
case, the radiation pressure tensor can be written as
\begin{equation}
  \label{pab.thin} P_{\alpha \beta} = \frac{E }{F_{\mu} F^{\mu}} F_{\alpha}
  F_{\beta} .
\end{equation}
We remark that, differently from the optically thick limit, the
optically thin limit is not unique. It is instead determined by the
global geometry of the radiation field. This choice of the optically
thin limit is also responsible for the appearance of ``radiation
shocks'' in M1 calculations. These artifacts emerge when radiation beams
from different directions intersect. In these cases the M1 method will
force radiation to stream in the direction of the total (weighted and
averaged) radiation flux causing neutrinos to interact in an unphysical
manner. To quantify the impact of such artifact, we perform calculations
in which the optically thick closure is used throughout the simulation
domain. This is the so-called \textit{Eddington closure}. It is not
affected by radiation shocks, since it preserves the linearity of the
transport operator. However, it predicts a maximum propagation speed of
neutrinos of $\frac{c}{\sqrt{3}}$ and leads to substantial artificial
diffusion (radiation can diffuse past obstacles that would otherwise
cause shadows to appear).

\subsection{Minerbo closure}

The Minerbo closure combines the optically thin and optically thick limits as
\begin{equation}
  \label{pab.closure} P _{\alpha \beta} = \frac{3 \chi - 1}{2} P_{\alpha
  \beta}^{\tmop{thin}} + \frac{3 (1 - \chi)}{2} P_{\alpha
  \beta}^{\tmop{thick}},
\end{equation}
where $\chi \in \left[ \frac{1}{3}, 1 \right]$ is the so-called Eddington
factor, which is taken to be
\begin{equation}
  \label{eddington.factor} \chi (\xi) = \frac{1}{3} + \xi^2 \left( \frac{6 - 2
  \xi + 6 \xi^2}{15} \right),
\end{equation}
where
\begin{equation}
  \label{flux.factor} \xi^2 = \frac{H_{\alpha} H^{\alpha}}{J^2} .
\end{equation}
In the optically thick regions of the flow $H_{\alpha} \simeq 0$ and $\chi
\simeq \frac{1}{3}$, so $P_{\alpha \beta} \simeq P_{\alpha
\beta}^{\tmop{thick}} .$ Conversely, in the optically thin regions $\xi \simeq
1$ and $\chi \simeq 1$, so $P_{\alpha \beta} \simeq P_{\alpha
\beta}^{\tmop{thin}} .$ It is important to remark that $\xi$ is computed
using $H_{\alpha}$ and $J$, instead of $F_{\alpha}$ and $E$. This is because
$F_{\alpha}$ is not guaranteed to be small in the optically thick limit if
the background flow is moving. On the other hand, the knowledge of the M1
evolved quantities, $E$ and $F^{\alpha}$, is not immediately sufficient to
calculate $H_{\alpha}$: it is necessary to also know $P_{\alpha \beta}$. Eqs.
\eqref{pab.closure}, \eqref{eddington.factor}, and \eqref{flux.factor} need to
be solved numerically for $\chi$ using a root finding scheme. To this purpose,
we adopt the Brent-Dekker method as implemented in the GNU Scientific Library
{\citep{GSL}}.

\subsection{Neutrino number density}

Weak reactions conserve the total lepton number of the system, but they can
alter the electron fraction of the matter. For this reason, it is desirable to
also evolve the number density of neutrinos. To this aim, we follow the
phenomenological approach proposed by {\citet{Foucart:2016rxm}} and, for each
neutrino species, we introduce a neutrino number current $N^{\alpha}_{(\nu)}$,
with $\nu  \in \{ \nu_e, \bar{\nu}_e , \nu_x \}$. The neutrino number
density in the fluid frame is
\begin{equation}
  \label{neutrino.density} n = - N^{\alpha} u_{\alpha},
\end{equation}
where we have suppressed once again the index $\cdummy_{(\nu)} .$The
continuity equation for neutrinos reads
\begin{equation}
  \label{divN} \nabla_{\alpha} N^{\alpha} = \sqrt{- g}  (\eta^0 - \kappa_a^0
  n),
\end{equation}
where $g$ is the determinant of the spacetime metric and $\kappa_a^0$ and
$\eta^0$ are the neutrino number absorption and emission coefficients. Eq.
\eqref{divN} is exact, but like the neutrino energy and momentum equations
\eqref{3p1.system}, it is also not closed. The closure we adopt for Eq.
\eqref{divN} is
\begin{equation}\label{eq:closure.number}
  N^{\alpha} = nf^{\alpha} = n \left( u^{\alpha} + \frac{H^{\alpha}}{J}
  \right) .
\end{equation}
Since $H^{\alpha} u_{\alpha} = 0$, this closure is consistent with Eq.
\eqref{neutrino.density}. The closure assumes that the neutrino number
and energy flux are aligned. While this closure would be exact if
neutrinos had a single energy, it is not for the energy-integrated
fluxes in general. The closure on the neutrino number flux
\eqref{eq:closure.number} and neutrino pressure tensor
\eqref{pab.closure}, the simplified treatment of the energy dependence
of neutrino absorption and scattering opacities
(Sec.~\ref{sec:opacity.correction}), and the fact that we neglect
neutrino oscillations are the only modeling assumptions in
\texttt{THC\_M1}. The closure assumes that the neutrino number and
energy flux are aligned.

In 3+1 form Eq. \eqref{divN} becomes
\begin{equation}
  \partial_t  \left( \sqrt{\gamma} n \Gamma \right) + \partial_i  \left(
  \alpha \sqrt{\gamma} nf^i \right) = \alpha \sqrt{\gamma}  (\eta^0 -
  \kappa_a^0 n),
\end{equation}
where
\begin{align}
  \Gamma = \alpha f^0 = W - \frac{1}{J} H^{\alpha} n_{\alpha}, &  & f^i = W
  \left( \upsilon^i - \frac{\beta^i}{\alpha} \right) + \frac{H^i}{J} .
  \label{nnu.quantities}
\end{align}
When computing $\Gamma$, we follow {\citet{Foucart:2016rxm}} and rewrite it as
\begin{equation}
  \Gamma = - f^{\alpha} n_{\alpha} = W \left( \frac{E - F_{\alpha}
  \upsilon^{\alpha}}{J} \right),
\end{equation}
where we have used the fact that
\begin{equation}
  - H^{\mu} n_{\mu} = W (E - J - F^{\alpha} \upsilon_{\alpha}) .
\end{equation}

% ------------------------------------------------
\section{Numerical Implementation}
\label{sec:methods}
% ------------------------------------------------

The M1 equations can be summarized as
\begin{equation}
  \label{compressed} \partial_t \tmmathbf{U}+ \partial_i \tmmathbf{F}^i
  (\tmmathbf{U}) =\tmmathbf{G} (\tmmathbf{U}) +\tmmathbf{S} (\tmmathbf{U}),
\end{equation}
where:
\begin{equation}
  \tmmathbf{U}= \left(\begin{array}{c}
    \sqrt{\gamma} n  \Gamma\\
    \sqrt{\gamma} E\\
    \sqrt{\gamma} F_k
  \end{array}\right), \qquad
\end{equation}
\begin{equation}
  \tmmathbf{F}^i = \left(\begin{array}{c}
    \alpha \sqrt{\gamma} nf^i\\
    \sqrt{\gamma} [\alpha F^i - \beta^i E]\\
    \sqrt{\gamma}  \left[ \alpha P^i_{\phantom{i} k} - \beta^i F_k \right]
  \end{array}\right), \\
\end{equation}
\begin{equation}
  \tmmathbf{S}= \left(\begin{array}{c}
    \alpha \sqrt{\gamma}  [\eta^0 - \kappa_a^0 n]\\
    - \alpha \sqrt{\gamma} \mathcal{S}^{\mu} n_{\mu}\\
    \alpha \sqrt{\gamma} \mathcal{S}^{\mu} \gamma_{k \mu}
  \end{array}\right),
\end{equation}
and
\begin{equation}
 \tmmathbf{G}= \left(\begin{array}{c}
     0\\
     \alpha \sqrt{\gamma}  [P^{i k} K_{i k} - F^i \partial_i \log \alpha]\\
     \sqrt{\gamma}  \left[ F_i \partial_k \beta^i - E \partial_k \alpha +
     \frac{\alpha}{2} P^{i j} \partial_k \gamma_{j i} \right]
   \end{array}\right) .
\end{equation}
Among these terms, the coupling with matter $\tmmathbf{S}$ is stiff and
cannot be treated using an explicit time integration strategy. Since
$\mathcal{S}^{\mu}$ is a function of $(E, F^i)$ through the (nonlinear)
closure of the M1 scheme, the matter coupling is not only stiff, but
also nonlinear. Our code is the first M1 code in {\tmabbr{GR}} to treat
this term in full generality in the merger context. On the other
hand, if the opacity coefficients are kept fixed during the update of
the radiation quantities, the number density equation formally decouples
from the others, so it can be treated separately.

\tmtexttt{THC\_M1} integrates Eq. \eqref{compressed} using a semi-implicit
scheme. Given the solution $\tmmathbf{U}^{(k)}$ at time $t = k \Delta t$, we
compute the solution at the next timestep $\tmmathbf{U}^{(k + 1)}$ in two main
steps:
\begin{enumerate}
  \item $\displaystyle
    \frac{\tmmathbf{U}^{\ast} -\tmmathbf{U}^{(k)}}{\Delta t/2} = - \partial_i
    \tmmathbf{F}^i [\tmmathbf{U}^{(k)}] +\tmmathbf{G} [\tmmathbf{U}^{(k)}]
    +\tmmathbf{S} [\tmmathbf{U}^{\ast}],
  $
  \item $\displaystyle
  \frac{\tmmathbf{U}^{(k + 1)} -\tmmathbf{U}^{(k)}}{\Delta t} = -
  \partial_i \tmmathbf{F}^i [\tmmathbf{U}^{\ast}] +\tmmathbf{G}
  [\tmmathbf{U}^{\ast}] +\tmmathbf{S} [\tmmathbf{U}^{(k + 1)}] .
  $
\end{enumerate}
In particular, the advection terms and the metric sources are treated
explicitly, as discussed below, while the coupling with matter is
treated implicitly. Fluid quantities are kept fixed during the radiation
update until the end of the second step, when matter energy and momentum
densities, as well as the electron fraction, are updated according to
energy, momentum, and lepton number conservation. Conservation is also
enforced by limiting the changes in the radiation quantities that would
correspond to negative matter energy density, or to electron fractions
outside the boundaries of the \ac{EOS} table (typically $0 \leq Y_e \leq
0.6$). The treatment of the advective and source terms are discussed in
detail below. The derivative of the metric terms appearing in
$\tmmathbf{G}$ are discretized using standard 2\textsuperscript{nd} order finite
differencing.

\subsection{Radiation advection}
\label{sec:method.advection}

\tmtexttt{THC\_M1} uses a 2\textsuperscript{nd} order flux-limited conservative
finite-differencing scheme to evolve the radiation fields. In particular,
numerical fluxes are computed separately for each variable and
direction-by-direction. These are then combined in a directionally unsplit
fashion. For simplicity, we discuss the treatment of the radiation fluxes for
one of the evolved variables, say $u$, in the $x$-direction.

Let $u_i$ be the evolved quantity at the coordinate position $x_i$. Then,
\tmtexttt{THC\_M1} approximates the derivative of the flux $f (u)$ at the
location $x_i$ as
\begin{equation}
  \partial_x f (u) \simeq \frac{F_{i - 1 / 2} - F_{i + 1 / 2}}{\Delta x},
\end{equation}
where $F_{i - 1 / 2}$ and $F_{i + 1 / 2}$ are numerical fluxes defined at $x_i
\mp \frac{\Delta x}{2}$, respectively. The fluxes are constructed as linear
combination of a non diffusive 2\textsuperscript{nd} order flux $F^{\tmop{HO}}$ and a diffusive
1\textsuperscript{st} order correction $F^{\tmop{LO}}$:
\begin{equation}
  F_{i + 1 / 2} = F_{i + 1 / 2}^{\tmop{HO}} - A_{i + 1 / 2} (1 - \varphi_{i + 1 /
  2})  (F_{i + 1 / 2}^{\tmop{HO}} - F_{i + 1 / 2}^{\tmop{LO}}) .
\end{equation}
The term $\varphi_{i + 1 / 2}$ is the so called {\tmem{flux limiter}}
\citep{leveque}, while $A_{i + 1 / 2}$ is a coefficient introduced to
switch off the diffusive correction at high optical depth (more below).
The role of the flux limiter is to introduce numerical dissipation in
the presence of unresolved features in the solution $u$ and ensure the
nonlinear stability of the scheme. In particular, if $A_{i + 1 / 2}
(1 - \varphi_{i + 1 / 2}) = 0$ the second order flux is used, while if $A_{i +
1 / 2} (1 - \varphi_{i + 1 / 2}) = 1$, then the low order flux is used. A
standard 2\textsuperscript{nd} order non diffusive flux is used for $F^{\tmop{HO}}$, while
the Lax-Friedrichs flux is used for $F^{\tmop{LO}}$:
\begin{align}
  &F^{\tmop{HO}}_{i + 1 / 2} = \frac{f (u_i) + f (u_{i + 1})}{2}, \\
  &F^{\tmop{LO}}_{i + 1 / 2} = \frac{1}{2}  [f (u_i) + f (u_{i + 1})] -
  \frac{c_{i + 1 / 2}}{2}  [u_{i + 1} - u_i] .
\end{align}
The characteristic speed in the Lax-Friedrichs flux $c_i$ is taken to be the
maximum value of the speed of light between the right and left cells
\begin{equation}
  c_{i + 1 / 2} = \max_{a \in \{ i, i + 1 \}}  \left\{ \left| \alpha_a
  \sqrt{\gamma_a^{x x}} \pm \beta_a^x \right| \right\} .
\end{equation}
We remark that it is known that the M1 system can, in some circumstances, lead
to acausal (faster than light) propagation of neutrinos in {\tmabbr{GR}}
{\citep{Shibata:2011kx}}. For this reason, one might argue that a better choice
of the characteristic velocity for the Lax-Friedrichs formula would have been
given by the eigenvalue of the jacobian of $\tmmathbf{F}$. These values are
known analytically {\citep{Shibata:2011kx}}, however in our preliminary tests
we found that the use of the full eigenvalues resulted did not improve on the
stability or accuracy of the M1 solver.

The flux limiter is computed using a standard \tmtexttt{minmod} approach:
\begin{equation}
  \varphi_{i + 1 / 2} = \max \left\{ \min \left[ 1, \min \left( \frac{u_i - u_{i - 1}}{u_{i
  + 1} - u_i}, \frac{u_{i + 2} - u_{i + 1}}{u_{i + 1} - u_i} \right)
  \right], 0 \right\}.
\end{equation}
The resulting scheme is formally 2\textsuperscript{nd} order accurate away from shocks or
extrema in the solution.

The coefficient $A_{i + 1 / 2}$ is computed as
\begin{equation}
  A_{i + 1 / 2} = \min \left( 1, \frac{1}{\Delta x \kappa_{\tmop{ave}}}
  \right),
\end{equation}
where
\begin{equation}
  \kappa_{\tmop{ave}} = \frac{1}{2} [(\kappa_a)_i + (\kappa_a)_{i + 1} +
  (\kappa_s)_i + (\kappa_s)_{i + 1}] .
\end{equation}
In particular, $A_{i + 1 / 2} = 1$ in optically thin regions, while $A_{i + 1 /
2} \ll 1$ at high optical depths ($\Delta x \kappa_{\tmop{ave}}$ is the optical
distance between $x_i$ and $x_{i + 1}$). In the optically thick limit $F_{i + 1
/ 2} \simeq F^{\tmop{HO}}_{i + 1 / 2}$ and the scheme reduces to a centered 2nd
order scheme, which is {\tmem{asymptotic preserving}}
{\citep{2002IJNMF..40..479R}}. This means that \tmtexttt{THC\_M1} can capture
the optically thick limit without having to artificially replace the advective
terms with the flux obtained from the diffusion equation, which is known to be
ill posed\footnote{Here we say that a mathematical problem is ``ill posed''
if it is not well posed according to Hadamard. That is if it does not 1) admit a
single solution that 2) depends continuously on the initial/boundary data.} in
special and general relativity \citep{Hiscock:1985zz, Andersson:2011fd}.

This can be shown easily for an optically thick stationary medium in flat
spacetime. To keep our notation simple, we also restrict ourselves to the
discussion of the 1D case, however the generalization to 3D is
straightforward. In this case, the radiative transfer equations reduce to
\begin{equation}
\begin{split}
  \label{toy.diffusion}
    &\partial_t E + \partial_x F^x = \kappa_a  (B - E),\\
    &\partial_t F^x + \frac{1}{3} \partial_x E = - (\kappa_a + \kappa_s) F^x .
\end{split}
\end{equation}
where $B$ is the blackbody function. In the limit of $L (\kappa_a + \kappa_s)
\gg 1$, where $L$ is a characteristic length scale of the system, the
radiation flux becomes
\begin{equation}
  F_x = \frac{- 1}{3 (\kappa_a + \kappa_s)} \partial_x E,
\end{equation}
and the energy equation reduces to the heat diffusion equation:
\begin{equation}
  \partial_t E - \partial_x  \left( \frac{1}{3 (\kappa_a + \kappa_s)}
  \partial_x  E \right) = \kappa_a  (B  - E),
\end{equation}
A similar derivation applied to the \tmtexttt{THC\_M1} discretization of Eq.
\eqref{toy.diffusion}, shows that the numerical discretization of the
radiation energy flux reduces to a finite differencing scheme for the heat
equation:
\begin{equation}
  \label{asymptotic.limit} [\partial_x F ^x]_i \simeq \frac{F^x_{i + 1} - F^x_{i
  - 1}}{2 \Delta x} = \frac{- 1}{3 (\kappa_a + \kappa_s)} \left( \frac{E_{i +
  2} - 2 E_i + E_{i - 2}}{(2 \Delta x)^2} \right) .
\end{equation}
In the last step we have also assumed the absorption coefficients to be
constant in space for simplicity. However, a valid scheme for the diffusion
equation is also obtained for non-constant coefficients.

Although the scheme described by Eq. \eqref{asymptotic.limit} is a valid
discretization of the heat equation, it can suffer from an odd-even decoupling
instability, as evident from the fact that the solution at $x_i$ does not
depend on the solution at $x_{i - 1}$ and $x_{i + 1}$. To suppress this
instability, we check if
\begin{eqnarray*}
  (u_i - u_{i - 1})  (u_{i + 1} - u_i) < 0 & \text{and} & (u_{i + 1} - u_{i})
  (u_{i + 2} - u_{i + 1}) < 0.
\end{eqnarray*}
If this condition is satisfied, we set $A_{i + 1 / 2} = 1$. We find this to be
sufficient to obtain stable evolution in the scattering dominated regime.

\subsection{Radiation-matter coupling}

The implicit update of the neutrino number densities does not pose particular
challenges and reads (in the first substep of the method):
\begin{equation}
  N^{\ast} = \frac{N^{(k)} - \Delta t \partial_i  \left[ \alpha \sqrt{\gamma}
  n^{(k)} f^i \right] + \Delta t \left[ \alpha \sqrt{\gamma} \eta^0 \right]}{1
  + \Delta t \alpha \kappa_a^0 \Gamma^{- 1}},
\end{equation}
where $N = \sqrt{\gamma} n \Gamma$, and $\Gamma$ is given by Eq.
\eqref{nnu.quantities}. $n^{\ast}$ is obtained from $N^{\ast}$ using the
$\Gamma$ recomputed with the updated neutrino fields $(E, F_i)$. The flux
terms are computed as discussed in the previous section.

The implicit part of the time update for the radiation energy quantities is
significantly more complex, since it involves the solution of a $4 \times 4$
system of nonlinear equations. These are in the form
\begin{equation}
  \label{implicit.update} \tmmathbf{U}^{\ast} =\tmmathbf{W}+^{} \Delta
  t\tmmathbf{S} [\tmmathbf{U}^{\ast}],
\end{equation}
where $\tmmathbf{W}$ contains the explicit terms of the scheme. For example,
in the first substep of the update
\begin{equation}
  \tmmathbf{W}=\tmmathbf{U}^{(k)} + \Delta t (- \partial_i \tmmathbf{F}^i
  [\tmmathbf{U}^{(k)}] +\tmmathbf{G} [\tmmathbf{U}^{(k)}]) .
\end{equation}
We employ the Powell's Hybrid method as implemented in the GNU scientific
library {\citep{GSL}} \ to solve \eqref{implicit.update}. This algorithm
requires the evaluation of the Jacobian of the system as well as a suitable
initial guess. Both are discussed in detail below. Before diving into the
details, we remark that Eq. \eqref{implicit.update} requires the solution of a
nested nonlinear equation for the closure. \tmtexttt{THC\_M1} is the first GR
code to treat these terms without approximations and it is thus able to
correctly captured the trapping of neutrinos in optically thick rapidly moving
media. In some rare situations, the nonlinear solver can fail to converge to
the desired accuracy. This typically happens in the optically thick limit,
since the source term become stiff only in this limit. In such cases, we
linearize the equations by fixing $\chi = 1 / 3$. Finally, we
remark that, to save computational resources, we treat the source term
explicitly in the optically thin (non-stiff) limit.

\subsubsection{Source Jacobian}

The undensitized collisional source terms $\tmmathbf{S} [\tmmathbf{U}]$ are
composed of the projections
\begin{align}
  & - \alpha n_{\alpha} \mathcal{S}^{\alpha} = \alpha W \big[ \eta + \kappa_s J -
  \kappa_{\tmop{as}}  (E - F_i \upsilon^i) \big], \\
  & + \alpha \gamma_{i \alpha} \mathcal{S}^{\alpha} = \alpha W (\eta -
  \kappa_a J) \upsilon_i - \alpha \kappa_{\tmop{as}} H_i .
\end{align}
where $\kappa_{\tmop{as}} = \kappa_a + \kappa_s$. For the computation of the
Jacobian matrix $\mathcal{J}_{a b} = \partial \tmmathbf{S}_a / \partial
\tmmathbf{U}_b$ ($a, b = 0, ..., 3$) the density and momentum in the
laboratory frame must be expressed in terms of the Eulerian quantities:

\begin{align}
  & J (E, F_i) = B_0 + d_{\tmop{thin}} B_{\tmop{thin}} + d_{\tmop{thick}}
  B_{\tmop{thick}}, \\
  & H_i  (E, F_i) = - (a_{\upsilon \hspace{0.17em} 0} + d_{\tmop{thin}} a_{\upsilon
  \hspace{0.17em} \tmop{thin}} + d_{\tmop{thick}} a_{\upsilon \hspace{0.17em}
  \tmop{thick}}) \upsilon_i, \\
  &\qquad- d_{\tmop{thin}} a_{f \hspace{0.17em} \tmop{thin}}
  \hat{f}_i - (a_{F \hspace{0.17em} 0} + d_{\tmop{thick}} a_{F \hspace{0.17em}
  \tmop{thick}}) F_i,
\end{align}
with $\hat{f}_i = F_i / \sqrt{F_k F^k} = F_i / F$ , the definitions
\begin{equation}
  d_{\tmop{thick}} = \frac{3}{2} (1 - \chi), \quad
  d_{\tmop{thin}} = 1 - d_{\tmop{thick}},
\end{equation}
and the coefficients
\begin{align}
  B_{_0} &=  W^2 [E - 2 (\upsilon \cdot F)],  \\
  B_{\tmop{thin}} &=  W^2 E (\upsilon \cdot \hat{f})^2,  \\
  B_{\tmop{thick}} &=  \frac{W^2 - 1}{2 W^2 + 1} [4 W^2 (\upsilon \cdot F) + (3 - 2
  W^2) E],  \\
  a_{\upsilon 0} &=  W^3 [E - 2 (\upsilon \cdot F)] = WB_0,  \\
  a_{\upsilon \tmop{thin}} &=  W^3 E (\upsilon \cdot \hat{f})^2 = WB_{\tmop{thin}},  \\
  a_{\upsilon \tmop{thick}} &=  W \frac{W^2 - 1}{2 W^2 + 1} [4 W^2 (\upsilon \cdot F) + (3 -
  2 W^2) E]   \nonumber\\
  &   + \frac{W}{2 W^2 + 1} [(2 W^2 - 1) (\upsilon \cdot F) + (3 - 2 W^2) E] & \\
  &=  W B_{\tmop{thick}} + \frac{W}{2 W^2 + 1} [(2 W^2 - 1) (\upsilon \cdot F) + (3 -
      2 W^2) E],   \nonumber\\
  a_{f \tmop{thin}} & = W E (\upsilon \cdot \hat{f}),  \\
  a_{F 0} & = - W,  \\
  a_{F \tmop{thick}} & =  W \upsilon^2 .
\end{align}
The contractions between the fluid's velocity and the radiation momentum are
shortly indicated as, e.g., $F_i \upsilon^i = \upsilon \cdot F$. The Jacobian is then given
by
\begin{align}
  \mathcal{J}_{00} & = - \alpha W \left( \kappa_{\tmop{as}} - \kappa_s
  \frac{\partial J}{\partial E} \right), \\
  \mathcal{J}_{0 j} & = \alpha W \kappa_s  \frac{\partial J}{\partial F_j}
  + \alpha W \kappa_{\tmop{as}} \upsilon^j, \\
  \mathcal{J}_{i 0} & = - \alpha \left( \kappa_{\tmop{as}}  \frac{\partial
  H_i}{\partial E} + W \kappa_a  \frac{\partial J}{\partial E} \upsilon_i \right), \\
  \mathcal{J}_{ij} & = - \alpha \left( \kappa_{\tmop{as}}  \frac{\partial
  H_i}{\partial F_j} + W \kappa_a \upsilon_i  \frac{\partial J}{\partial F_j} \right) .
\end{align}
The necessary derivatives are
\begin{align}
  \frac{\partial J}{\partial E} & = W^2 + d_{\tmop{thin}}  (\upsilon \cdot \hat{f})^2
  W^2 + d_{\tmop{thick}}  \frac{(3 - 2 W^2) (W^2 - 1)}{(1 + 2 W^2)}, \\
  \frac{\partial J}{\partial F_j} & = J_F^\upsilon \upsilon^j + J_F^f  \hat{f}^j, \\
  \frac{\partial H_i}{\partial E} & = H_E^\upsilon \upsilon_i + H_E^f  \hat{f}_i, \\
  \frac{\partial H_i}{\partial F_j} & = H_F^{\delta} \delta_i^j + H_F^{\upsilon \upsilon} \upsilon_i
  \upsilon^j + H_F^{ff}  \hat{f}_i  \hat{f}^j + H_F^{vf} \upsilon_i  \hat{f}^j + H_F^{fv}
  \hat{f}_i \upsilon^j,
\end{align}
where the factors $X_Y^z$ in the derivatives $\partial X / \partial Y$ are the
common terms multiplying the terms with indexes $z_i^j$. Specifically, they
are
\begin{align}
  J_F^\upsilon & =  2 W^2 \left( - 1 + d_{\tmop{thin}} \frac{E (\upsilon \cdot \hat{f})}{F}
  + 2 d_{\tmop{thick}} \frac{W^2 - 1}{1 + 2 W^2} \right), \\
  J_F^f & =  - 2 d_{\tmop{thin}}  \frac{W^2 E (\upsilon \cdot \hat{f})^2}{F},  \\
  H_E^\upsilon & =  W^3 \left( - 1 - d_{\tmop{thin}}  (\upsilon \cdot \hat{f})^2 +
  d_{\tmop{thick}}  \frac{2 W^2 - 3}{1 + 2 W^2} \right), \\
  H_E^f & =  - d_{\tmop{thin}} W (\upsilon \cdot \hat{f}), \\
  H_F^{\delta} & =  W \left( 1 - d_{\tmop{thick}} \upsilon^2 - d_{\tmop{thin}}
  \frac{E (\upsilon \cdot \hat{f})}{F} \right) \\
  H_F^{\upsilon \upsilon} & = 2 W^3 \left[ 1 -
  d_{\tmop{thin}}  \frac{^{} E (\upsilon \cdot \hat{f})}{F} - d_{\tmop{thick}}
  \left( \upsilon^2 + \frac{1}{2 W^2 (1 + 2 W^2)} \right) \right], \\
  H_F^{ff} & =  2 d_{\tmop{thin}} \frac{W E (\upsilon \cdot \hat{f})}{F}, \\
  H_F^{vf} & =  2 d_{\tmop{thin}}  \frac{W^3 E (\upsilon \cdot \hat{f})^2}{F}, \\
  H_F^{fv} & =  - d_{\tmop{thin}} \frac{W E}{F}
\end{align}
The calculation of the above terms proceed as follows. The Eulerian multipole
$(E, F_i)$ enter the term $- \alpha n_{\alpha} \mathcal{S}^{\alpha}$ both
directly and via the fluid frame multipoles $(J, H_i)$. In
particular, $F_i$ enters only via combinations $\upsilon \cdot F$ (directly and in
$B_0$, $B_{\tmop{thin}}$) and $(\upsilon \cdot \hat{f})^2$ (in $B_{\tmop{thin}}$).
The relevant derivatives are
\begin{align}
  & \frac{\partial F_i}{\partial F_j} = \delta_{\phantom{j} i}^j \\
  & \frac{\partial (\upsilon \cdot F)}{\partial F_j} = \upsilon^j \\
  & \frac{\partial \hat{f}_i}{\partial F_j} = \frac{1}{F}
  \delta_i^j - \frac{1}{F}  \hat{f}_i  \hat{f}^j \\
  & \frac{\partial (\upsilon \cdot \hat{f})}{\partial F_j} = \frac{1}{F}
   \upsilon^j - \frac{(\upsilon \cdot \hat{f})}{F}  \hat{f}^j \\
  & \frac{\partial (\upsilon \cdot \hat{f})^2}{\partial F_j} = 2 \frac{(\upsilon \cdot
  \hat{f})}{F} \upsilon^j - 2 \frac{(\upsilon \cdot \hat{f})^2}{F} \hat{f}^j
\end{align}
Consequently, the derivatives $\partial J / \partial F_j$ have terms
proportional to $\upsilon^j$ and to $\hat{f}^j$. The Eulerian multipoles
$(E, F_i)$ do not enter directly the terms $\alpha \gamma_{i \alpha}
\mathcal{S}^{\alpha}$. The dependence on $F_i$ of $H_i $ is either in
terms proportional to $(\upsilon \cdot \hat{f})^2$ (in $a_{\upsilon
\hspace{0.17em} \tmop{thin}}$), $\upsilon \cdot \hat{f}$ (in $a_{f
\hspace{0.17em} \tmop{thin}}$), $\upsilon \cdot F$ (in $a_{\upsilon
\hspace{0.17em} \tmop{thick}}$), or in the direct terms explicitly
indicated in Eq.(15). Consequently, the derivatives $\partial H_i /
\partial F_j$ have terms proportional to $\delta_{\phantom{j} i}^j$, to
$\upsilon_i \upsilon^j$ and to $F_i F^j$.

A particular cases of the above calculation is the linearization around
the zero state $\tmmathbf{U}_0 = 0$ and the zero fluid's velocity limit
$\upsilon_i = 0$. For the former case, the undensitized collisional term
is
\begin{equation}
  \tmmathbf{S} (0) = [\alpha \eta W, \hspace{0.17em} \alpha \eta W\upsilon_i],
\end{equation}
and the Jacobian matrix simplifies: since $\hat{f}_i = 0$, the first column
and first row are proportional to $\upsilon_i$ and $\upsilon^j$ respectively, while the
spatial block has a term proportional to $\delta_{ij}$ and a term proportional
to $\upsilon_i \upsilon^j$. A simple analytical inversion can be calculated with any
computer algebra software. For a static fluid $\upsilon^i = 0$ ($E = J$ and $F_i =
H_i$), one obtains
\begin{equation}
  \tmmathbf{S} (\tmmathbf{U}_0) = [\alpha \eta - \kappa_a E, - \alpha
  \kappa_{\tmop{as}} F_i],
\end{equation}
and the Jacobian matrix is diagonal
\begin{align}
  \mathcal{J}_{00} & = - \alpha \kappa_a, \\
  \mathcal{J}_{ij} & = - \alpha \kappa_{\tmop{as}} \delta_{ij} .
\end{align}

The \tmtexttt{THC\_M1} implementation is different from most of the
other M1 schemes in general relativity. In particular, \texttt{THC\_M1}
and \citet{2020ApJ...900...71A} are the only codes fully treating the
nonlinear terms in the radiation matter coupling. In
\citet{Roberts:2016lzn} the linearization is performed about the zero
state and only retains some of the $(\upsilon \cdot \hat{f})$ terms in
the Jacobian matrix. In {\citet{Foucart:2015vpa}} and
{\citet{Foucart:2016rxm}} the linearization is also performed about the
zero state and the angle between the velocity and the neutrino flux is
kept fixed, i.e., $(\upsilon \cdot F) = \tmop{const}$ and $(\upsilon
\cdot \hat{f}) = \tmop{const}$. In {\citet{Weih:2020wpo}} the
linearization is also performed about the zero state and $P^{ij}$ is
assumed to be independent from $\tmmathbf{U}$. Hence the projections of
$P^{ij}$ appear explicitly in the $\tmmathbf{U}$ terms, but the $P^{ij}$
(closure) is not included in the Jacobian matrix.

\subsubsection{Blackbody function}
\label{sec:blackbody}

Emissivity, absorption, and scattering coefficients are kept fixed throughout
the implicit time integration. This can cause the numerical scheme to
oscillate if matter is thrown out of equilibrium over a small timescale
compared with $\Delta t$. To avoid this problem, first we compute the
blackbody function for neutrinos in two ways.
\begin{enumerate}
  \item When the radiation-matter equilibration time $\tau = (c
  \sqrt{\kappa_a  (\kappa_a + \kappa_s)})^{-1}$ is larger than $\Delta t$, then
  we set
  \begin{equation}
    \label{blackbody.function} B_{\nu} = \frac{4 \pi}{(ch)^3}  (k_B T)^4 F_3
    (\eta_{\nu}),
  \end{equation}
  where $F_3$ is the Fermi function of order 3
  \begin{equation}
    F_k (\eta) = \int_0^{\infty} \frac{x^k}{e^{x - \eta} + 1} \mathd x
  \end{equation}
  and $\eta_{\nu} = \mu_{\nu} / (k_B T)$ is the degeneracy
  parameter of the neutrinos. The equilibrium number density of
  neutrinos is computed as
  \begin{equation}
    \mathcal{B}_{\nu} = \frac{4 \pi}{(ch)^3}  (k_B T)^3 F_2 (\eta_{\nu}) .
  \end{equation}
  The temperature $T$ is taken to be the fluid temperature, while the
  neutrino chemical potential are evaluated at equilibrium using the EOS
  at the fluid density, temperature, and electron fraction $Y_e$,
  separately for each neutrino flavor. In particular, $\mu_{\nu_e}=
  \mu_e + \mu_p - \mu_n$, $\mu_{\bar{\nu}_e} = - \mu_{\nu_e}$, and
  $\mu_{\nu_x}=0$..

  \item If $\tau$ is smaller than $\Delta t / 2$, then the blackbody function
  is computed again using \eqref{blackbody.function}, but now $T$ and $Y_e$
  are taken to be the equilibrium temperature and electron fraction that
  matter would achieve under the assumption of weak equilibrium with
  neutrinos, and lepton number and energy conservation
  \citep{Perego:2019adq}. In particular, we solve the following equations
  \begin{align}
    Y_l & = Y_{e, \tmop{eq}} + Y_{\nu_e} (Y_{e, \tmop{eq}}, T_{\tmop{eq}}) -
    Y_{\bar{\nu}_e} (Y_{e, \tmop{eq}}, T_{\tmop{eq}}), \\
    u & = e (Y_{e, \tmop{eq}}, T_{\tmop{eq}}) + \frac{\rho}{m_{\mathrm{b}}}
    \Big[ Z_{\nu_e} (Y_{e, \tmop{eq}}, T_{\tmop{eq}}) \\ \nonumber
    &\qquad + Z_{\bar{\nu}_e}
    (Y_{e, \tmop{eq}}, T_{\tmop{eq}}) + 4 Z_{\nu_x} (T_{\tmop{eq}}) \Big],
  \end{align}
  where $Y_l$ is the total lepton fraction, inferred from both
  fluid and radiation quantities, $u$ is the total (matter and
  neutrino-radiation) energy density, and $Z_{\times}$ denotes the
  energy fraction of the species $\times$. These equations are
  solved for $T_{\rm eq}$ and $Y_{e, {\rm eq}}$ under the assumption of
  weak equilibrium, that is $T_{\rm matter} = T_{\rm \nu_{x}} = T_{\rm
  eq}$, $Y_e = Y_{e, {\rm eq}}$, $\mu_{\nu_e} = - \mu_{\bar{\nu_e}}$,
  $\mu_{\nu_{x}} = 0$, and $\mu_{\nu_e} = \mu_e + \mu_p - \mu_n$. The
  rationale for this choice is that it captures the correct equilibrium
  distribution for neutrinos, while the blackbody function of point 1)
  is valid for a mixture of matter and radiation in a thermal and lepton
  bath, or for short times compared with the equilibration time.

  \item For intermediate values of $\tau$ we linearly interpolate between the
  prescriptions from points (i) and (ii).
\end{enumerate}
Given the blackbody functions, we compute the $\nu_e$ and $\bar{\nu}_e$
emission coefficients and the $\nu_x$ absorption coefficients using
Kirchhoff's law. That is, we set
\begin{equation}
  \eta_{\nu_e} = \kappa_{a, \nu_e} B_{\nu_e},
  \qquad \eta_{\bar{\nu}_e} = \kappa_{a, \bar{\nu}_e} B_{\bar{\nu}_e},
  \qquad \kappa_{a, \nu_x} = \frac{B_{\nu_x}}{\eta_{\nu_x}} .
\end{equation}
We apply the same treatment to the neutrino number emissivities and opacities,
but using $\mathcal{B}$ instead of $B$.

\subsubsection{Opacity correction}
\label{sec:opacity.correction}

Following {\citet{Foucart:2016rxm}}, we correct absorption and scattering
opacities by a factor
\[ \left( \frac{\varepsilon_{\nu}}{\varepsilon_{\nu, \tmop{eq}}} \right)^2, \]
where $\varepsilon_{\nu}$ is the average incoming neutrino energy and
$\varepsilon_{\nu, \tmop{eq}}$ is the average neutrino energy at the
thermodynamic equilibrium (computed as in the previous section). This
correction is applied prior to the imposition of Kirchhoff's law, to ensure
the preservation of the correct equilibrium.

\subsubsection{Initial guess}
\label{sec:methods.matter.iniguess}

In order to initialize the implicit solver for Eq. \eqref{implicit.update} we
proceed as follows.
\begin{enumerate}
  \item We update the radiation fields according to the non-stiff part of the
  equations. For the first substep this update reads:
  \begin{equation}
    \tilde{\tmmathbf{U}} =\tmmathbf{U}^{(k)} + \Delta t (\tmmathbf{G}
    [\tmmathbf{U}^{(k)}] - \partial_i \tmmathbf{F}^i [\tmmathbf{U}^{(k)}]) .
  \end{equation}
  A similar formula is used for the second substep, but using
  $\tmmathbf{U}^{\ast}$ to evaluate the terms in the parenthesis.

  \item The M1 closure is updated and quantities are transformed to the fluid
  frame to obtain $\tilde{J}$ and $\widetilde{H_i}$.

  \item We compute new values $\hat{J}$ and $\widehat{H_i}$ in the fluid rest
  frame according to
  \begin{align}
    \hat{J} & = \tilde{J} + \frac{\Delta t}{W}  (\eta - \kappa_a  \hat{J}), \\
    \widehat{H_i} & = \widetilde{H_i} - \frac{\Delta t}{W} (\kappa_a + \kappa_s)  \widehat{H_i} .
  \end{align}
  $\widehat{H_0}$ is obtained from the requirement that $\hat{H}_{\alpha}
  u^{\alpha} = 0$.

  \item Finally, the initial guess for Eq. \eqref{implicit.update} is obtained
  by transforming the radiation quantities to the laboratory frame. For this
  transformation we take $\chi = 1 / 3$, since the initial guess becomes
  important only in the optically thick limit.
\end{enumerate}
It is important to remark that $\hat{J}$ and $\hat{H}_{\alpha}$ are exact
solution only at leading order in $v / c$, when $u^{\alpha}
\partial_{\alpha} \simeq W \partial_t$. It would be incorrect to take the
obtained $\hat{E}$ and $\hat{F}_i$ as the updated solution, even if we
were to update the closure before boosting back the solution to the
laboratory frame. However, \tmtexttt{THC\_M1} only uses $\hat{E}$ and
$\hat{F}_i$ as initial guesses for the full non-linear solver. An
exception, is the test in Sec.~\ref{sec:diff.moving}, where we show that
using the $\hat{E}$ and $\hat{F}_i$ as the new states for the radiation
fields, instead of performing a nonlinear solve, result in large errors
in the case of moving media.

% ------------------------------------------------
\section{Test Problems}
\label{sec:tests}
% ------------------------------------------------
We validate \texttt{THC\_M1} by performing a series of demanding tests
meant to independently verify different components of the code. This
section describes the most representative tests we have performed. Most
of the tests discussed here are fairly standard and have been considered
by a number of authors, although with some differences in the setup
\citep[e.g.][]{Audit:2002hr, 2011JQSRT.112.1323V, Radice:2012tg,
McKinney:2013txa, Foucart:2015vpa, Skinner:2018iti, Weih:2020wpo,
2020ApJ...900...71A}.

\subsection{Optically Thin Advection Through a Velocity Jump}
As a first test we consider the propagation of beam of radiation in an
optically thin medium. We assume slab geometry and consider initial data
with $E(t=0, z) = H(z + \frac{1}{2})$ (arbitrary units), where $H$ is
the Heaviside function, and $F^z = E$. The background fluid velocity is
chosen to be:
\[
  \upsilon^z(z) = \begin{cases}
    0.87, & z < 0, \\
    -0.87, & z > 0.
  \end{cases}
\]
That is, the medium is moving with Lorentz factor $W = 2$ in the grid
frame and the two parts of the domain with $z < 0$ and $z > 0$ have a
relative Lorentz factor of $7$. The fluid is taken to be transparent. We
set $\Delta z = 0.01$. The time step is chosen so as to have a CFL of
$0.5$. It is important to emphasize that, although matter and radiation
do not interact in this test, because our closure is defined in the
fluid frame (Eq.~\ref{flux.factor}), the equations become stiff in the
limit in which $W \gg 1$, so this is actually a rather demanding test.

\begin{figure}
  \includegraphics[width=\columnwidth]{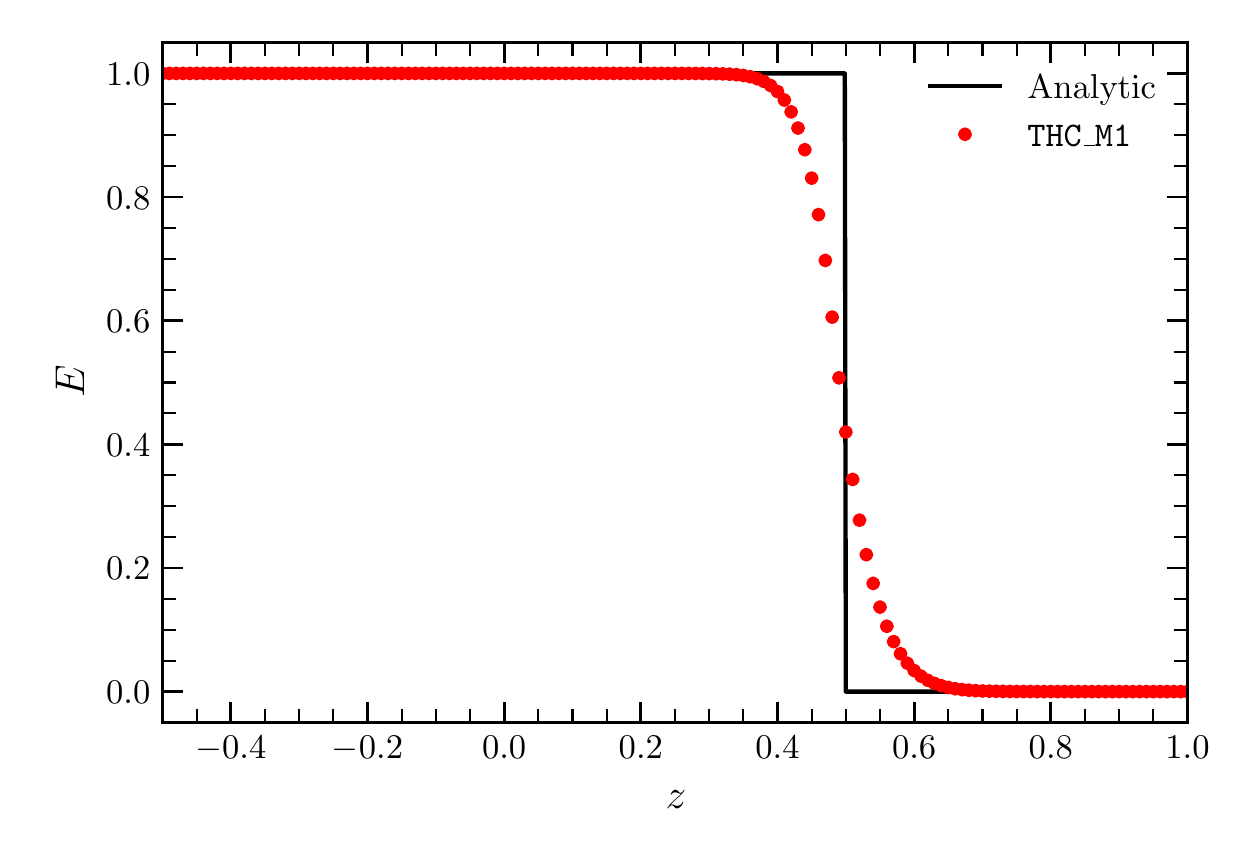}
  \caption{Optically thin advection of radiation through a large
  velocity discontinuity. The frame in which we compute the closure has
  a velocity $0.87\ {\rm c}$ for $z < 0$ and a velocity of $-0.87\ {\rm
  c}$ for $z > 0$ (the relative Lorentz factor between left and right
  state is 7). No artifact appears as \texttt{THC\_M1} advects a pulse
  of radiation through the interface at $z = 0$.}
  \label{fig:beam1d}
\end{figure}

Figure~\ref{fig:beam1d} shows the radiation energy density profile at
time $t = 1$, after the beam has propagated through the velocity jump at
$z = 0$. As it can be seen from the figure, \texttt{THC\_M1} transports
the radiation front through the shock without creating artificial
oscillations. The discontinuity is spread over many grid cells, since
\texttt{THC\_M1} uses a rather dissipative central scheme to handle the
transport operator in the M1 formalism
(Sec.~\ref{sec:method.advection}). However, since neutrino sources do
not switch on abruptly, a sharp preservation of the radiation front is
not a critical modeling requirement for our applications. Being able to
handle transport through fast moving media is, instead, critical for
\ac{NS} merger applications, since the outflows produced in these events
can be mildly relativistic ($W \lesssim 2$) \citep{Hotokezaka:2018gmo,
Nedora:2021eoj}. This test demonstrates that \texttt{THC\_M1} meets this
requirement.

\subsection{Diffusion Limit}

Another requirement for the modeling of \ac{NS} mergers is to correctly
handle the diffusion of neutrinos from the central remnant on secular
timescales. As discussed in Sec.~\ref{sec:method.advection},
\texttt{THC\_M1} uses a numerical scheme designed to correctly capture
the scattering dominated limit.

To validate it, we consider a purely scattering medium of constant
density $\rho = 1$ (arbitrary units) and with scattering opacity
$\kappa_s = 10^3$. As in the previous test, we assume slab geometry, so
this is effectively a 1D problem. Initially, radiation is concentrated
in the region $[-0.5, 0.5]$ and is spatially homogeneous and isotropic
in this region.  That is, $E(t=0, z) = H(z + 0.5) - H(z - 0.5)$ and $F^i
= 0$. In these conditions, when considering timescales longer than the
equilibration time, the radiative transfer equation can be well
approximated by the diffusion equation:
\begin{equation}\label{eq:diffusion}
  \partial_t E = \frac{1}{3 \kappa_s} \partial_x^2 E.
\end{equation}
\texttt{THC\_M1} solves the equations in hyperbolic form
\eqref{3p1.system}. Typical hyperbolic solvers have numerical diffusion
with an effective diffusion coefficient $\nu_{\rm num} \sim (\Delta
z)^{-1}$. In essence, this means that standard numerical schemes fail to
predict the correct diffusion of radiation in a scattering dominated
region, unless the mean free path of the neutrinos (or photons) is well
resolved on the grid \citep{2002IJNMF..40..479R, 2008JCoPh.227.9711M}.
Given that the mean free path of neutrinos at the center of a \ac{NS}
merger remnant is of the order of a few meters or less, the resolution
requirements for merger simulations would be extremely demanding. To
work around this issue, \texttt{THC\_M1} uses a special numerical scheme
for which $\nu_{\rm num} \to 0$ when $\kappa_s (\Delta z) \gtrsim 1$
(see Sec.~\ref{sec:method.advection}). In this respect, our approach is
different from that of \citet{Foucart:2015vpa}, which instead solve the
heat diffusion equation in the scattering regime.

\begin{figure}
  \includegraphics[width=\columnwidth]{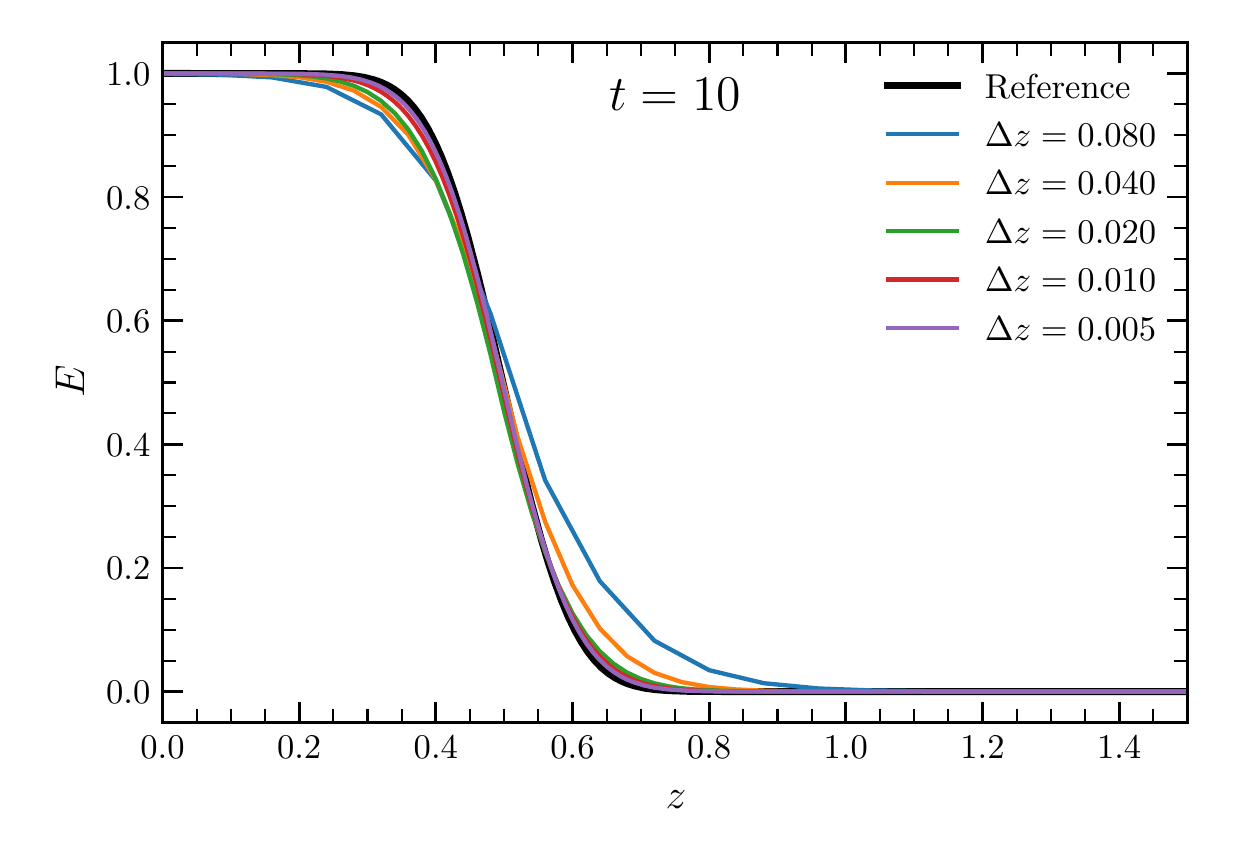}
  \caption{Diffusion of radiation in a purely scattering
  medium. Initially the radiation has a square profile. The reference
  profile is a semi-analytic solution of the diffusion equation.
  \texttt{THC\_M1} correctly capture the diffusion limit. We measure
  first order convergence for this test problem.}
  \label{fig:diff.profile}
\end{figure}

Figure \ref{fig:diff.profile} shows the radiation energy density profile
at time $t = 10$ at different resolution.  The CFL is set to $0.625$ in
all calculations. The reference solution is a semi-analytic solution of
Eq.~\eqref{eq:diffusion}. We find that \texttt{THC\_M1} captures the
correct diffusion rate for radiation even when $\kappa_s (\Delta z) \gg
1$. The numerical solutions are non oscillatory, even though the initial
radiation profile is discontinuous and slope limiting is essentially
disabled in the scattering dominated regime. We measure 1\textsuperscript{st} order
convergence in this test, which is the expected order of convergence
given that the initial data is discontinuous.

\subsection{Diffusion Limit in a Moving Medium}
\label{sec:diff.moving}

Matter in \ac{NS} mergers is not only optically thick, but also moving
at mildly relativistic velocities. Correctly capturing the advection of
trapped radiation in moving media is one of the key challenges in
radiation hydrodynamics and is of crucial importance for both mergers
and core-collapse supernovae \citep{Nagakura:2014nta, Chan:2020quo}.
This requires a careful treatment of the radiation matter coupling in
the stiff limit.

To demonstrate that our code can handle this, we consider a constant
density, purely scattering medium with $\rho = 1$ and $\kappa_s = 10^3$,
which we take to be moving towards the right with velocity $\upsilon^z =
0.5$. Once again, we assume slab geometry. We setup a Gaussian pulse of
radiation centered around the origin:
\begin{equation}
  E(t=0, z) = e^{-9 z^2}.
\end{equation}
To initialize the radiation flux, we use Eqs.~\eqref{tmunu.thick},
\eqref{energy.thick}, and \eqref{flux.thick} under the assumption of
fully trapped radiation ($H^\alpha = 0$) to write
\begin{equation}
  J = \frac{3 E}{4 W^2 - 1}, \qquad
  F_i = \frac{4}{3}\, J\, W^2\, \upsilon_\alpha.
\end{equation}
The exact solution corresponds to a slowly diffusing and translating
pulse of radiation. The baseline grid spacing adopted for this problem
is $\Delta z = 0.01$ and the CFL is fixed to $0.625$.

\begin{figure}
  \includegraphics[width=\columnwidth]{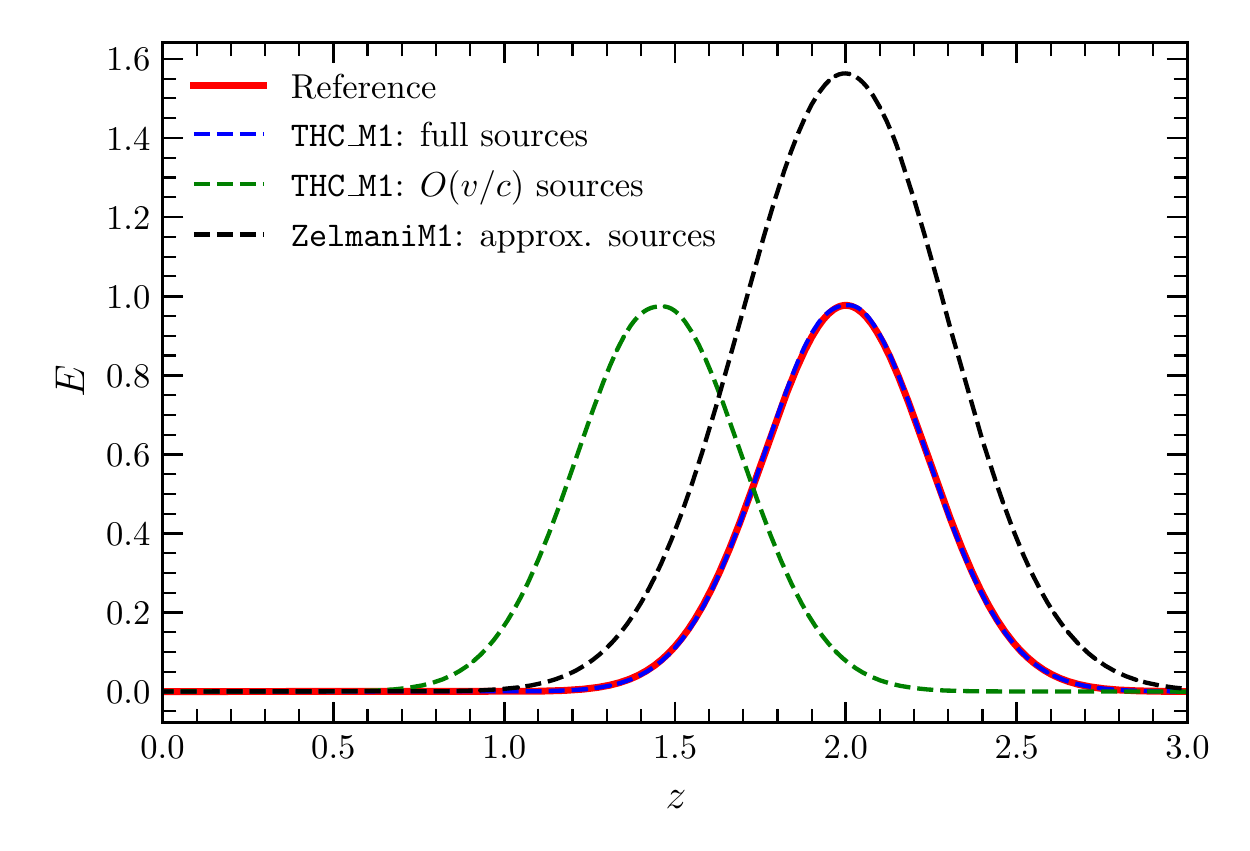}
  \caption{Diffusion and advection of Gaussian pulse of radiation in a
  purely scattering moving medium. The medium is moving with velocity
  $\upsilon = 0.5$. The reference profile is a translated semi-analytic
  solution of the diffusion equation. Our results show that it is
  essential to properly treat all of the source terms in the M1
  equations to correctly capture the advection of trapped radiation.}
  \label{fig:diff.moving}
\end{figure}

Figure \ref{fig:diff.moving} shows the results obtained using different
schemes. The reference profile is a semi-analytic solution of the
diffusion equation advected along the background fluid velocity. We find
that \texttt{THC\_M1} reproduce the correct solution when all the
nonlinear terms in the sources are consistently treated. This ensures
that neutrinos will not be ``left behind'' as the \ac{NS} merger
remnant, typically deformed into a bar \citep{Shibata:2005xz},
rotates.

We remark that the solution of the full nonlinear source term is
computationally expensive, however cheaper alternatives fail to capture
the correct behaviour of the trapped radiation. A first order in $v/c$
approach obtained by Lorentz transforming the radiation moments to and
from the reference frame as discussed in
Sec.~\ref{sec:methods.matter.iniguess} produces a stable evolution, but
predicts the wrong advection speed for the radiation energy
(Fig.~\ref{fig:diff.moving}). Even worse, this approach predicts
different advection speeds for the neutrino number densities (not shown)
and the radiation energy density which produce large errors in the
average neutrino energies.

The treatment of the optically thick limit of \texttt{ZelmaniM1}
\citep{Roberts:2016lzn}, which is similar to the approach used in SpEC
\citep{Foucart:2015vpa}, is also problematic and affected by two
important issues. First, the diffusive fluxes corrected using the
(acausal) heat diffusion equation significantly overestimate the rate of
diffusion for the radiation, resulting in a significant broadening of
the radiation pulse. Minor improvements in the diffusion rate can be
obtained by implementing a better treatment using modified HLLE fluxes
following \citet{Skinner:2018iti}. Second, because of the approximation
in the source terms, the \texttt{ZelmaniM1} solution violates energy
conservation and the radiation energy density increases with time
(Fig.~\ref{fig:diff.moving}). The violation of energy conservation is
exacerbated in this problem, because there is no back reaction of the
radiation onto the matter. In a more realistic setting,
\texttt{ZelmaniM1} would enforce energy conservation, so the increase in
the radiation energy density would come at the expense of the fluid
kinetic energy. That is matter would experience an unphysical drag force
driving it to rest in the grid frame.

\begin{figure}
  \includegraphics[width=\columnwidth]{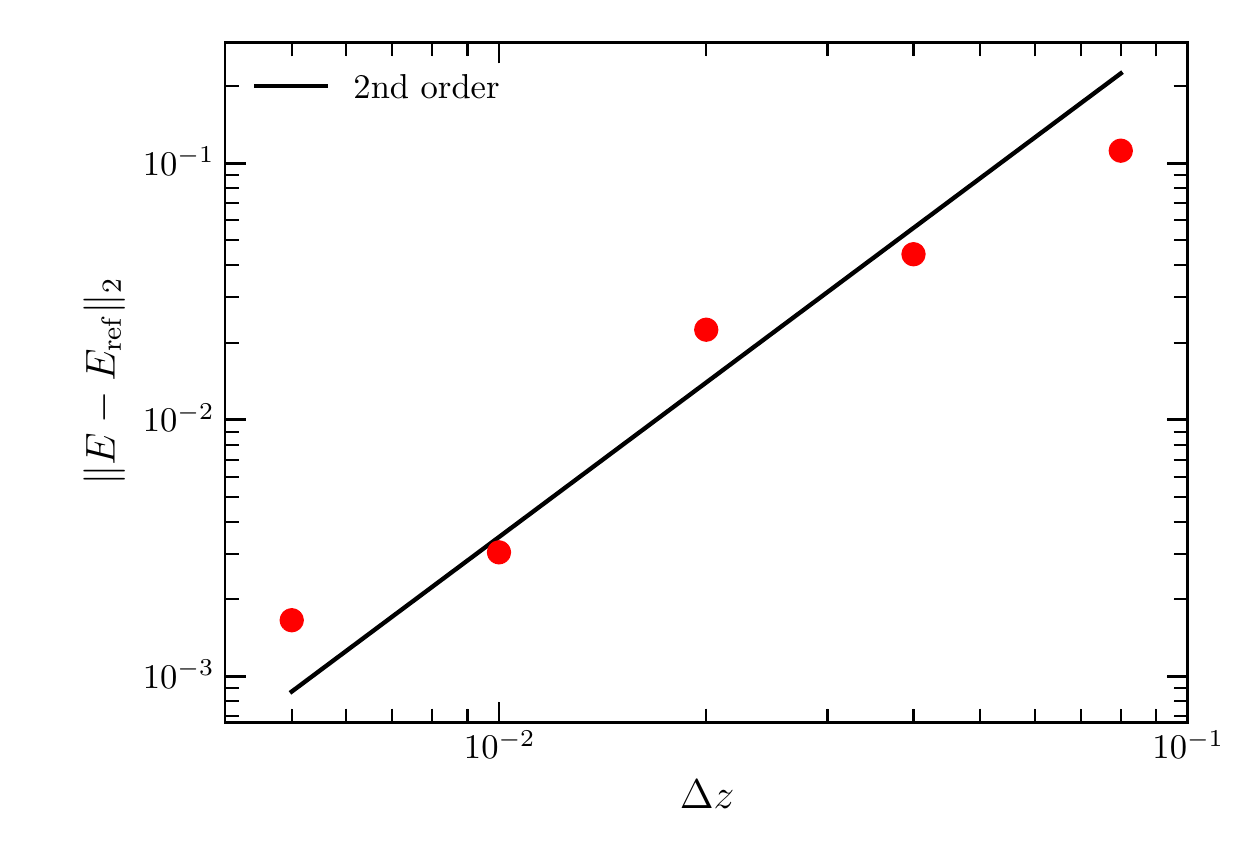}
  \caption{Convergence of the \texttt{THC\_M1} to the reference solution
  for the diffusion test in a moving medium. We find an approximate 2\textsuperscript{nd}
  order convergence.}
  \label{fig:diff_moving_conv}
\end{figure}

We perform additional simulations with $\Delta z = 0.16, 0.08, 0.04$,
and $0.02$ in addition to $\Delta z = 0.01$. The $L^2$ norm of the
difference between the \texttt{THC\_M1} solution with the complete
treatment of the radiation-matter source terms and the semi-analytic
solution is presented in Fig.~\ref{fig:diff_moving_conv}. Overall, we
find second order convergence for \texttt{THC\_M1} in this test.

\subsection{Shadow Test}

\begin{figure}
  \includegraphics[width=\columnwidth]{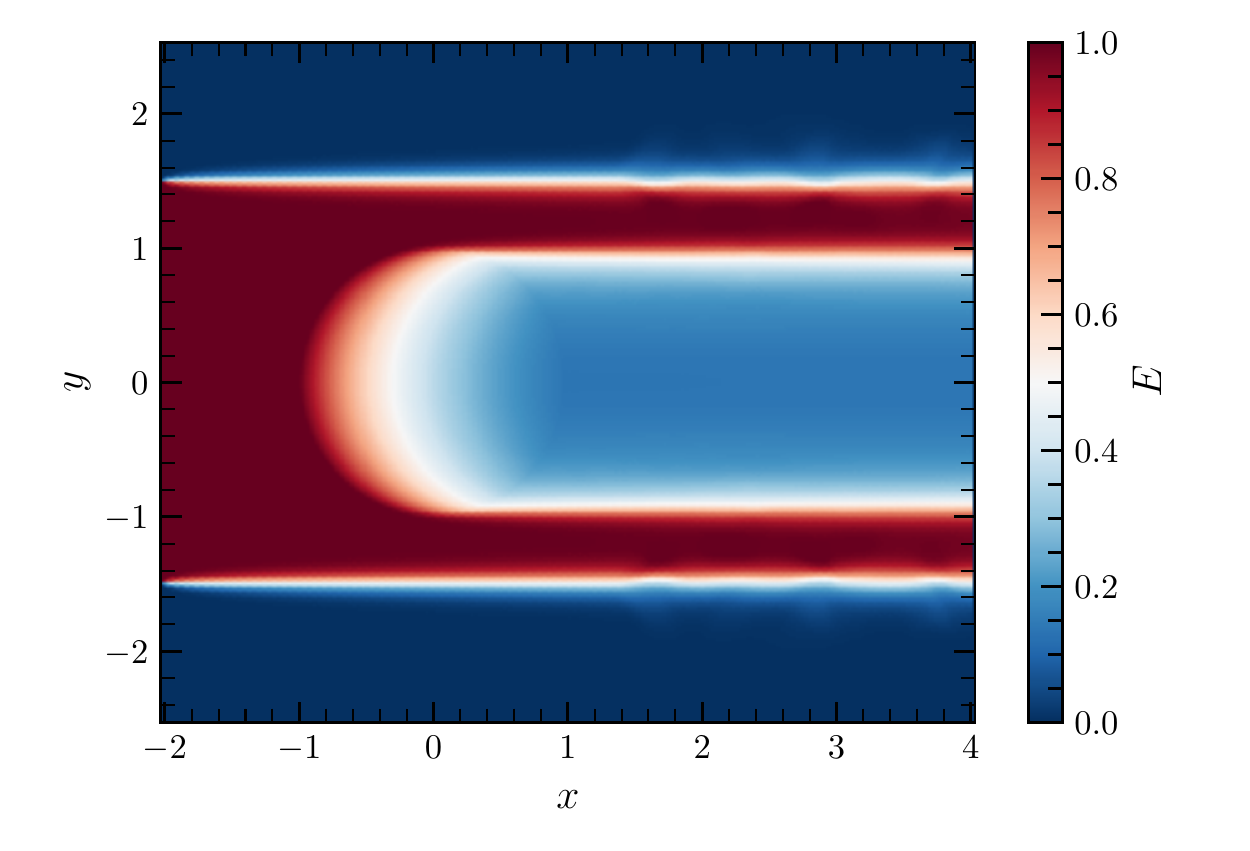}
  \caption{Shadow cast by an absorbing cylinder illuminated by a beam of
  radiation propagating from the left to the right. \texttt{THC\_M1}
  correctly captures the formation of the shadow.}
  \label{fig:shadow}
\end{figure}

As a first multidimensional test, we consider the problem of a beam of
radiation interacting with a semi-transparent cylinder with radius
$\varpi = 1$ centered at the origin. The absorption opacity in the
cylinder is set to $\kappa_a = 1$ and the density to $1$. Absorption is
zero elsewhere. The scattering opacity $\kappa_s$ is set to zero. We
initialize the radiation fields to zero and inject a beam of radiation
from the left of the domain with $F^x = E = 1$. The grid spacing used in
this test is $\Delta x= \Delta y = 0.0125$ and the CFL is set to $0.4$.

Figure~\ref{fig:shadow} shows the radiation energy density at time $t =
10$, when the solution has achieved steady state.  We observe some
lateral diffusion of radiation and the formation of small unsteady
oscillation in the radiation field in the wake of the cylinder. The
latter are artefacts caused by the nonlinear nature of the Minerbo
closure. Nevertheless, \texttt{THC\_M1} correctly captures the overall
solution. The attenuation of radiation in the cylinder and the formation
of a shadow behind it agree with the analytic solution for this problem.

\subsection{Homogeneous Sphere}

\begin{figure}
  \includegraphics[width=\columnwidth]{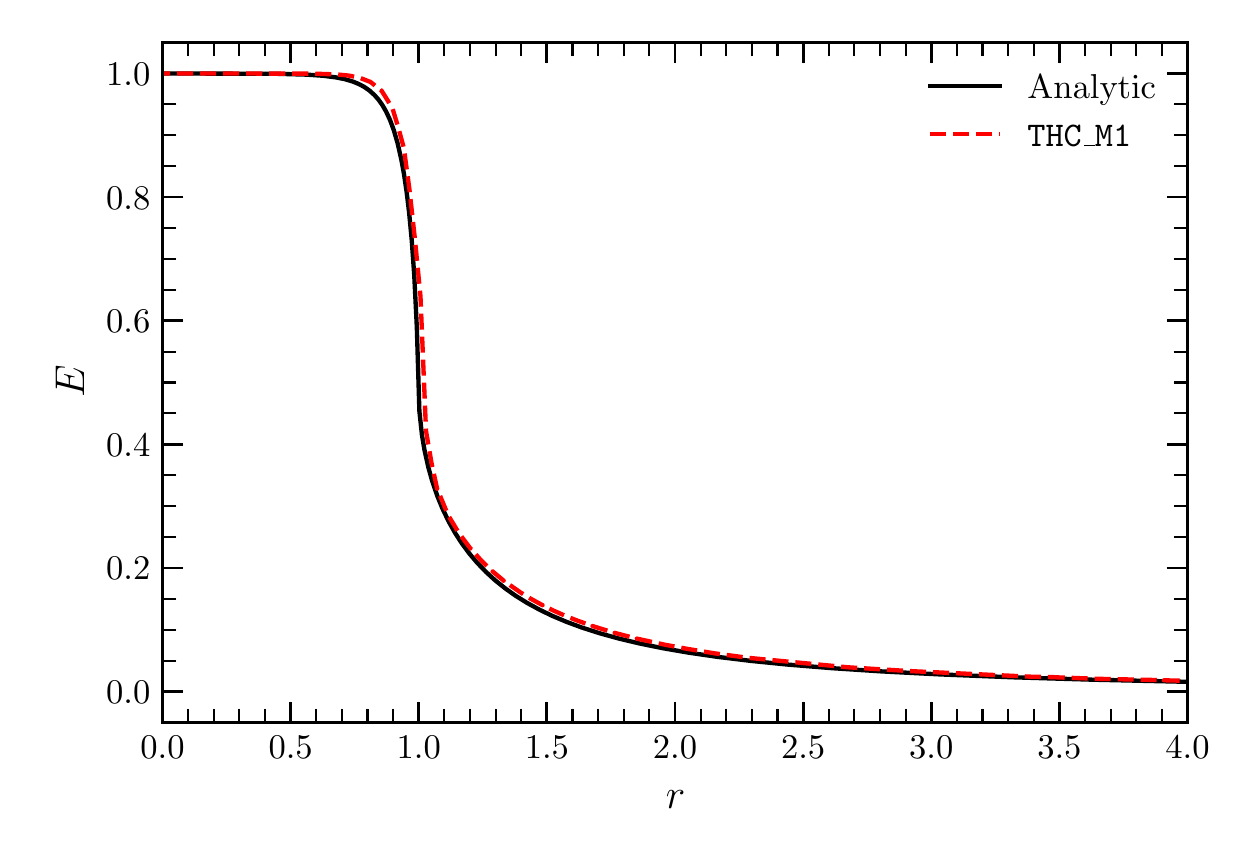}
  \caption{Radiation from an homogeneous absorbing and emitting sphere
  of radius one. The reference solution is obtained by solving the
  radiative transfer equation. The \texttt{THC\_M1} solution agrees well
  with the analytical solution. Small deviations are expected, since the
  adopted closure is not expected to reproduce the solution to the
  angle-dependent radiative transfer equations.}
  \label{fig:sphere}
\end{figure}

The homogeneous sphere test has been considered by many authors, since
it reproduces the typical geometry encountered in astrophysical
applications. In this test an homogeneous sphere, which we take to be of
radius $r = 1$, emits and absorbs radiation at a constant rate $\eta =
\kappa_a = 10$. Scattering is neglected in this problem, so it is
possible to compute an exact solution of the radiative transfer
equations by numerical quadrature. This is an extremely idealized model
of a hot protoneutron star or a neutron star merger remnant emitting
neutrinos. We perform this test in full 3D and in Cartesian coordinates. The
resolution adopted for this test is $\Delta x = \Delta y = \Delta z =
0.0125$. This corresponds to about 80 points along the radius of the
``star'', a typical resolution for production neutron star merger
simulations. To reduce the computational costs, we impose reflection
symmetry across the $xy$, $xz$, and $yz$ planes and only simulate the
part of the domain with $x,y,z \geq 0$. The CFL is set to $0.3$.

Figure \ref{fig:sphere} shows the radiation energy density as a function
of radius in the diagonal direction at time $t = 10$, when the solution
has reached steady state.  \texttt{THC\_M1} does not solve the full
radiative transfer equations, so the numerical solution is not expected
to converge to the exact solution. Nevertheless, the \texttt{THC\_M1}
solution shows excellent agreement with the analytic solution and even
compares favorably with the full-Boltzmann $F\!P_N$ solution presented
in \citet{Radice:2012tg} for modest angular resolutions.

\subsection{Gravitational Light Bending}

\begin{figure}
  \includegraphics[width=\columnwidth]{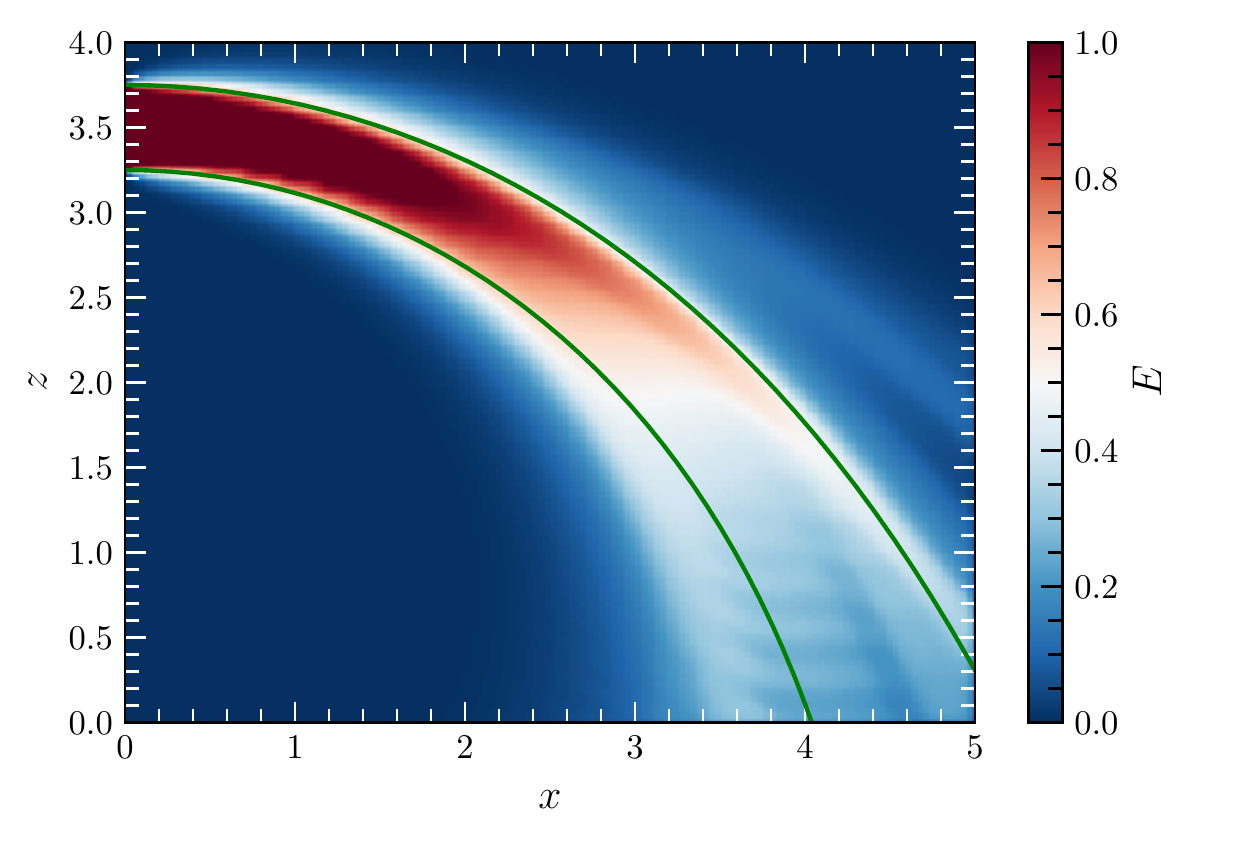}
  \caption{Beam of radiation propagating in the meridional plane of a
  nonrotating black hole (BH) in Kerr-Schild coordinates. The mass of the
  BH is set to be one. The green lines show null geodesics.
  \texttt{THC\_M1} correctly captures the bending of the beam of
  radiation in the strong gravitational field of the BH.}
  \label{fig:kerrschild}
\end{figure}

Finally, we present a test validating the implementation of spacetime
curvature source terms in \texttt{THC\_M1}. We study the propagation of
a beam of radiation in a \ac{BH} spacetime described by the Kerr-Schild
metric. The \ac{BH} spin is set to zero and its mass to one (in
geometrized units). The computational grid is centered at the location
of the \ac{BH}. We only consider the region near the meridional plane $y
= 0$ and $x, z > 0$. We simulate a beam of radiation injected at the
location $x = 0$ and $z = 3.5$ propagating towards the positive
$x$-direction (see Fig.~\ref{fig:kerrschild}). In particular, we set $E
= 1$ at the beam injection location. The fluxes $F_i$ are set so that
$\alpha F^i - \beta^i E$ is along the $x$-axis and $F_i F^i = 0.99 E^2$.
The resolution used for this test is $\Delta x = \Delta z = 0.025$ and
the CFL is set to $0.2$.

Figure~\ref{fig:kerrschild} shows the \texttt{THC\_M1} solution at time
$t = 20$, after steady state has been achieved. We also plot two
analytically predicted trajectories for photons (null geodesics) in the
same metric. \texttt{THC\_M1} correctly captures the bending of
radiation due to the \ac{BH}, indicating that curvature terms have
been implemented correctly. The \texttt{THC\_M1} solution shows lateral
diffusion of radiation comparable to other \ac{GR} M1 codes
\citep{McKinney:2013txa, Foucart:2015vpa, Weih:2020wpo}. This later
diffusion is a numerical artefact. However, we do not consider this to
be as a significant issue for our approach, because isolated beams of
radiation are never found in the astrophysical systems we intend to
model.

% ------------------------------------------------
\section{Neutron Star Mergers}
\label{sec:nsmerger}
% ------------------------------------------------
As a first application of \texttt{THC\_M1}, we consider the late
inspiral and merger of a binary of two $1.364\ M_\odot$ \acp{NS}. We
adopt the SRO(SLy4) \ac{EOS} (SLy for brevity in the rest of the text;
\citealt{Douchin:2001sv, Schneider:2017tfi}). To ease the comparison
with previous results, we use the same set of reactions and opacities as
in \citet{Radice:2018pdn}. We construct initial data with an initial
separation of 45~km using the \texttt{Lorene} pseudo-spectral code
\citep{Gourgoulhon:2000nn}. We have already considered this initial data
in \citet{Endrizzi:2019trv} and \citet{Nedora:2020hxc}, to which we
refer for more details.  The evolution grid employs 7 levels of
\ac{AMR}, with the finest grid having finest grid spacing of $h = 0.25\
G M_\odot/c^2, 0.167\ G M_\odot/c^2$ and $0.125\ G M_\odot/c^2$,
respectively denoted as VLR, LR, and SR setups. For this purpose, we use
the \texttt{Carpet} \ac{AMR} driver \citep{Schnetter:2003rb,
Reisswig:2012nc} of the \texttt{Einstein Toolkit} \citep{Loffler:2011ay,
EinsteinToolkit:2021_05}. \texttt{Carpet} implements the Berger-Oilger
scheme with refluxing \citep{Berger:1984zza, 1989JCoPh..82...64B}.
\texttt{THC} can make use of this infrastructure to ensure mass and
energy conservation as matter flows between different refinement levels.
However, since the current implementation of refluxing in
\texttt{Carpet} is memory intensive, we do not employ it for the
radiation variables.
To have a baseline for comparison, in addition to the
simulations with \texttt{THC\_M1}, we perform three simulations using
the M0+Leakage neutrino scheme \citep{Radice:2016dwd, Radice:2018pdn}.
This is the current methodology employed for neutrino transport in
production simulations with the \texttt{THC} general-relativistic
hydrodynamics code \citep{Radice:2012cu, Radice:2013hxh, Radice:2013xpa,
Radice:2015nva}. However, we have updated the M0 scheme to compute
neutrino opacities using the approach discussed in
Sec.~\ref{sec:blackbody}. Although \texttt{THC} has the ability to model
subgrid-scale viscous angular momentum transport using the GRLES
formalism \citep{Radice:2017zta, Radice:2020ids}, we do not employ it in
the simulations presented here.

\subsection{Merger Dynamics}

Our simulations span the last ${\sim}4$ orbits of the binary prior to
merger, the merger, and extend to ${\sim}10$ milliseconds after the
merger. After the star come into contact, the remnant experiences one
centrifugal bounce before collapsing to \ac{BH}. We use the
\texttt{AHFinderDirect} \cite{Thornburg:2003sf} thorn of the
\texttt{Einstein Toolkit} to locate an apparent horizon. Both the M0
scheme and \texttt{THC\_M1} excise the region inside the apparent
horizon. Both codes handle \ac{BH} formation well, but, due to the low
resolution, the \ac{BH} experiences an unphysical drift starting from
${\sim}5{-}10$ milliseconds after merger. The drift is particularly large
for the M0 runs and eventually the code fails when the \ac{BH} leaves the
finest refinement level in the grid. The M1 runs, instead, experience
smaller drifts. The M1 LR run fails at ${\sim}12$ milliseconds after
merger, while the M1 VLR and SR runs remain stable for the entire
duration of the simulation. Such drifts are known to be the result of
issues in the shift gauge condition, they are often seen in simulations,
and some fixes have been proposed \citep{Bruegmann:2006ulg, Most:2021ytn,
Shibata:2021xmo}. We remark that such drifts are also seen in
purely-hydrodynamics simulations, so this issue does not appear to
be connected with the neutrino treatment. Since we are not interested in
evolving the system for long times after \ac{BH} formation, we do not
attempt to address this issue here.

\begin{figure}
  \includegraphics[width=\columnwidth]{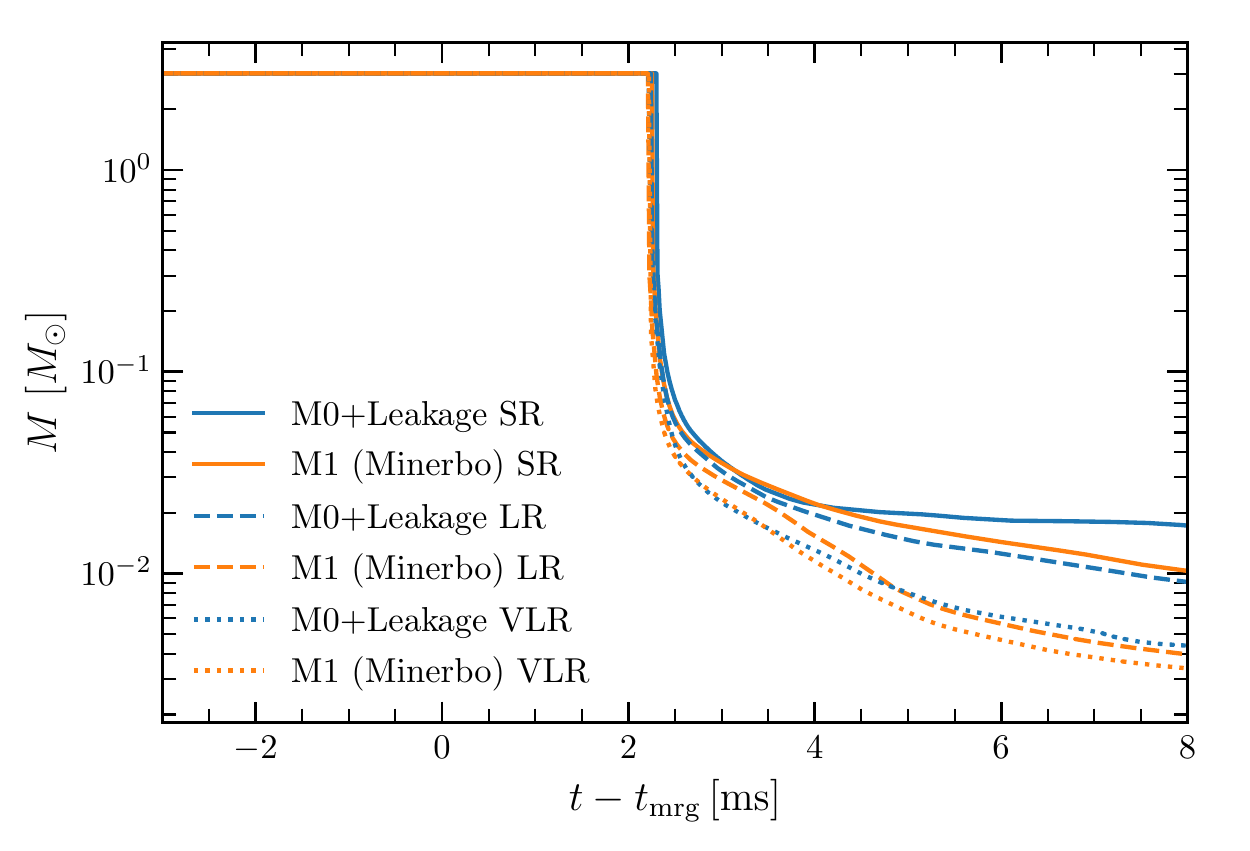}
  \caption{Rest mass outside of the apparent horizon for the SLy4 $1.364\ M_\odot -
  1.364\ M_\odot$ binary as a function of time. The figure shows the
  results for three resolutions (VLR, LR, and SR) and two radiation
  transport methods. There is a persistent trend of increasing disk mass
  with resolution, but we find good agreement between M0 and M1 results.}
  \label{fig:bns.1364.dens}
\end{figure}

Figure~\ref{fig:bns.1364.dens} shows the amount of (rest) mass
outside of the \ac{BH} apparent horizon as a function of time. The mass
of the accretion disk increases monotonically with resolution and does
not appear to have fully converged even at the highest resolution
considered in this study. There are differences of up to ${\sim}50\%$ in
the disk mass 8~ms after merger between M1 and M0. However, such
differences are smaller than the overall uncertainty due to finite
resolution effects, suggesting that neutrino transport is not the
dominant source of uncertainty in the merger dynamics over these
timescales.

\begin{figure}
  \includegraphics[width=\columnwidth]{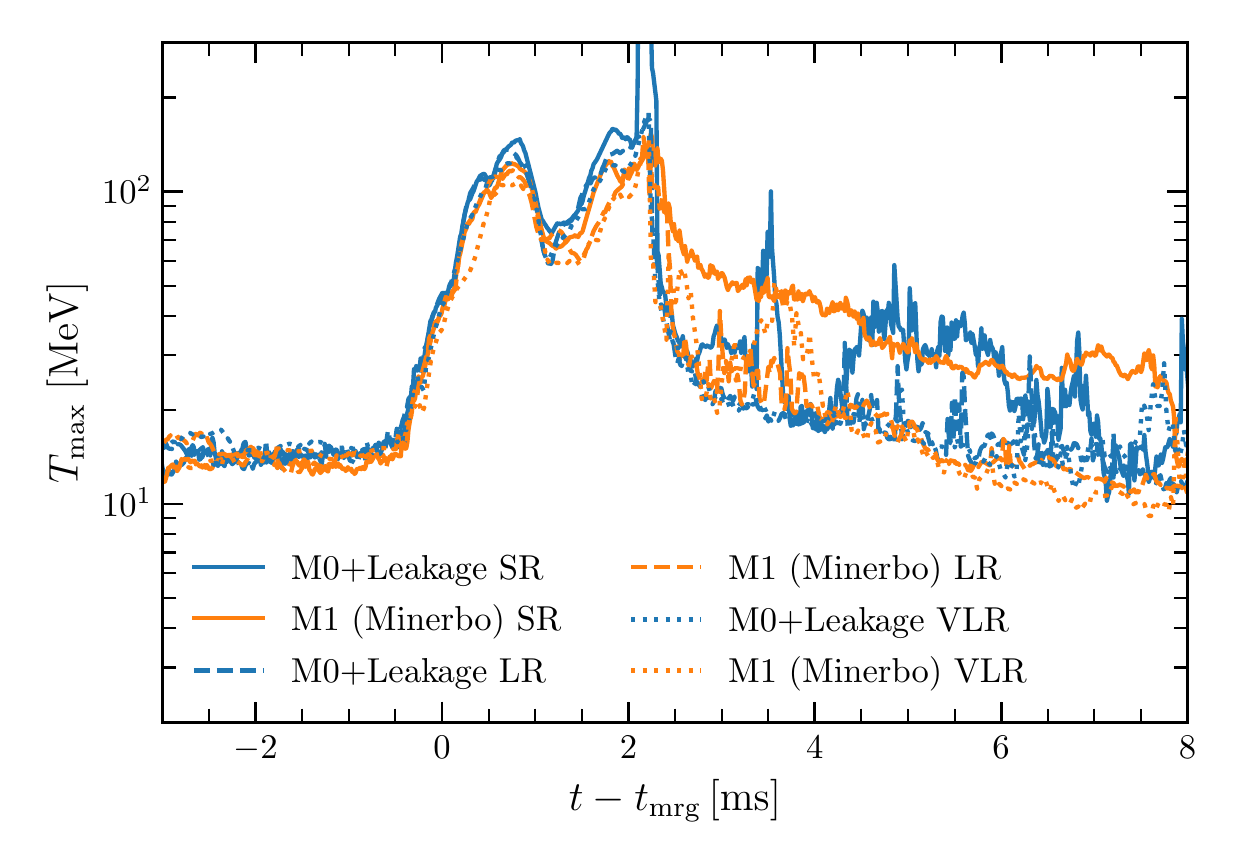}
  \caption{Maximum temperature for the SLy4 $1.364\ M_\odot -
  1.364\ M_\odot$ binary as a function of time. The figure shows the
  results for three resolutions (VLR, LR, and SR) and two radiation
  transport methods.  We find consistent results among all the
  simulations, even after BH formation, demonstrating the robustness of
  our new M1 solver.}
  \label{fig:bns.1364.tmax}
\end{figure}

Figure~\ref{fig:bns.1364.tmax} shows the maximum temperature outside the
apparent horizon. During the inspiral, the surface of the stars is
artificially heated to temperatures exceeding $10\ {\rm MeV}$
\citep{Hammond:2021vtv}. Equal mass systems with soft \acp{EOS}, such as
the one considered here, experience the most violent mergers
\citep{Radice:2020ddv}. Indeed, we observe the temperature to raise to
values in excess of $120\ {\rm MeV}$. This leads to the production of a
dense trapped neutrino gas. This is a very challenging test for a neutrino
radiation-hydrodynamics code, since matter is thrown out of weak
equilibrium and the radiation-matter coupling becomes very stiff.  Our
leakage schemes circumvent this problem by using effective source terms
that are not stiff, but does not capture the correct thermodynamic
conditions of matter in the remnant \citep[see][for a discussion on the
implications]{Perego:2019adq}.  \texttt{THC\_M1}, instead, captures the
correct weak equilibrium of matter inside the star, but at the price of
having to solve a stiff set of equations.

After $t - t_{\rm mrg} \simeq 2\ {\rm ms}$, an apparent horizon is found
and Fig.~\ref{fig:bns.1364.tmax} shows the maximum temperature in the
accretion stream outside of the horizon. Since the highest temperatures
are reached close to the horizon, this data is rather sensitive to
resolution. It also has large excursions when new grid cells are tagged
as being inside the horizon, or when the converse happens. Overall, we
find good agreement between the M0+Leakage and the M1 simulations. This
test demonstrate that \texttt{THC\_M1} can handle even the most
demanding conditions encountered in \ac{NS} mergers.

\begin{figure}
  \includegraphics[width=\columnwidth]{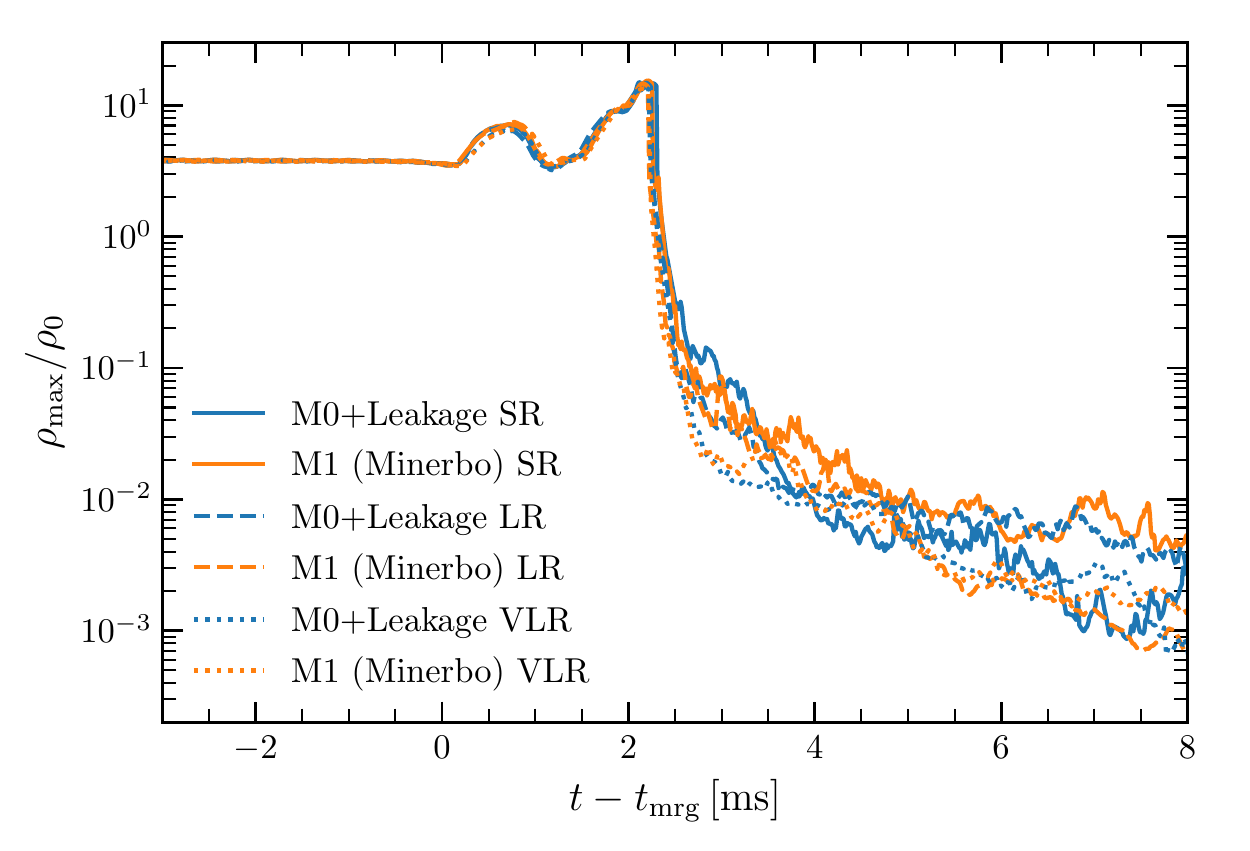}
  \caption{Maximum density for the SLy4 $1.364\ M_\odot -
  1.364\ M_\odot$ binary as a function of time. The figure shows the
  results for three resolutions and three radiation transport methods.
  We find consistent results among all the simulations, even after BH
  formation, demonstrating the robustness of our new M1 solver.}
  \label{fig:bns.1364.rhoc}
\end{figure}

A complementary view on the dynamics of the system can be obtained from
Fig.~\ref{fig:bns.1364.rhoc} which shows the maximum density outside the
apparent horizon. We observe a large oscillation in the maximum density
corresponding to the merger and a subsequent centrifugal bounce,
followed by the collapse. After $t - t_{\rm mrg} \simeq 2$, the figure
shows the maximum density reached in the accretion disk as a function of
time. This figure shows that all simulations are in excellent agreement
in the description of the bulk motion of matter in the system.

\begin{figure}
  \includegraphics[width=\columnwidth]{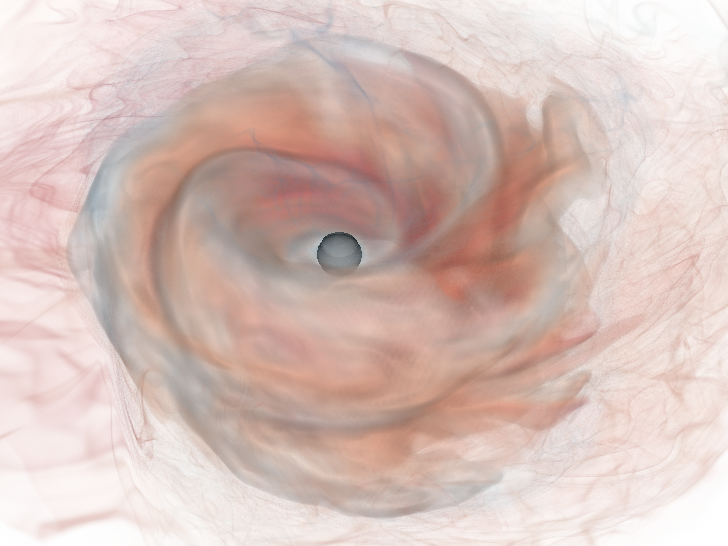}
  \caption{Remnant BH + torus system for the SLy $1.364\ M_\odot -
  1.364\ M_\odot$ M1 (Minerbo) SR binary at $t - t_{\rm mrg} \simeq 2.5\
  {\rm ms}$.  The color code represents $Y_e$ (red: $Y_e < 0.25$; blue:
  $Y_e > 0.25$). The inner grey surface shows the approximate location
  of the apparent horizon ($\alpha = 0.3$; \citealt{Bernuzzi:2020txg}).
  The transparency is set to show only matter with density $\rho \gtrsim
  5 \times 10^{10}\ {\rm g}\ {\rm cm}^{-3}$. The visualization shows the
  data in a box of diameter $118\ {\rm km}$ centered at the origin of
  the coordinate system used in the simulation. We find that the torus
  is in a turbulent state and is far from axisymmetric.}
  \label{fig:bns.1364.ye.3d}
\end{figure}

Figure~\ref{fig:bns.1364.ye.3d} shows the composition of the remnant
accretion torus formed in the highest-resolution M1 binary shortly after
\ac{BH} formation. The disk is primarily composed of matter expelled
from the inner part of the remnant at the time of merger. The accretion
flow is turbulent. The torus has a large $\ell = 2$ deformation, an
imprint of the geometry of the remnant shortly after merger
\citep{Radice:2018pdn}. We find that the bulk of the torus is very
neutron rich, but that its surface layers have higher $Y_e \gtrsim
0.25$ (blue color in the figure).

\subsection{Neutrino Luminosities}

\begin{figure*}
  \includegraphics[width=\textwidth]{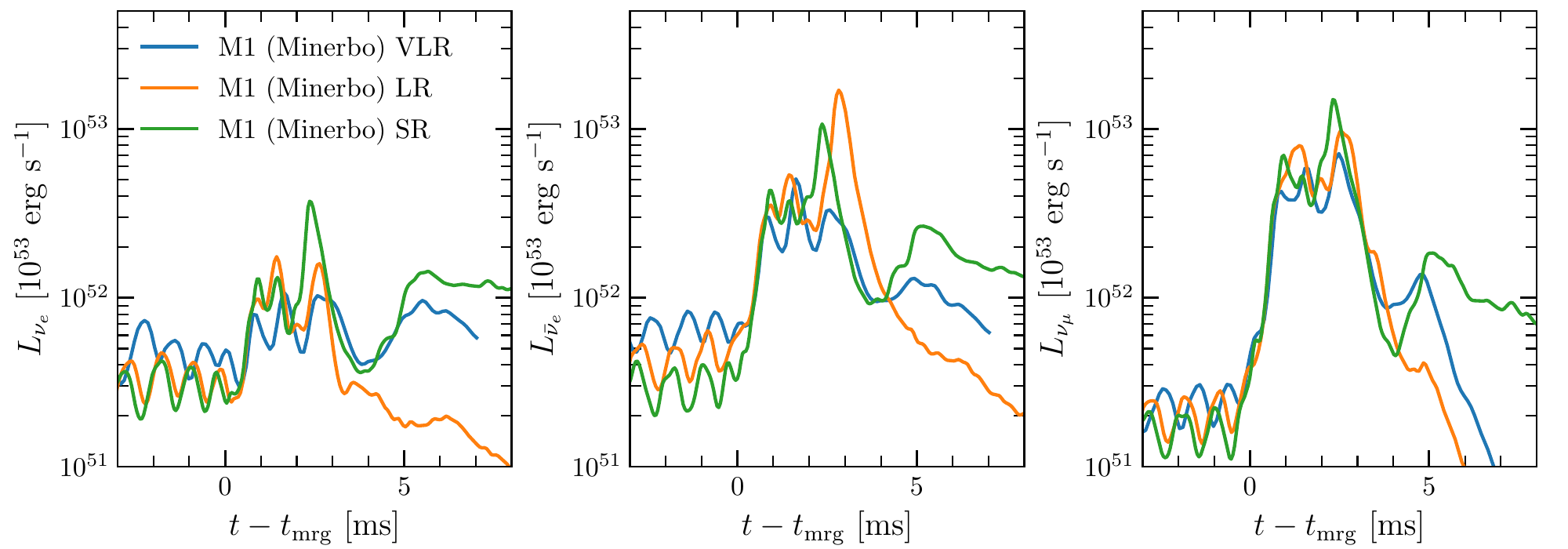}
  \caption{Neutrino luminosity for the SLy $1.364\ M_\odot -
  1.364\ M_\odot$ binary computed with \texttt{THC\_M1} at three
  resolutions. The simulations are in good qualitative agreement at the
  peak of the neutrino burst, but diverge after \ac{BH} formation,
  indicating that the collapse phase is not well resolved in these
  simulations.}
  \label{fig:bns.1364.Lnu}
\end{figure*}

We compute the emergent neutrino luminosities on a coordinate sphere
with radius $r = 300\ G M_\odot /c^2 \simeq 443\ {\rm km}$. The results
are shown in Fig.~\ref{fig:bns.1364.Lnu}. The curves are time shifted to
approximately take into account the time of flight of the neutrinos from
the remnant to the detection sphere. The neutrino luminosity is
artificially large prior to the merger, due to the spurious heating of
the stellar surface discussed in the previous subsection. This effect is
less severe at higher resolutions. The luminosity peaks shortly after
merger and sharply drops following \ac{BH} formation. As dense neutron
rich material is decompressed and heated during merger, it tends to
protonize. As a result, the anti-electron neutrino luminosity is the
highest among all species, while the electron neutrino luminosity is
partially suppressed and is the smallest among all species. Overall, the
different resolutions are in good qualitative and quantitative
agreement, particularly before \ac{BH} formation. Discrepancies are
found after \ac{BH} simulation, likely because of the low resolution
adopted in this study.

The binary considered here has not yet been simulated by other groups,
so detailed comparisons with the literature are not possible.
However, the overall neutrino luminosities are in good qualitative
agreement with those reported by \citet{Foucart:2016rxm},
\citet{Vincent:2019kor}, and \citet{Foucart:2020qjb} for similar
binaries. An important difference is that our luminosities peak at the
time of merger and then drop rapidly after \ac{BH} formation, while the
luminosities shown in the aforementioned works increase monotonically,
since no \ac{BH} is formed in those cases. Moreover, those works report
the neutrino luminosity only for $t - t_{\rm mrg} > 0$. The
luminosities predicted by \texttt{THC\_M1} are in good agreement with
the M0 luminosities, but the M0 data (not shown in
Fig.~\ref{fig:bns.1364.Lnu} to avoid overcrowding the figure) is
truncated shortly after \ac{BH} formation. A more quantitative
comparison between M1 and M0 is discussed in
Sec.~\ref{sec:bns.130.nulum}. The luminosities predicted by M1 are a
factor of several smaller than those predicted by the leakage scheme alone
(not taking into account reabsorption; \textit{cf.}
\citealt{Sekiguchi:2011zd, Palenzuela:2015dqa, Radice:2016dwd,
Lehner:2016lxy}). Our luminosities are also a factor of several smaller
than those predicted by the M1+Leakage scheme of
\citet{Sekiguchi:2015dma, Sekiguchi:2016bjd}.

\begin{figure*}
  \includegraphics[width=\textwidth]{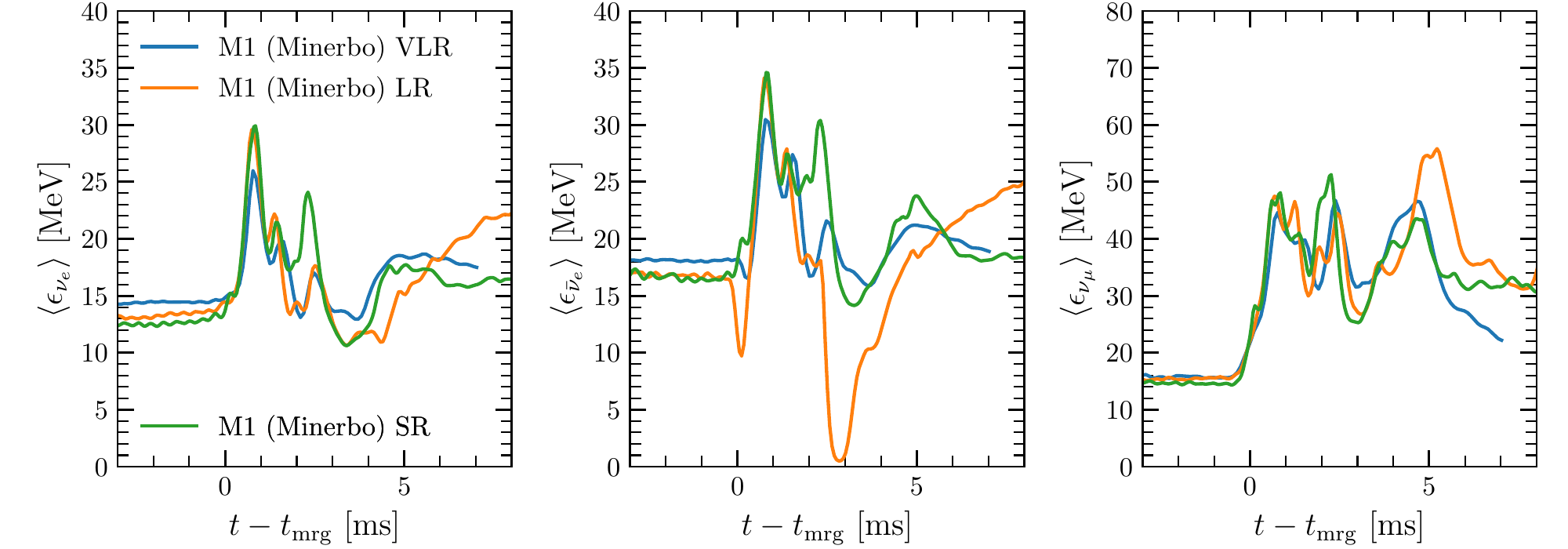}
  \caption{Average neutrino energies for the SLy $1.364\ M_\odot -
  1.364\ M_\odot$ binary computed with \texttt{THC\_M1} at three
  resolutions. We find good qualitative and quantitative agreement
  between the three resolutions. The dip in the average $\bar{\nu}_e$
  for the LR resolution is due to a burst of highly redshifted radiation
  originating in the vicinity of the BH.}
  \label{fig:bns.1364.Enu}
\end{figure*}

The average neutrino energies are also computed on a coordinate sphere
of radius $r = 300\ G M_\odot /c^2 \simeq 443\ {\rm km}$ and are shown
in Fig.~\ref{fig:bns.1364.Enu}. With the exception of the average energy
anti-electron neutrinos in the LR resolution simulation, we find
excellent quantitative agreement between the simulations. The average
energies satisfy the expected hierarchy $\langle \epsilon_{\nu_\mu}
\rangle > \langle \epsilon_{\bar{\nu}_e} \rangle > \langle
\epsilon_{\nu_e} \rangle$ \citep{Ruffert:1998vp, Foucart:2016rxm,
Endrizzi:2019trv, Cusinato:2021zin} and are in good quantitative
agreement with the Monte Carlo simulations of \citet{Foucart:2020qjb},
with the caveat that we are not considering the same binary
configuration. At $t - t_{\rm mrg} \simeq 2.5\ {\rm ms}$ we observe the
formation of a shock in the collapsing remnant of the LR simulation,
just outside the apparent horizon. This generates a burst of neutrinos
that is responsible for the peak in $L_{\bar{\nu}_e}$. Because the
radiation is highly redshifted this results in a dip in $\langle
\epsilon_{\bar{\nu}_e} \rangle$. This feature is absent in the other
resolutions.

\subsection{Dynamical Ejecta}
\label{sec:bns.1364.ejecta}

Material is ejected dynamically during the merger by tidal torques and
shocks \citep{Shibata:2019wef}. We monitor this dynamical ejecta by
computing the flux of matter on a coordinate sphere of radius $r = 300\
G M_\odot /c^2 \simeq 443\ {\rm km}$. We consider a fluid element to be
unbound if its velocity is larger than the escape velocity from the
system ($- u_t > 1$). This is the so-called geodesic criterion
\citep[\textit{e.g.,}][]{Kastaun:2014fna}.

\begin{figure}
  \includegraphics[width=\columnwidth]{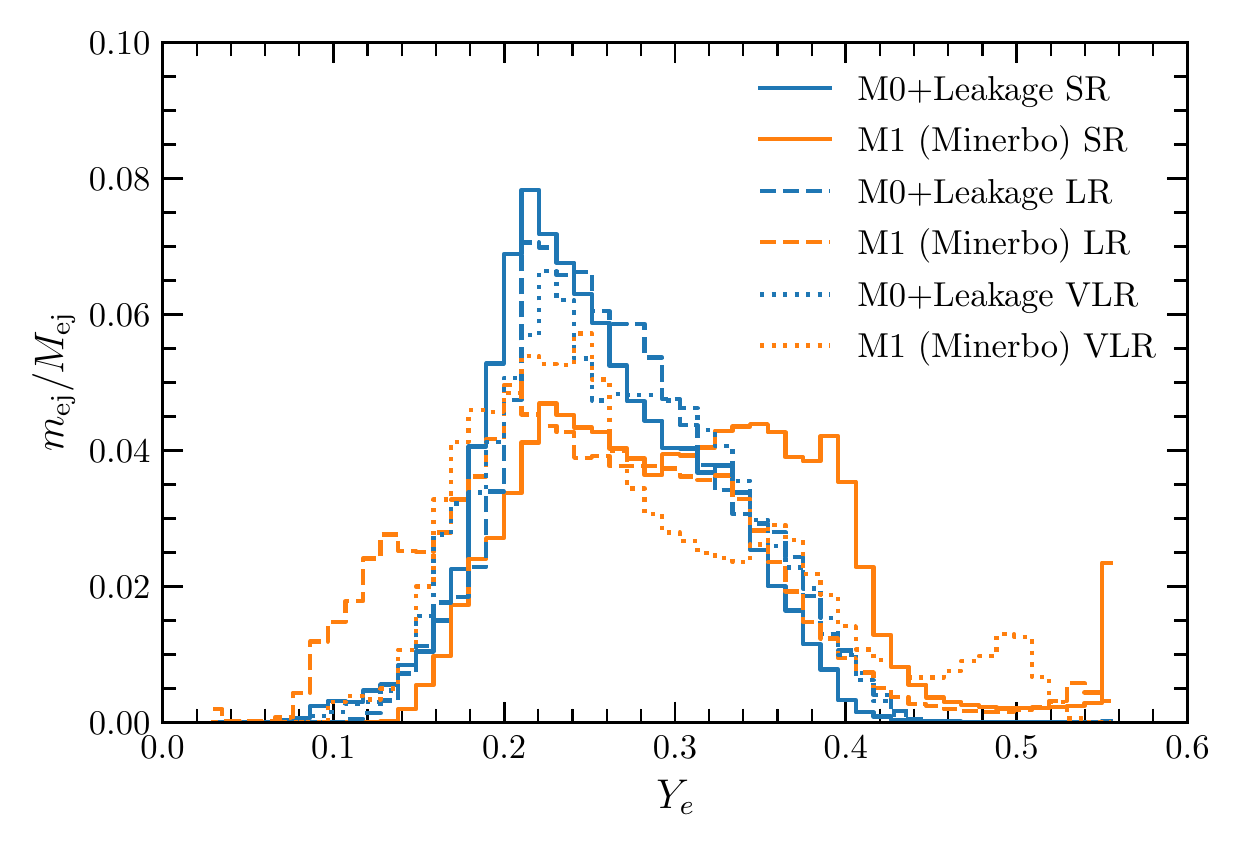}
  \caption{Electron fraction distribution for the dynamical ejecta from
  the SLy4 $1.364\ M_\odot - 1.364\ M_\odot$ binary. M1 predicts a
  broad distribution in $Y_e$ extending to $Y_e \simeq 0.4$. The M0
  ejecta distribution, instead, clearly peaks at $Y_e \simeq 0.2$. The
  M1 results appear to be more sensitive to resolution.}
  \label{fig:bns.1364.ye}
\end{figure}

Neutrino irradiation is known to have a strong impact on the composition
of the dynamical ejecta from NS mergers \citep{Sekiguchi:2015dma,
Foucart:2015gaa, Radice:2016dwd, Foucart:2016rxm, Perego:2017wtu,
Foucart:2020qjb}, which, in turn, has a strong impact on their
nucleosynthesis yields \citep{Lippuner:2015gwa, Thielemann:2017acv,
Cowan:2019pkx, Perego:2021dpw}. Not surprisingly, we find that the
composition of the dynamical ejecta, shown in
Fig.~\ref{fig:bns.1364.ye}, is sensitive to the adopted neutrino
transport scheme. In particular, the M0+Leakage simulations show a
characteristic peak in the $Y_e$ distribution at $Y_e\simeq 0.2$, while
the SR M1 shows a broader distribution extending to $Y_e \simeq 0.4$. It
also predict the presence of a proton rich component of the ejecta with
$0.55 < Y_e < 0.6$. This component is lumped in the highest $Y_e$ bin in
our analysis and is responsible for the bump in the histogram at $Y_e
\simeq 0.55$. That said, while the $Y_e$ distribution of the M0 runs is
consistent across all resolutions, the $Y_e$ distribution for M1 vary
significantly with resolution. The VLR results are in better agreement
with the M0 calculations, apart from the presence of a high-$Y_e$ peak
for $Y_e \simeq 0.5$. The LR M1 simulations, instead, predict a lower
$Y_e$ than the M0 simulations.

\begin{figure}
  \includegraphics[width=\columnwidth]{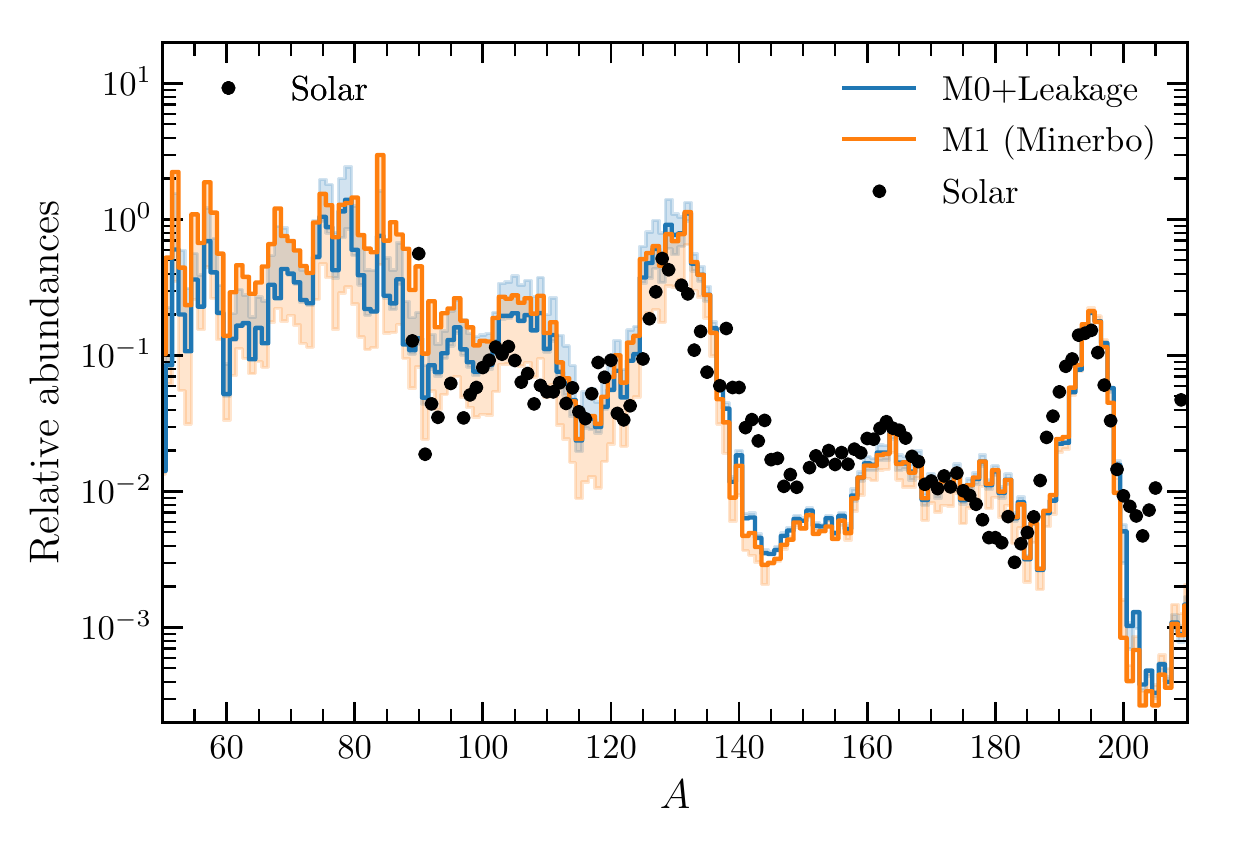}
  \caption{Nucleosynthesis yields for the SLy4 $1.364\ M_\odot - 1.364\
  M_\odot$ binary. Compared to the Solar abundance pattern from
  \citep{Arlandini:1999an}, this binary overproduces r-process elements
  with $A \simeq 100$, or, equivalently, underproduces second and
  third peak elements, according to all schemes. Despite the qualitative
  differences in the $Y_e$ distribution, well resolved M0 and M1
  simulations produce similar abundance patterns.}
  \label{fig:bns.1364.yields}
\end{figure}

The differences in composition are reflected in the final abundances
after r-process nucleosynthesis, shown in
Fig.~\ref{fig:bns.1364.yields}. The abundances are obtained using a grid
of precomputed trajectories with \texttt{SkyNet}
\citep{Lippuner:2017tyn}, as discussed in detail in
\citet{Radice:2018pdn}. We normalize the relative abundances by fixing
the height of the third r-process peak ($A \simeq 190$). We also report
Solar r-process abundances from \citep{Arlandini:1999an} in the same
figure.  However, we emphasize that even if \ac{NS} mergers were the
sole contributor of r-process elements, there is no reason to expect
that every merger should produce ejecta with relative abundances close
to Solar. Indeed, variability between the yields of different mergers is
required to explain observed abundances in metal poor stars
\citep{Holmbeck:2019xnd}. Overall, the simulations span a factor
${\sim}2$ in the ratio of $A \simeq 100$ to third r-process
peak. However, the difference between the M0 and M1 at the SR
resolution, which is the resolution we use for production simulation,
are modest compared to the systematic uncertainties from the unknown
\ac{NS} \ac{EOS} and to the variability due to the binary mass ratio
\citep{Radice:2018pdn, Nedora:2020hxc}. Clearly, strong conclusions
cannot be drawn from this limited study alone, but our simulations
suggest that the uncertainties in the yields from mergers arising from
neutrino radiation treatment are modest. This is also supported by the
results of \citet{Foucart:2020qjb}. They compared M1 and Monte Carlo
neutrino transport in the context of \ac{NS} mergers and reported only a
modest ${\sim}10\%$ difference in the $Y_e$ of the ejecta between the
two schemes. Interestingly, they reported that M1 systematically
overestimates the $Y_e$ of the ejecta, so we cannot exclude that the
M0+Leakage results are actually more accurate than the results obtained
with \texttt{THC\_M1}. That said, it is important to emphasize that this
comparisons has only been made for the dynamical ejecta and not for the
secular ejecta, which we discuss in Sec~.\ref{sec:secular.ejecta}.

% ------------------------------------------------
\section{Long-Term Postmerger Evolutions}
\label{sec:postmerger}
% ------------------------------------------------

The main application we envision for \texttt{THC\_M1} is to simulate the
diffusion of neutrinos out of the merger remnant and the production of
winds on secular time scales after merger. These winds are currently
thought to constitute the bulk of the outflow from binary mergers
\citep{Siegel:2019mlp, Shibata:2019wef, Radice:2020ddv, Nedora:2020hxc}.
In this section, we demonstrate the viability of this approach by
performing long-term postmerger simulations for a binary producing a
long-lived remnant. In particular, we consider the merger of two
identical $1.3\ M_\odot$ \acp{NS} simulated with the SLy \ac{EOS}.
Initial data produced with the \texttt{Lorene} code are prepared at an
initial separation of 45~km, and have already been considered in
\citet{Breschi:2019srl}. We perform simulations with \texttt{THC\_M1}
with both the Eddington and Minerbo closures. Additionally, we perform a
simulation with the M0+Leakage scheme used in production simulations
with \texttt{THC}. The M1 (Eddington) simulation is discontinued shortly
after \ac{BH} formation ($t - t_{\rm mrg} \simeq 55\ {\rm ms}$), while
the M0+Leakage and the M1 (Minerbo) simulations are carried out until $t
- t_{\rm mrg} \simeq 77\ {\rm ms}$. The simulation setup is the same as
in that of the calculations presented in the previous section.  However,
due to the large computational costs, we only present results with the
VLR grid spacing.

\subsection{Qualitative Dynamics}

\begin{figure}
  \includegraphics[width=\columnwidth]{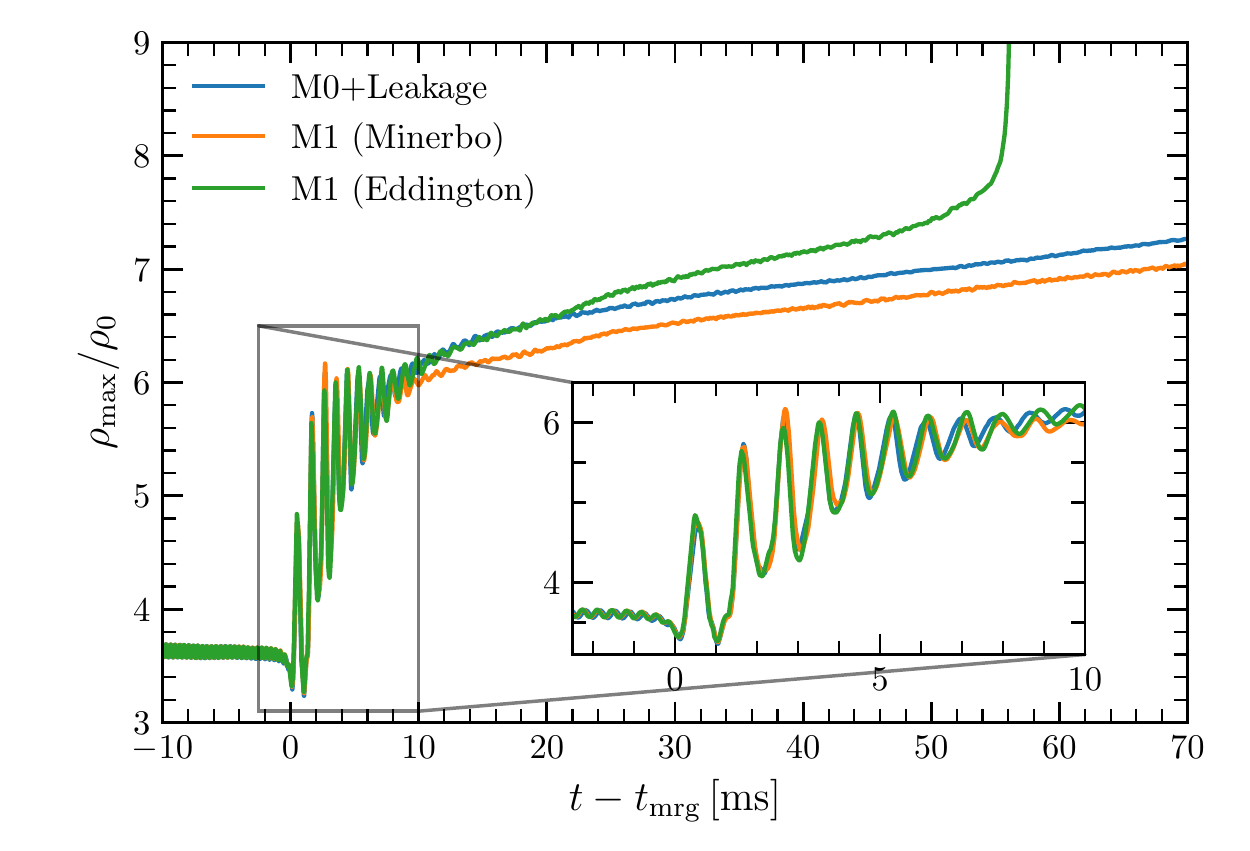}
  \caption{Maximum density as a function of time in millisecond from
  the merger for the SLy $1.3\ M_\odot - 1.3\ M_\odot$
  binary. Small differences in the evolution of the merger remnant are
  seen starting from ${\sim}10\ {\rm ms}$ after merger.}
  \label{fig:bns.130.rhoc}
\end{figure}

Figure~\ref{fig:bns.130.rhoc} shows the maximum density for the three
simulations. These are in good agreement, especially during the first 10
milliseconds after the merger. Systematic differences appear at later
times. In particular, the M1 simulation with Eddington closure collapses
to \ac{BH} at $t - t_{\rm mrg} \simeq 55\ {\rm ms}$, while the other
remnants remain stable for the entire simulation time. That said, we
caution the reader that the collapse time of the remnant \ac{NS} is known to
be sensitive to resolution and small perturbations, so these differences
might not be related to the different neutrino treatment. A detailed
investigation of possible neutrino effects on the evolution of the
remnant would require many more simulations at higher resolution, so it
is outside of the scope of this work.

It has been proposed that out-of-weak-equilibrium effects in the
postmerger could give raise to an effective bulk viscosity
\citep{Alford:2017rxf, Alford:2020lla, Most:2021zvc, Hammond:2021vtv}.
Such effect cannot be captured with leakage schemes, but can be captured
with \texttt{THC\_M1}, since our code does not assume thermodynamic
equilibrium between matter and neutrinos. Our M1 simulations do not show
evidences of enhanced damping of the radial oscillations of the remnant
compared to the M0 runs. This suggests that the impact of bulk viscosity
cannot be too large. That said, higher resolution simulations with a
variety of possible \acp{EOS} would be required to draw firm
conclusions. We also leave this to future work.

\begin{figure*}
  \includegraphics[width=\textwidth]{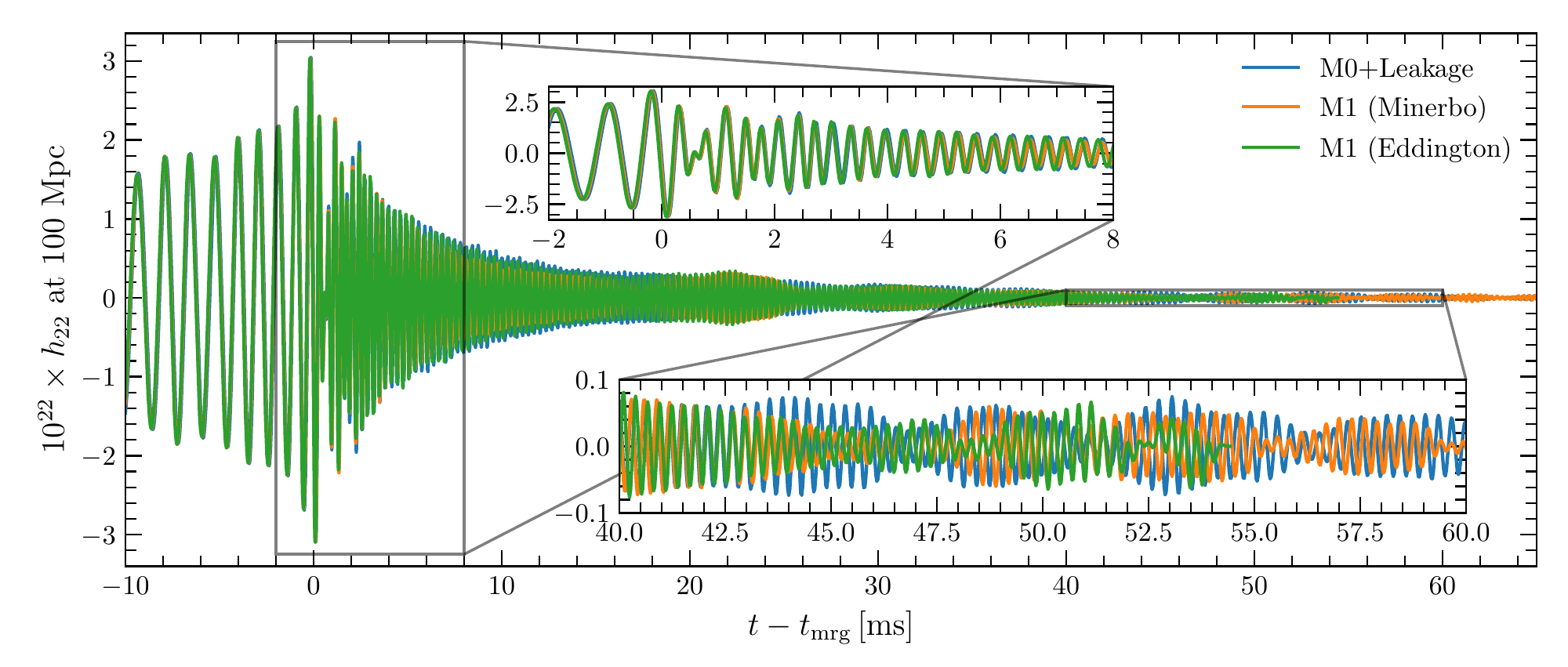}
  \caption{Real part of the $\ell = 2, m = 2$ dominant GW strain mode
  for the SLy $1.3\ M_\odot - 1.3\ M_\odot$ binary. Small differences
  in the evolution of the merger remnant are seen starting from
  ${\sim}10\ {\rm ms}$ after merger.}
  \label{fig:bns.130.h22}
\end{figure*}

The dynamics of the binary is imprinted in the \ac{GW} signal. We show
the dominant $\ell = 2, m = 2$ component of the strain in
Fig.~\ref{fig:bns.130.h22}. As for the maximum density, we find that the
strain from the three simulations agree both qualitatively and
quantitatively. There is a small dephasing between the three waveforms
in the postmerger, as can be observed in the figure inset.  However,
this dephasing is well within the estimated uncertainties in the
postmerger signal at this resolution \citep{Radice:2016rys,Breschi:2019srl}. The most
substantial difference between the waveforms is that the M1 (Eddington)
\ac{GW} emission abruptly shuts off at the time of \ac{BH} formation.
Overall, our results show that leakage simulations are adequate to study
the \ac{GW} emission and the early evolution of binary \ac{NS} remnants.
This is not surprising, given the typical neutrino cooling timescale for
the remnant is of a few seconds \citep{Sekiguchi:2011zd}, while most of
the \ac{GW} energy is radiated within ${\sim}20\ {\rm ms}$ of the merger
\citep{Bernuzzi:2015opx, Zappa:2017xba}.

\subsection{Dynamical Ejecta}

\begin{figure*}
  \includegraphics[width=\textwidth]{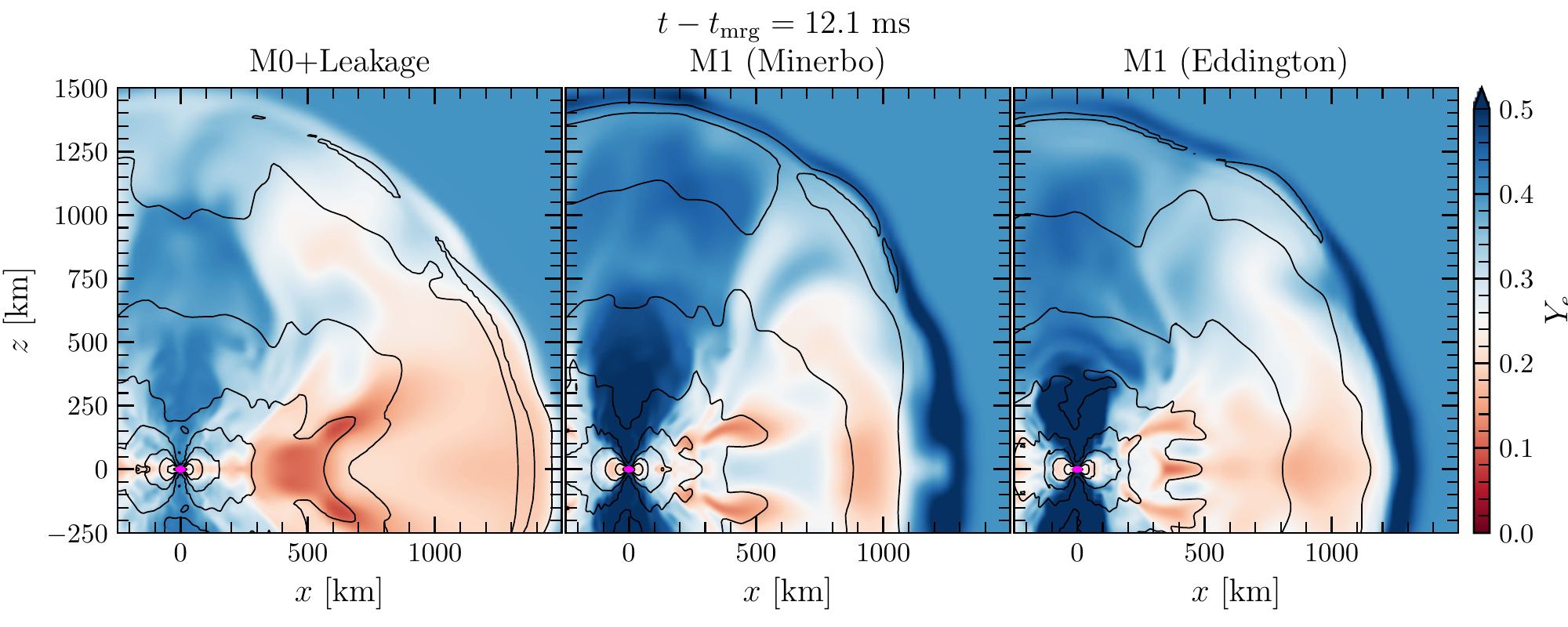}
  \caption{Electron fraction (color) of the dynamical ejecta cloud
  formed for the SLy $1.3\ M_\odot - 1.3\ M_\odot$ binary. The black
  lines are isodensity contours of $\rho = 10^{5}, 10^{6}, 10^7, 10^8,
  10^9, 10^{10}, 10^{11}$, and $10^{12}\ {\rm g}\ {\rm cm}^{-3}$. The
  purple contour shows corresponds to $\rho = 10^{13}\ {\rm g}\ {\rm
  cm}^{-3}$ and denotes the approximate location of the surface of the
  merger remnant. M0 and M1 results are in good qualitative agreement,
  but M1 predicts higher electron fractions for both the polar and
  equatorial ejecta.}
  \label{fig:bns.130.ye.profile1}
\end{figure*}

Figure~\ref{fig:bns.130.ye.profile1} shows the electron fraction of the
ejecta in the meridional plane of the binary about $12$ milliseconds
after the merger. Overall, we find that the M0+Leakage scheme tends to
underestimate the proton fraction in the ejecta, when compared to the M1
scheme. This is consistent with our findings in Sec.~\ref{sec:nsmerger},
but the 2D plot reveals two interesting systematic differences.

First, the M1 simulations find pockets of moderate $Y_e$ material also
in the equatorial regions. This is material that is shock heated and
irradiated as the tidal tail and the shocked ejecta collide. The M0
simulations also exhibits an interaction between the tidal tail and the
shocked ejecta, however the material remains very neutron rich $Y_e
\lesssim 0.2$. A possible explanation for this difference is that the
irradiating neutrinos are not propagating radially, so they are not
correctly treated by the M0 scheme. This is suggested by the fact that
there is a strong density and temperature gradient in the ejecta along
the azimuthal direction. This effect is more prominent in the M1
simulation with the Minerbo closure, likely because the Eddington
closure limits the propagation velocity of free streaming neutrinos to
$c/\sqrt{3}$. This implies that neutrinos interact with the ejecta at
larger radii, where they are more diluted.

Second, the M1 simulations predict the formation of a tenuous, but
rapidly expanding neutrino driven wind with $Y_e \simeq 0.5$ starting
few milliseconds after the merger. A similar wind also develops in the
M0 case, but with a delay of ${\sim}10-15$ milliseconds from the merger.
The properties of the neutrino driven winds are discussed in more detail
in Sec.~\ref{sec:remnant} and in \citet{Nedora:2020hxc}. We
speculate that the reason for this discrepancy is that the M0 scheme
only models neutrino heating in optically thin
regions\footnote{Absorption is included also at high optical-depth, but
is suppressed with a factor $O(e^{-\tau})$, $\tau$ being the optical
depth, to be consistent with the effective sources of the leakage
scheme.} and might not be able to capture the sharp transition from
optically thick to thin conditions along the spin axis of the remnant.  As
a result, the wind needs to be bootstrapped by the presence of a
sufficient amount of of low density material ($\rho \lesssim 10^{11}\
{\rm g}\ {\rm cm}^{-3}$) in the polar region of the remnant. This
speculation is tentatively confirmed by the fact that the M0
luminosities for electron-flavor neutrinos are larger by a factor of a
few compared to the M1 luminosities (see Sec.~\ref{sec:bns.130.nulum}
and Fig.~\ref{fig:bns.130.Lnu}), as expected if neutrinos do not entrain
baryons in their way out.
We remark that the wind is present with both the Minerbo and
Eddington closure, so it is not the result of the well known
beam-crossing artifact of nonlinear M1 closures \citep{frank:2007}.

\subsection{Remnant Structure}
\label{sec:remnant}

\begin{figure*}
  \includegraphics[width=\textwidth]{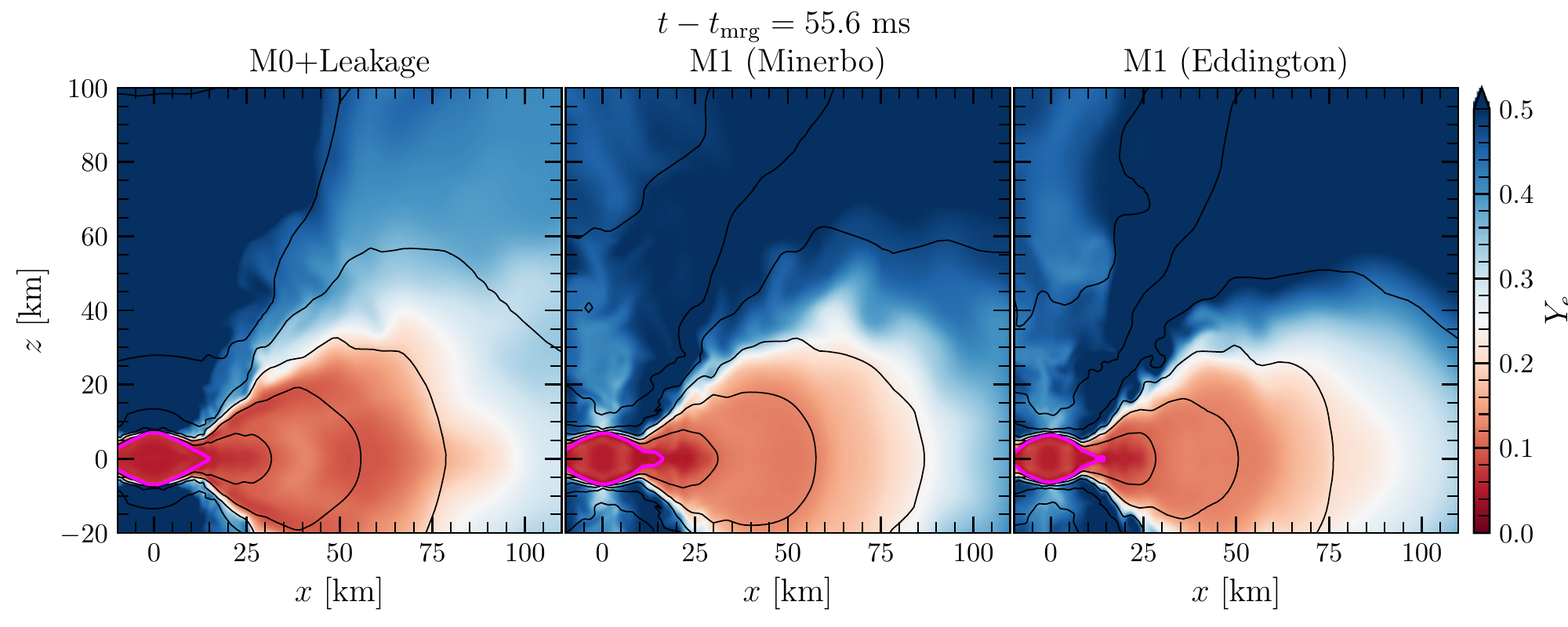}
  \caption{Electron fraction (color) of the of the SLy $1.3\ M_\odot -
  1.3 M_\odot$ merger remnant ${\sim}55$ milliseconds after merger.  The
  black lines are isodensity contours of $\rho = 10^{5}, 10^{6}, 10^7,
  10^8, 10^9, 10^{10}, 10^{11}$, and $10^{12}\ {\rm g}\ {\rm cm}^{-3}$.
  The purple contour shows corresponds to $\rho = 10^{13}\ {\rm g}\ {\rm
  cm}^{-3}$ and denotes the approximate location of the surface of the
  merger remnant. M0 and M1 results are in good qualitative agreement,
  but M1 predicts higher electron fractions in the disk corona and
  somewhat smaller electron fraction in the neutrino driven wind along
  the rotation axis.}
  \label{fig:bns.130.ye.profile2}
\end{figure*}

Figure~\ref{fig:bns.130.ye.profile2} shows the structure and composition
of the merger remnant ${\sim}55$ milliseconds after the merger. We find
good qualitative agreement between the three numerical schemes. In
particular, all simulations predict a very neutron rich composition
($Y_e \lesssim 0.2$) for the accretion torus and the presence of a
high-$Y_e$ wind at high latitudes. They also predict a shift to higher
$Y_e$ at densities below $10^{11}\ {\rm g}\ {\rm cm}^{-3}$, where
thermal electron-type neutrinos are expected to decouple
\citep{Endrizzi:2019trv}. However, there are important quantitative
differences.

First, the M0+Leakage scheme systematically underpredicts the $Y_e$ in
the corona of the disk. This is because M0 only transports neutrinos
radially, so it cannot model the irradiation of the corona by neutrinos
emerging from the disk below, while \texttt{THC\_M1} does not have this
limitation.

Second, there are small, but important differences in the $Y_e$ of the
remnant. These differences arise because our leakage scheme does not
model the presence of a trapped component of $\bar{\nu}_e$ in the
remnant. \texttt{THC\_M1}, instead, correctly captures the protonization
of the region of the remnant around ${\rho} = 10^{14}\ {\rm g}\ {\rm
cm}^{-3}$ and the creation of a trapped component of anti-electron
neutrinos, in agreement with the predictions of \citet{Perego:2019adq}.
This trapped neutrino component can impact the pressure at the several
percent level \citep{Perego:2019adq}, which might be sufficient to
impact the remnant stability \citep{Radice:2018xqa}.

Third, the M1 simulations produce a denser neutrino driven wind, as can
be seen from the isodensity contours in
Fig.~\ref{fig:bns.130.ye.profile2}. This wind also entrains material
from the outer layers of the central remnant, so it is more neutron rich
than that predicted by the M0 simulation. This difference could have
been anticipated, because the M0+Leakage scheme only models the
transport and reabsorption of free streaming neutrinos, while M1 can
also capture the heating of the outer layers of the remnant due to the
diffusion of neutrinos along the steep density and temperature gradient
along the rotational axis of the binary. In particular, because the
opacity in the M0+Leakage scheme is weighted with the optical depth,
this scheme systematically underestimates heat deposition for optical
depths $\tau \gtrsim 1$.

\begin{figure*}
  \includegraphics[width=\textwidth]{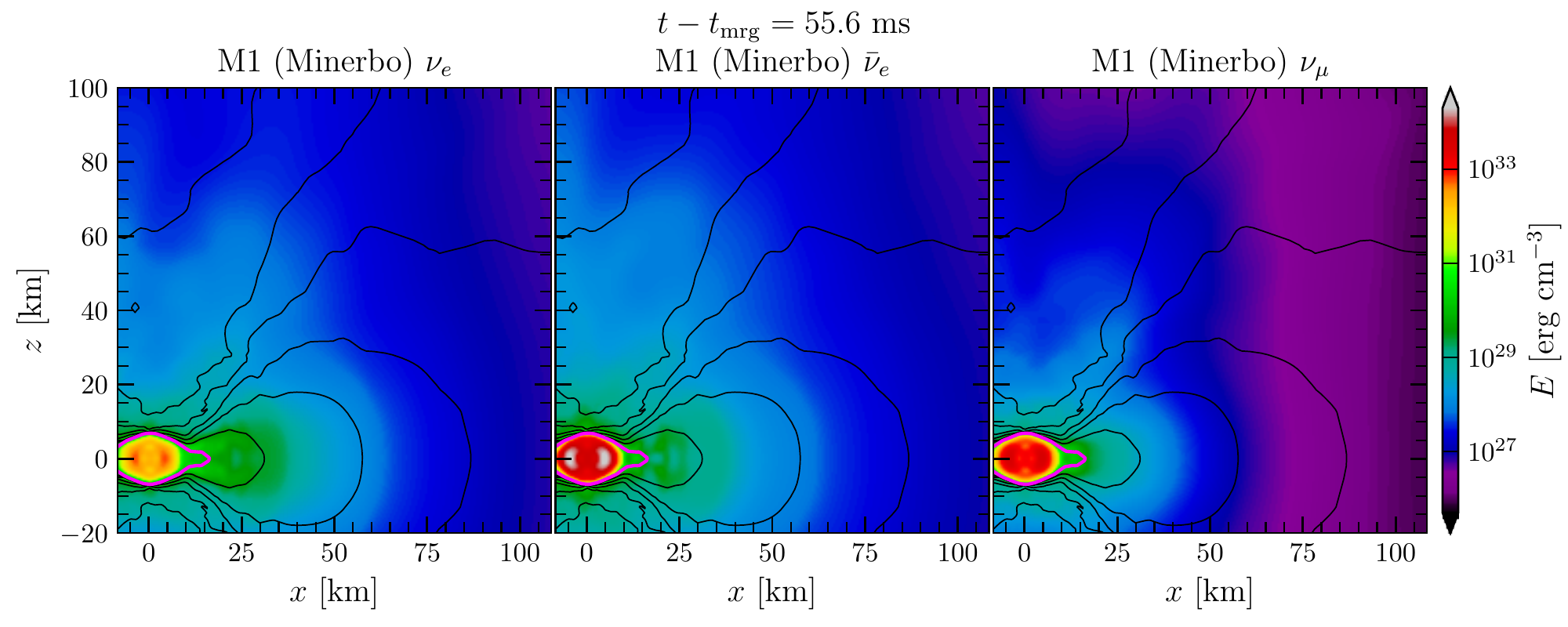}
  \caption{Neutrino radiation energy density (color) for the SLy $1.3\
  M_\odot - 1.3 M_\odot$ binary ${\sim}55$ milliseconds after
  merger.  The black lines are isodensity contours of $\rho = 10^{5},
  10^{6}, 10^7, 10^8, 10^9, 10^{10}, 10^{11}$, and $10^{12}\ {\rm g}\
  {\rm cm}^{-3}$. The purple contour shows corresponds to $\rho =
  10^{13}\ {\rm g}\ {\rm cm}^{-3}$ and denotes the approximate location
  of the surface of the merger remnant. Due to the geometry of the
  postmerger, radiation is preferentially focused in the polar regions.}
  \label{fig:bns.130.rE.profile}
\end{figure*}

\begin{figure*}
  \includegraphics[width=\textwidth]{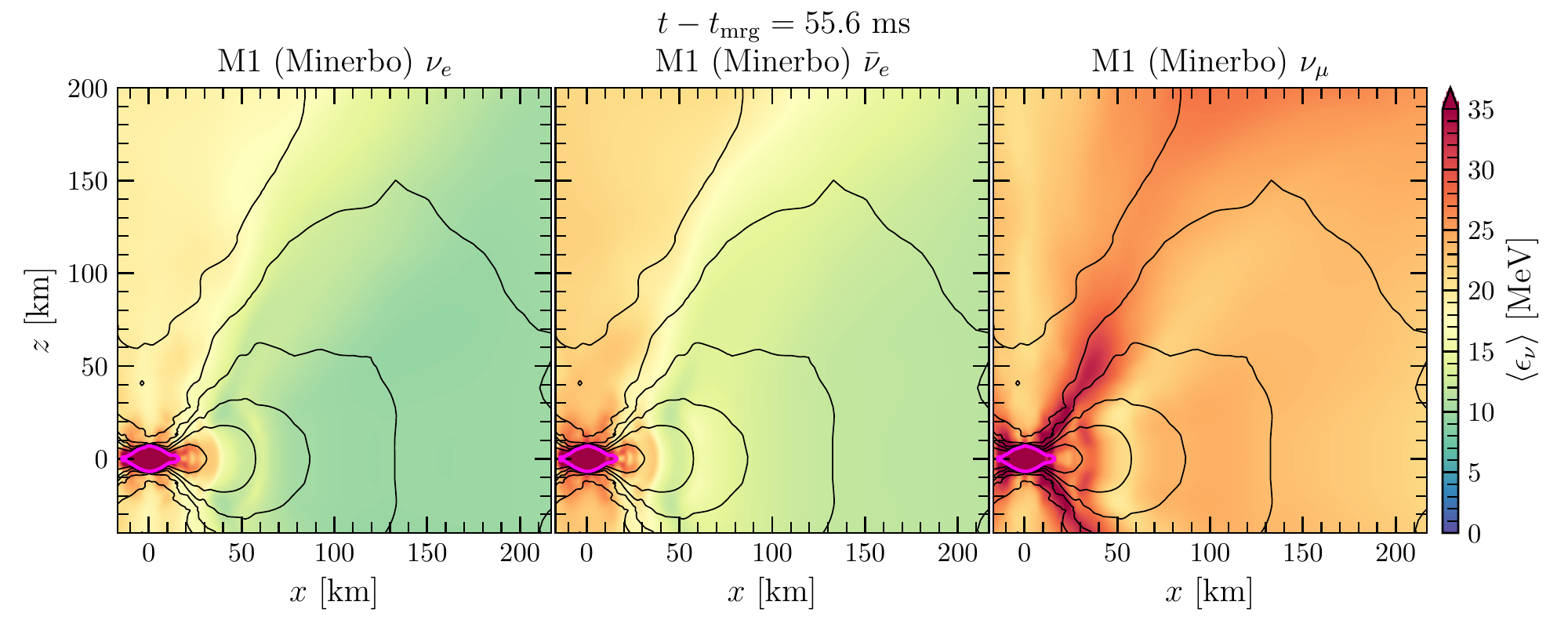}
  \caption{Average neutrino energies (color) for the SLy $1.3\ M_\odot -
  1.3 M_\odot$ binary ${\sim}55$ milliseconds after merger.  The black
  lines are isodensity contours of $\rho = 10^{5}, 10^{6}, 10^7, 10^8,
  10^9, 10^{10}, 10^{11}$, and $10^{12}\ {\rm g}\ {\rm cm}^{-3}$. The
  purple contour shows corresponds to $\rho = 10^{13}\ {\rm g}\ {\rm
  cm}^{-3}$ and denotes the approximate location of the surface of the
  merger remnant. The average neutrino energy is highly anisotropic,
  especially for electron-type neutrinos, since the disk is optically
  thick to high energy neutrinos.}
  \label{fig:bns.130.eave.profile}
\end{figure*}

Figure \ref{fig:bns.130.rE.profile} shows the neutrino energy
density for the M1 (Minerbo) simulation ${\sim}55$ milliseconds
after the merger. This is a representative time for the neutrino field
in the postmerger. However, we emphasize that the neutrino energy
density oscillates and shows quasi-periodic bursts, especially shortly
after merger. The M1 (Eddington) neutrino radiation energy densities
are qualitatively and quantitatively similar. We observe the formation
of a trapped component of neutrinos. As previously discussed,
$\bar{\nu}_e$ are the dominant neutrino species in the inner part of the
remnant. However, we find trapped neutrinos of all flavors in the
central part of the remnant and in the accretion disk. Radiation is
geometrically focused in the polar direction and most intense
${\sim}10{-}20\ {\rm km}$ above the surface of the massive \ac{NS}.
Equatorial outflows are shielded from the intense neutrino radiation
from the inner part of the remnant by the torus, but they are instead
irradiated by neutrinos produced directly in the disk.

There are effectively two sources of electron-flavor neutrinos. The
massive \ac{NS} at the center and the disk. Neutrinos from the massive
\ac{NS} have ${\sim} 50\%$ higher average energies (see
Fig.~\ref{fig:bns.130.eave.profile}), so their interaction cross section
with matter is ${\sim}3$ times larger. However, only material outflowing in
the polar direction is directly exposed to these neutrinos. The
neutrinos from the disk are less energetic, but fill a significantly
larger area (Fig.~\ref{fig:bns.130.eave.profile}). The net effect is to
enhance the differences in the $Y_e$ of polar and equatorial ejecta and
to increase the anisotropic character of the resulting kilonova emission
\citep{Perego:2017wtu, Kawaguchi:2019nju, Korobkin:2020spe}.

\begin{figure*}
  \includegraphics[width=\textwidth]{fig22.pdf}
  \caption{Average neutrino energies (color) for the SLy $1.3\ M_\odot -
  1.3 M_\odot$ binary ${\sim}55$ milliseconds after merger.  The black
  lines are isodensity contours of $\rho = 10^{5}, 10^{6}, 10^7, 10^8,
  10^9, 10^{10}, 10^{11}$, and $10^{12}\ {\rm g}\ {\rm cm}^{-3}$. The
  purple contour shows corresponds to $\rho = 10^{13}\ {\rm g}\ {\rm
  cm}^{-3}$ and denotes the approximate location of the surface of the
  merger remnant. This figure should be contrasted with
  Fig.~\ref{fig:bns.130.eave.profile}, which shows the same profiles
  with obtained with the Minerbo closure.}
  \label{fig:bns.130.edd.eave.profile}
\end{figure*}

Figure~\ref{fig:bns.130.edd.eave.profile} shows the average neutrino
energy obtained with the Eddington closure. There are small differences
with the Minerbo closure in the location of the separatix between the
stream of neutrinos emerging from the massive \ac{NS} and the disk. This
is because, on the one hand, the Minerbo closure artificially prevents
different neutrino streams from mixing. On the other hand, the Eddington
closure tends to smooth out structures in the radiation energy density
profile. Most notably, the x-shaped feature present in the M1 (Minerbo)
run for both $E_{\nu_\mu}$ and $\langle \epsilon_{\nu_\mu} \rangle$
close to the massive \ac{NS} is absent in the Eddington simulations.
This suggests that this feature is likely to be an artifact of the
Minerbo closure. That said, Minerbo and Eddington closure are broadly
consistent with each other, suggesting that the results discussed so far
are robust.

\subsection{Neutrino Emission}
\label{sec:bns.130.nulum}

\begin{figure*}
  \includegraphics[width=\textwidth]{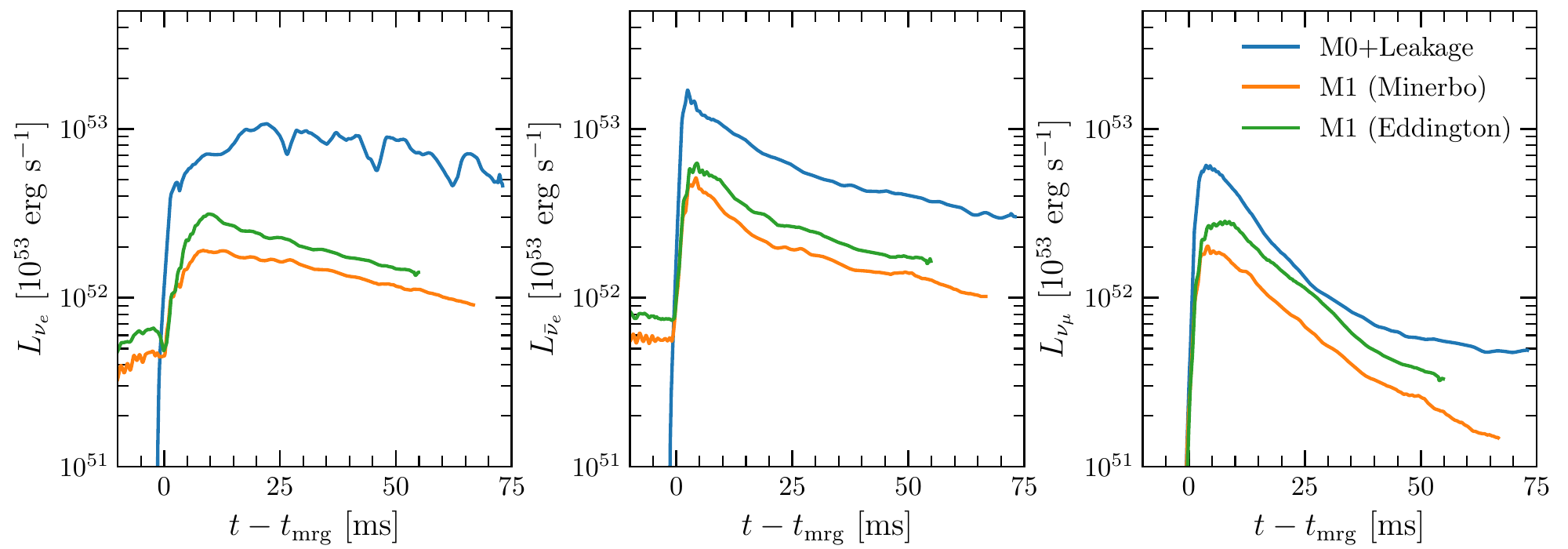}
  \caption{Neutrino luminosity for the SLy $1.3\ M_\odot - 1.3\ M_\odot$
  binary. The data is smoothed using a rolling average with
  width of 1~ms. We find that at this resolution M0 systematically
  overestimates the $\nu_e$ and $\bar{\nu}_e$ luminosities by about a
  factor of two. The M1 Eddington and Minerbo luminosities are in good
  agreement.}
  \label{fig:bns.130.Lnu}
\end{figure*}

We show the angle integrated neutrino luminosities for the SLy $1.3\
M_\odot - 1.3\ M_\odot$ binary in Fig.~\ref{fig:bns.130.Lnu}. The
neutrino luminosities for the M1 (Minerbo) and M1 (Eddington)
simulations are extracted on a coordinate sphere of radius $r = 300\ G
M_\odot /c^2 \simeq 443\ {\rm km}$. The luminosities of the M0+Leakage
scheme are computed at the outer boundary of the M0 spherical grid $512\
G M_\odot /c^2 \simeq 756\ {\rm km}$. All data is time shifted to
account for the neutrino time of flight. As anticipated, we find that
M0+Leakage systematically overestimates the luminosity of
electron-flavor neutrinos. Good agreement is found for heavy-lepton
neutrons, instead. In all cases, the luminosities peak within a few
milliseconds of the merger, in contrast to \citet{Vincent:2019kor}, and
then decay exponentially. The oscillations in the luminosity are not due
to a numerical artifact, but are associated with the oscillations of the
massive \ac{NS} remnant \citep{Cusinato:2021zin}. As was the case for
the SLy $1.364\ M_\odot - 1.364\ M_\odot$ binary, $L_{\bar{\nu}_e} >
L_{\nu_\mu} > L_{\nu_e}$, showing that the remnant is protonizing.

\begin{figure*}
  \includegraphics[width=\textwidth]{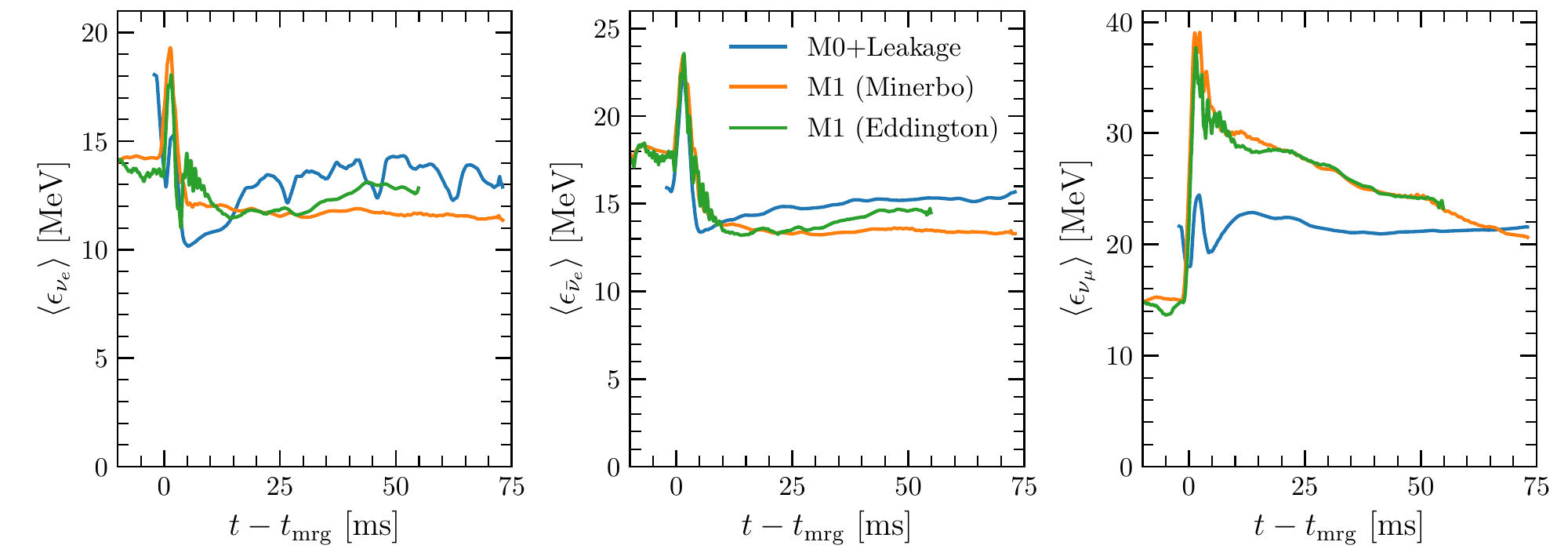}
  \caption{Average neutrino energies for the SLy $1.3\ M_\odot - 1.3\
  M_\odot$ binary. The data is smoothed using a rolling average with
  width of 1~ms. We find excellent agreement in the average
  neutrino energies for electron type neutrinos. The M0 scheme predicts
  smaller average energies for heavy-lepton flavor neutrinos.}
  \label{fig:bns.130.Enu}
\end{figure*}

The average neutrino energies for the SLy $1.3\ M_\odot - 1.3\ M_\odot$
binary are shown in Fig.~\ref{fig:bns.130.Enu}. We find excellent
agreement in the average electron-flavor neutrino energies for all
schemes. The M0+Leakage scheme predicts a ${\sim}50\%$ smaller average
energy for heavy-lepton neutrinos. Moreover, the M0+Leakage scheme
predicts a nearly constant heavy-lepton neutrino energy as a function of
time.  This is because M0 does not properly diffuse neutrinos through
the remnant. Instead, each part of the remnant cools at a rate that
depends on its optical depth. In contrast, \texttt{THC\_M1} models the
diffusion of neutrinos to the neutrino spheres and their thermalization.
Also in this case, we find that $\langle \epsilon_{\nu_\mu} \rangle >
\langle \epsilon_{\bar{\nu}_e} \rangle > \langle \epsilon_{\nu_e}
\rangle$, as expected.

\subsection{Secular Ejecta}
\label{sec:secular.ejecta}

\begin{figure}
  \includegraphics[width=\columnwidth]{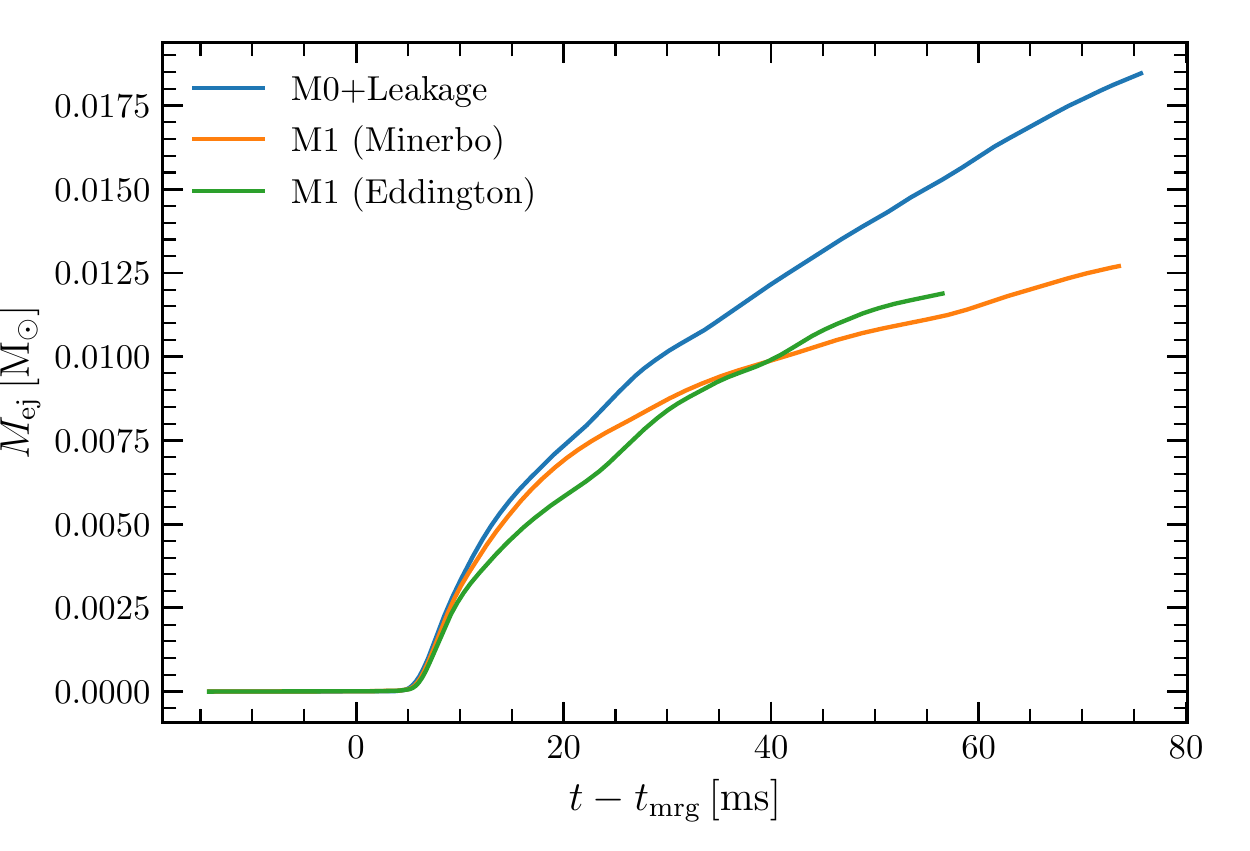}
  \caption{Ejecta mass for the SLy $1.3\ M_\odot - 1.3\ M_\odot$ binary.
  The M0 simulations slightly overestimate the outflow rate for the
  postmerger wind compared to M1. The results of the Minerbo and
  Eddington closures are consistent with each other.}
  \label{fig:bns.130.outflow1.ber.ejectedmass}
\end{figure}

We observe the emergence of an outflow driven by hydrodynamics torques:
the so-called spiral-wave wind \citep{Nedora:2019jhl, Nedora:2020hxc},
in addition to the aforementioned neutrino-driven wind.  This secular
ejecta is extracted at the same extraction radius of the dynamical
ejecta ($r = 300\ G M_\odot /c^2 \simeq 443\ {\rm km}$), but we use the
Bernoulli criterion ($- h u_t > 1$), which is more appropriate for a
steady wind. See \citet{Foucart:2021ikp} for a recent discussion of the
issues connected to the discrimination between gravitationally bound and
unbound outflows. The time-integrated outflow rate is shown in
Fig.~\ref{fig:bns.130.outflow1.ber.ejectedmass}. We find that the
leakage+M0 simulation produces a more robust wind with a larger
$\dot{M}$, while the two M1 simulations are in good agreement with each
other. However, we warn the reader that, at this resolution, the
numerical uncertainties in the outflow is $\gtrsim 50\%$
\citep{Nedora:2020hxc}, so these differences might not be particularly
meaningful. In particular, our previous simulations at higher resolution
\citep{Breschi:2019srl}, but with simpler neutrino physics, suggest that
this binary might form a \ac{BH} few tens of milliseconds after merger.
Since the spiral wave wind ceases with \ac{BH} formation
\citep{Nedora:2019jhl}, the uncertainty in the \ac{BH} formation time is
likely to dominate the overall error budget on the total ejecta mass for
this binary.

\begin{figure}
  \includegraphics[width=\columnwidth]{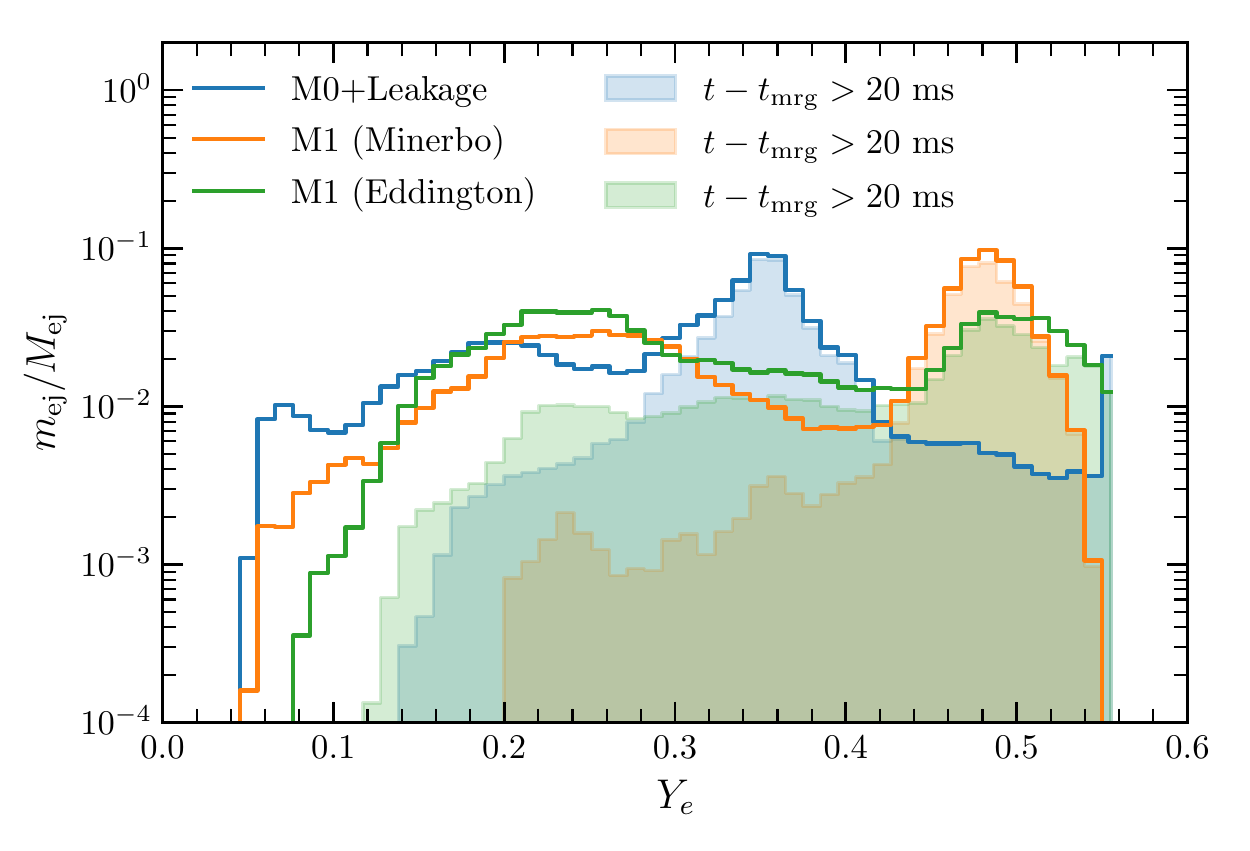}
  \caption{Ejecta $Y_e$ for the SLy $1.3\ M_\odot - 1.3\ M_\odot$
  binary. The shaded regions show the contributions to the $Y_e$
  histogram due to the material ejected after the first 20~ms of the
  merger. The ejecta distribution peaks at significantly larger $Y_e$
  in the M1 runs. The Eddington and Minerbo results are in good
  agreement, but the Eddington simulations produce a smaller amount of
  very neutron rich ejecta ($Y_e \sim 0.1$).}
  \label{fig:bns.130.outflow1.ber.histye}
\end{figure}

Figure~\ref{fig:bns.130.outflow1.ber.histye} shows the composition of
the overall ejecta (dynamical + secular) for the SLy $1.3\ M_\odot -
1.3\ M_\odot$ binary. We find that all schemes produce a wide
distribution in $Y_e$. The results are qualitatively consistent with our
previously published M0 simulations \citep{Nedora:2019jhl,
Nedora:2020hxc}. An important quantitative difference is that the M0
scheme predicts a peak in the electron fraction distribution at $Y_e
\simeq 0.3$. The outflows in the M1 simulations are, instead,
characterized by a peak in their electron fraction at ${\sim}0.5$. We
attribute this difference to the irradiation of outflows at intermediate
latitudes by neutrinos from the disk, an effect that is not captured by
the M0 scheme (see Fig.~\ref{fig:bns.130.ye.profile2}). Some differences
are also found in the low-$Y_e$ tail of the ejecta, which is primarily
of dynamical origin, as anticipated by
Fig.~\ref{fig:bns.130.ye.profile1}.

\begin{figure}
  \includegraphics[width=\columnwidth]{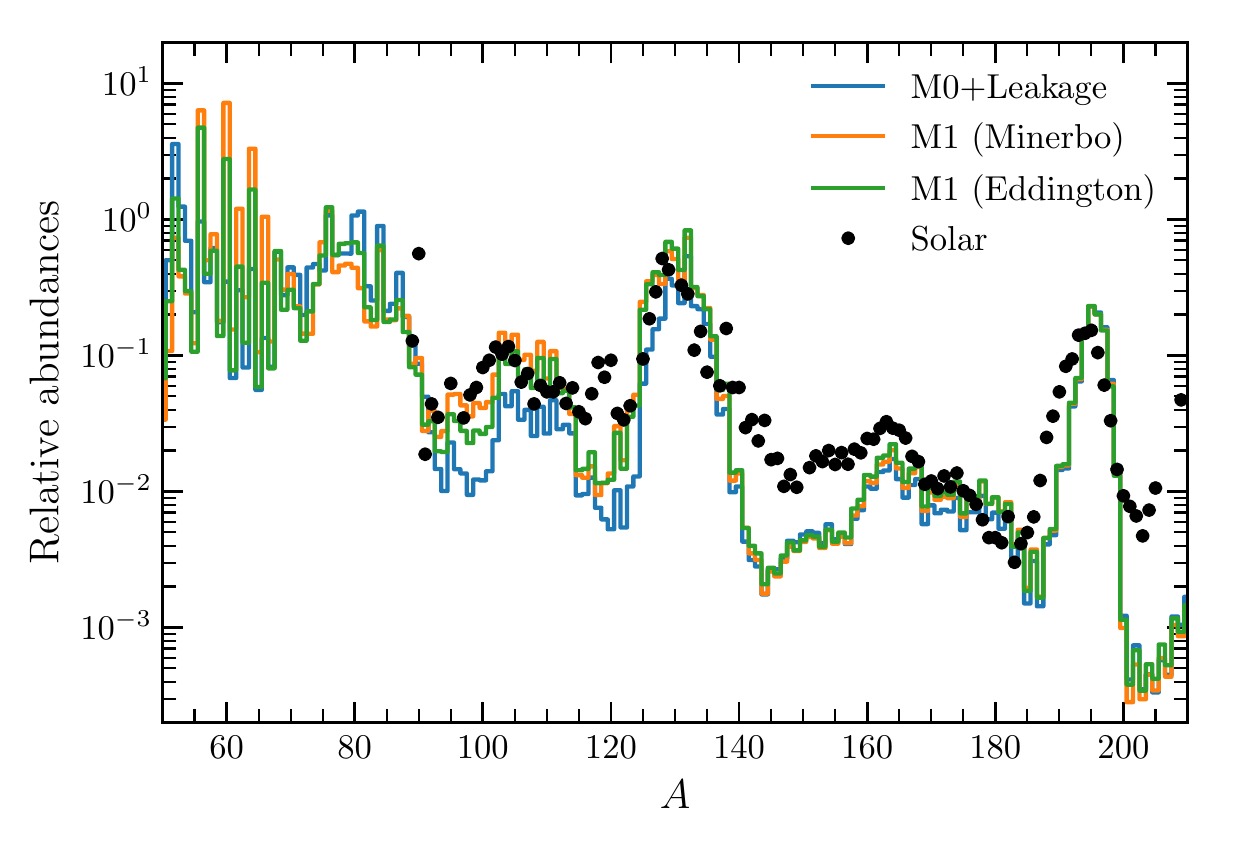}
  \caption{Normalized nucleosynthesis yield for for the SLy $1.3\
  M_\odot - 1.3\ M_\odot$ binary. The M1 runs predict elemental
  abundances that are in better agreement with the Solar pattern, while
  M0 underproduces r-process elements with $A \sim 110$.
  Overall, however, the differences between M0 and M1 are modest.}
  \label{fig:bns.130.yields.det1.ber}
\end{figure}

These changes in $Y_e$ do not contribute to very large differences in
the nucleosynthesis. This is because the main effect of M1 is to shift
the peak of the $Y_e$ distribution from $0.3$ to $0.5$, but both peaks
correspond to a regime in which only light r-process elements are
produced. The integrated nucleosynthesis yields for the three SLy $1.3\
M_\odot - 1.3\ M_\odot$ simulations are shown in
Fig.~\ref{fig:bns.130.yields.det1.ber}. The relative abundances of light
to heavy r-process peak elements differs by about a factor of two between
the M1 and the M0+Leakage runs. This is a significant, but not substantial
discrepancy, considering the large variabilities of the yields with
\ac{EOS} and mass ratio \citep{Radice:2018pdn, Nedora:2020hxc}. The
differences between the M1 (Minerbo) and M1 (Eddington) simulations are
below the level of finite resolution uncertainties (see
Sec.~\ref{sec:bns.1364.ejecta}).

% ------------------------------------------------
\section{Conclusions}
\label{sec:conclusions}
% ------------------------------------------------
We have presented \texttt{THC\_M1}, a new moment-based neutrino
transport code for numerical relativity simulations of merging
\acp{NS}. \texttt{THC\_M1} handles radiation advection using a
high-resolution shock capturing scheme that can capture both the free
streaming and the diffusive regimes. \texttt{THC\_M1} simultaneously
evolves the frequency-integrated energy and neutrino number density
equations. Ours is one of the first \ac{GR} radiation transport codes,
the first in the merger context, to include velocity dependent effects
at all orders in $\upsilon/c$. We have shown that this full treatment,
while technically more complex than that used in other codes, is
necessary to correctly capture neutrino trapping in relativistically
moving media, such as rotating \acp{NS} remnants.

After having validated our new code with a stringent series of tests, we
have coupled it with the \texttt{THC} relativistic hydrodynamics code to
perform merger simulations of two equal mass binaries: an intermediate
mass binary resulting in a short lived remnant that quickly collapses to
\ac{BH}, and a low-mass binary that produces a long-lived remnant. We
have studied numerical resolution effects using the first binary, while
we have performed long-term evolutions at a fixed resolution for the
second binary using two different closure relations, the so-called
Eddington and Minerbo closures. To have a baseline for comparison, we
have also simulated the same systems using a more approximated
M0+Leakage scheme, which is currently used for production simulations
with \texttt{THC}.

The intermediate mass binary experiences a violent merger. The remnant
is rapidly heated to temperatures exceeding $100\ {\rm MeV}$ following
the collision between the stars. The massive \ac{NS} formed in the
collision undergoes one centrifugal bounce that launches a shock in the
merger debris and drives a massive outflow, before collapsing to \ac{BH}.
This is one of the most challenging binary to model due to the high
temperatures and the \ac{BH} formation. We find that \texttt{THC\_M1} is
as robust, if not more robust, than our production leakage code. The
predicted neutrino luminosities and average energies are consistent with
theoretical expectations and other results from the literature.

The remnant of the low mass binary merger also experiences a series of
violent oscillations at birth, with maximum density jumping by more than
$50\%$ on a dynamical timescale. However, the remnant eventually settles
into a massive, differentially rotating \ac{NS} evolving on secular time
scales. Even though \texttt{THC\_M1} includes out-of-weak-equilibrium
effects which have been suggested to result in an effective bulk
viscosity \citep{Alford:2017rxf}, we do not find any evidence of
additional damping of the remnant oscillations in the M1 runs, compared
to simulations that do not model them. That said, simulations with a
more comprehensive set of reactions, with more \acp{EOS}, and at more
resolutions are needed before firm conclusions can be drawn.

We have performed simulations extending for over $70\ {\rm ms}$ after
the merger. For comparison, the longest published simulations performed
with a neutrino-transport scheme having comparable sophistication only
extended to $10\ {\rm ms}$ into the postmerger \citep{Vincent:2019kor}.
We find that the postmerger \ac{GW} signal is not sensitive to details
in the neutrino transport. However, the inner structure of the massive
\ac{NS} is modified by the presence of a trapped component of
anti-electron neutrinos. This could impact the stability of the remnant
of higher mass binaries. We find that, due to the geometry of the
system, neutrino radiation is most intense along the rotational axis of
the system. Matter at lower latitudes is shielded from the direct
irradiation from the massive \ac{NS} by the disk. Instead, it is
irradiated by lower energy neutrinos produced in the accretion disk.
Because neutrino absorption cross sections roughly scale with the
square of the incoming neutrino energy, this enhances the $Y_e$
difference between polar and equatorial ejecta and has implications for
the viewing angle dependency of kilonovae.

We have computed integrated neutrino luminosities and average neutrino
energies from our simulations. Consistently with previous studies, we
find that anti-electron neutrinos have the highest luminosity and that
heavy-lepton neutrinos have the highest average energies. Our M1 data is
in good qualitative and quantitative agreement with results published by
the SXS collaboration using \texttt{SpEC}. On the other hand, we find
that our older M0 neutrino scheme can overestimate electron-flavor
neutrino luminosities by as much as a factor two. Discrepancy with the
results from leakage calculations, either performed by us or by other
groups, are significantly larger and amount to factors of several. We
find an excellent agreement between M1 and M0+Leakage in the neutrino
average energies, instead.

Neutrino transport impacts the neutron richness of both the dynamical
and the secular ejecta in our simulations. In particular, we find that
there is a systematic tendency of M0+Leakage to underestimate the
electron fraction of the ejecta. This is because the M0 scheme does not
model the irradiation of material at intermediate latitudes with
neutrinos generated in the remnant accretion disk. However, because the
net effect is to reprocess material with $Y_e \simeq 0.2{-}0.35$ to $Y_e
\simeq 0.4{-}0.55$, this has only a modest impact on the final
abundances of the r-process nucleosynthesis.

\texttt{THC\_M1} represents a step forward in the modeling of neutrinos
in mergers, particularly over long timescales over which diffusion of
neutrinos from the inner part of the remnant needs to be taken into
account. However, the present study still has some important limitations
to be addressed. Most importantly, our work used a rather crude set of
weak reactions and accounted for the energy-dependence of
neutrino-matter cross sections in a simplistic way. We plan to update
the set of weak reactions included in our code and to use
Planck-averaged opacities that take into account the average incoming
neutrino energy. We also plan to perform a larger campaign of
simulations spanning a range of binary masses, mass ratios, and
\acp{EOS}, in order to understand the general features of neutrino
driven winds from \ac{NS} mergers and the role of non-equilibrium
effects in the postmerger. Finally, our work has neglected quantum
kinetic effects in the neutrino transport \citep{Zhu:2016mwa,
Deaton:2018ser, Richers:2019grc, George:2020veu, Richers:2021xtf,
Li:2021vqj}. Future work should quantify the importance of these
effects for mergers.

%%%%%%%%%%%%%%%%%%%%%%%%%%%%%%%%%%%%%%%%%%%%%%%%%%
\section*{Data Availability}
Data generated for this study will be made available upon reasonable
request to the corresponding authors.

\section*{Acknowledgements}
We wish to thank Luke Roberts for help with \texttt{ZelmaniM1}, and
Aviral Prakash and Francesco Zappa for having beta-tested an early
version of \texttt{THC\_M1}.
We also thank Xinyu Li and Patrick Cheong for having spotted some typos
in an earlier version of the manuscript.
DR acknowledges funding from the U.S. Department of Energy, Office of
Science, Division of Nuclear Physics under Award Number(s) DE-SC0021177
and from the National Science Foundation under Grants No. PHY-2011725,
PHY-2020275, PHY-2116686, and AST-2108467.
SB acknowledges support from the EU H2020 under ERC Starting Grant,
no.BinGraSp-714626, and from the Deutsche Forschungsgemeinschaft, DFG,
project MEMI number BE 6301/2-1.
RH acknowledges funding from the National Science Foundation under Grants
No. OAC-2004879, OAC-1550514, and ACI-1238993.
NR simulations were performed on %
ARA cluster at Friedrich Schiller University Jena, SuperMUC-NG at the
Leibniz-Rechenzentrum (LRZ) Munich, HPE Apollo Hawk
at the High Performance Computing Center Stuttgart (HLRS);
Marconi-CINECA (ISCRA-B project HP10BMHFQQ, INF20\_teongrav and
INF21\_teongrav allocation); Bridges, Comet, Stampede2 (NSF XSEDE
allocation TG-PHY160025), NSF/NCSA Blue Waters (NSF AWD-1811236),
Joliot-Curie at GENCI@CEA (PRACE-ra5202) supercomputers.
This research used resources of the National Energy Research
Scientific Computing Center, a DOE Office of Science User Facility
supported by the Office of Science of the U.S.~Department of Energy
under Contract No.~DE-AC02-05CH11231. Computations for this research
were also performed on the Pennsylvania State University's Institute for
Computational and Data Sciences' Roar supercomputer.
The ARA cluster is funded in part by DFG grants INST
275/334-1 FUGG and INST 275/363-1 FUGG, and ERC Starting Grant, grant
agreement no. BinGraSp-714626.
The authors acknowledge the Gauss Centre for Supercomputing
e.V. (\url{www.gauss-centre.eu}) for funding this project by providing
computing time on the GCS Supercomputer SuperMUC-NG at LRZ
(allocation {\tt pn68wi}).
%


\bibliographystyle{mnras}

\input{paper20230509.bbl}
%%%%%%%%%%%%%%%%%%%%%%%%%%%%%%%%%%%%%%%%%%%%%%%%%%

%%%%%%%%%%%%%%%
%\appendix

%%%%%%%%%%%%%%%%%%%%%%%%%%%%%%%%%%%%%%%%%%%%%%%%%%

\input{acronyms.inc}

% Don't change these lines
\bsp	% typesetting comment
\label{lastpage}
\end{document}

%% file: acronyms.inc
\acrodef{ADM}{Arnowitt-Deser-Misner}
\acrodef{AMR}{adaptive mesh-refinement}
\acrodef{BH}{black hole}
\acrodef{BBH}{binary black-hole}
\acrodef{BHNS}{black-hole neutron-star}
\acrodef{BNS}{binary neutron star}
\acrodef{CCSN}{core-collapse supernova}
\acrodefplural{CCSN}[CCSNe]{core-collapse supernovae}
\acrodef{CMA}{consistent multi-fluid advection}
\acrodef{CFL}{Courant-Friedrichs-Lewy}
\acrodef{DG}{discontinuous Galerkin}
\acrodef{HMNS}{hypermassive neutron star}
\acrodef{EM}{electromagnetic}
\acrodef{ET}{Einstein Telescope}
\acrodef{EOB}{effective-one-body}
\acrodef{EOS}{equation of state}
\acrodef{FF}{fitting factor}
\acrodef{GR}{general-relativistic}
\acrodef{GRLES}{general-relativistic large-eddy simulation}
\acrodef{GRHD}{general-relativistic hydrodynamics}
\acrodef{GRMHD}{general-relativistic magnetohydrodynamics}
\acrodef{GW}{gravitational wave}
\acrodef{ILES}{implicit large-eddy simulations}
\acrodef{LIA}{linear interaction analysis}
\acrodef{LES}{large-eddy simulation}
\acrodefplural{LES}[LES]{large-eddy simulations}
\acrodef{MRI}{magnetorotational instability}
\acrodef{NR}{numerical relativity}
\acrodef{NS}{neutron star}
\acrodef{PNS}{protoneutron star}
\acrodef{SASI}{standing accretion shock instability}
\acrodef{SGRB}{short $\gamma$-ray burst}
\acrodef{SPH}{smoothed particle hydrodynamics}
\acrodef{SN}{supernova}
\acrodefplural{SN}[SNe]{supernovae}
\acrodef{SNR}{signal-to-noise ratio}
\acrodef{ZAMS}{zero age main sequence}